\documentclass[a4paper,10pt]{article}
 
 \usepackage{microtype}
\usepackage[utf8]{inputenc}
\usepackage[T1]{fontenc}
\usepackage{lmodern}
\usepackage{amsmath,amsthm,amsfonts,amssymb,bm}
\usepackage{dsfont}			% indicator function, \mathds{1}
\usepackage{enumitem}
\usepackage{booktabs}
\usepackage{multirow}
\usepackage{siunitx}
\usepackage{graphicx}
\usepackage{booktabs}
\usepackage[bottom]{footmisc} %footnotes

\usepackage{titlesec}

\titleformat{\paragraph}
{\normalfont\normalsize\bfseries}{\theparagraph}{1em}{}
\titlespacing*{\paragraph}
{0pt}{3.25ex plus 1ex minus .2ex}{1.5ex plus .2ex}
\newcommand{\floor}[1]{\lfloor #1 \rfloor}

\usepackage{natbib}
\usepackage[right=2cm,left=2cm,top=2cm,bottom=3cm]{geometry}
\usepackage[font={small,it}]{caption}	% the options makes the font italic in floats
\usepackage[colorlinks,linkcolor=blue,citecolor=blue,urlcolor=blue]{hyperref}
\usepackage{multirow}
\usepackage{subfig}
\usepackage{quoting}
\quotingsetup{font=it}
\usepackage{url}
\usepackage{tikz}             %draw bayesian nw
\usepackage{algorithm}
\usepackage[noend]{algpseudocode}
\makeatletter
\def\BState{\State\hskip-\ALG@thistlm}
\makeatother

\usetikzlibrary{bayesnet}     

\usepackage{DejaVuSansMono}
\usepackage[scaled=0.8]{beramono}

 \usepackage{array}
 \newcolumntype{L}[1]{>{\raggedright\let\newline\\\arraybackslash\hspace{0pt}}m{#1}}
\newcolumntype{C}[1]{>{\centering\let\newline\\\arraybackslash\hspace{0pt}}m{#1}}
 \newcolumntype{R}[1]{>{\raggedleft\let\newline\\\arraybackslash\hspace{0pt}}m{#1}}

\title{ \textsc{Inferring food intake from multiple biomarkers \\using a latent variable model} }
\author{Silvia D'Angelo$^1$, Lorraine Brennan$^2$ and Isobel Claire Gormley$^1$}
\date{%
    $^1$School of Mathematics and Statistics, Insight Centre for Data Analytics,\\ University College Dublin, Belfield, Dublin 4.\\
    $^2$School of Agriculture and Food Science, Conway Institute,\\ University College Dublin, Belfield, Dublin 4.\\[2ex]%
    \today
}

\begin{document}
\maketitle
\begin{abstract}
Metabolomic based approaches have gained much attention in recent years due to their promising potential to deliver objective tools for assessment of food intake. In particular, multiple biomarkers have emerged for single foods. However, there is a lack of statistical tools available for combining multiple biomarkers to quantitatively infer food intake. Furthermore, there is a paucity of approaches for estimating the uncertainty around biomarker-based inferred intake.

Here, to estimate the relationship between multiple metabolomic biomarkers and food intake in an intervention study conducted under the A-DIET research programme, a latent variable model, multiMarker, is proposed. The multiMarker model integrates factor analytic and mixture of experts models: the observed biomarker values are related to intake which is described as a continuous latent variable which follows a flexible mixture of experts model with Gaussian components. The multiMarker model also facilitates inference on the latent intake when only biomarker data are subsequently observed.  A Bayesian hierarchical modelling framework provides flexibility to adapt to different biomarker distributions and facilitates inference of the latent intake along with its associated uncertainty.  

Simulation studies are conducted to assess the performance of the multiMarker model, prior to its application to the motivating application of quantifying apple intake.
\end{abstract}


\section{Introduction}
\label{sec:intro}
Dietary biomarkers have emerged in recent years as objective measures of food intake (\cite{Baldrick2011}). A dietary biomarker is a small molecule called a metabolite that can provide information on the level of intake of a food \citep{Gao2017}. By definition a dietary biomarker increases with food intake and is validated against a defined set of criteria \citep{Dragsted2018}. In recent years, many biomarkers have emerged for a range of foods.
The importance of such biomarkers stems from the fact that classical dietary assessment approaches rely on self-reported data which can be subjective and biased, and such issues are well documented in the literature (e.g. \cite{Bingham2002}, \cite{Kipnis2002}, \cite{Subar2003}, \cite{Lloyd2019}, \cite{Siddique2019}). As a consequence, dietary biomarkers have emerged as a potential objective tool to aid food intake assessment. 
As the biomarker field progressed and the analytical tools improved, the number of biomarkers identified as potential biomarkers of intake has increased. Indeed, there are now multiple biomarkers for individual foods. However, there is a paucity of statistical tools for modelling the relationship between multiple biomarkers and food intake. The work to date has employed biomarkers as panels to classify intake into categories, e.g. consumers and non-consumers.
For example, \cite{Garcia-Aloy2017} propose a panel of biomarkers for cocoa intake, which is employed to estimate cocoa consumption or non-consumption through a forward stepwise logistic regression model. \cite{Rothwell2014} propose a partial least squares discriminant analysis (PLS-DA) approach to distinguish between low and high consumption of coffee in a sample of individuals, using a panel of three coffee-specific biomarkers. Recent work by \cite{Vazquez2019} on discovery and validation of banana intake biomarkers also employs a PLS-DA approach to detect low, medium and high consumption of banana. The proposed panel of biomarkers proved to be effective in distinguishing between low and high consumers, while medium consumers were difficult to separate from the two extremal groups. \cite{Gurdeniz2016} present a PLS-DA approach to detect beer intake (consumers versus non-consumers) using a panel of aggregated biomarkers. While the aforementioned approaches avail of a panel of biomarkers, they provide only a categorical quantification of intake, not an estimation of the quantity of intake, nor its associated uncertainty. 

The A-DIET research programme (\url{www.ucdnutrimarkers.com/a-diet}) aims to identify new metabolomic biomarkers of dietary intake, and here provides the motivating context. Data from a panel of four novel metabolomic biomarkers were collected from an intervention study, where a group of participants consumed three different food quantities (of apple), over a three week period. The question of interest and associated statistical challenge is twofold: given the intervention study data, the aim is to estimate the relationship between the biomarkers and food intake. At a later stage, when biomarker data only have been collected, the objective is to infer intake from the panel of biomarkers alone, and to provide the associated uncertainty.

Here, to estimate the relationship between multiple biomarkers and food intake we have developed the ``multiMarker'' model. MultiMarker relies on a factor analytic latent variable construct \citep{Knott1999} to capture the relationship between the panel of observed biomarkers and the unobserved intake. The distribution of the latent factor (i.e. the unobserved food intake) is expressed through a flexible mixture model. Moreover, in order to ``refine'' inferred intakes, multiMarker avails of a mixture of experts framework (\cite{Jacobs1991}, \cite{Gormley2019}), such that the biomarker data first informs the inferred level of intake (relative to the most likely food quantity), with the more refined inferred intakes then resulting from the factor analytic aspect of multiMarker. The model is developed in a Bayesian framework, naturally allowing for uncertainty quantification and therefore providing more informed quantification of intake.

Traditional factor analysis models assume that the correlation structure between a collection of observed variables can be represented in terms of a linear combination of a lower number of latent variables, the factors \citep{Knott1999}. 
The latent variable(s) are often assumed to follow a Gaussian distribution, and many extensions have been proposed. Allowing an infinite number of factors to facilitate greater modelling flexibility has been proposed by \cite{Bhattacharya2011}, and \cite{Murphy2020} extend the framework to an infinite mixture of infinite factor analysers.
When in the presence of an heterogeneous population, \cite{Montanari2010} proposed an heteroscedastic factor mixture analysis model, where factors are distributed according to a mixture of multivariate Gaussian distributions.
Relaxations of the Gaussianity assumption for the factors have been considered by many authors, see for example \cite{McLachlan2007}, \cite{Murray2014} and \cite{Lin2016}. An extension of the factor mixture analysis approach to multivariate binary response data has been considered by \cite{Cagnone2012}. \cite{Galimberti2009} propose an approach to dimensionality reduction in factor mixture analysis models, where factor loadings are shrunk through a penalized likelihood approach. Robustification of factor analysis models was addressed by \cite{Pison2003}, who introduced a method to address outliers. A further issue with factor analytic models is non-identifiability, see for example \cite{Lopes2004}, \cite{Rockova2016} and \cite{FS2018}.

While much factor analytic research has focused on two common issues, the number of factors to employ and the distribution to adopt, attempting to model the scale of the latent factors has received little attention. Typically the latent factors are perceived as instrumental tools to achieve a lower-dimensional representation of the data at hand. On the contrary, motivated by the application, our approach is focused on a single latent
variable that is a proxy for the latent intake. Information on the scale of this latent variable is available and necessary in order to provide practically useful quantification of intake. While the availability of such information also ensures the multiMarker model is identifiable, unlike general factor analytic models, it adds complexity to inferring the latent variable which is no longer a simple instrument. Modelling the latent factor via a mixture model allows for a flexible framework, but introduces the issue of properly modelling mixture weights. Indeed, mixture components have the role of locating different regions in the intake range for the latent variable, where the order of such regions is relevant and should not be ignored. To this end, we embed the model for the latent variable in a mixture of experts framework, where the weights are informed by the observed biomarkers. Further, when modelling the weights we directly account for the ordinal nature of the food quantities via an ordinal regression model \citep{Agresti1999}, employing the robust Cauchit link function (\cite{Morgan1992}). Given subsequently observed biomarker data only, inference on the latent intake and its associated uncertainty is available through the latent variable's posterior predictive distribution. 

In what follows, Section \ref{sec:appleIntro} details the motivating application of inferring food (specifically apple) intake. Section \ref{sec:modello} outlines the multiMarker model, while Section \ref{sec:MCMC} develops an efficient Metropolis-within-Gibbs sampling strategy for Bayesian inference. Section \ref{sec:sim} provides details of thorough simulation studies exploring the performance of the proposed model across a series of realistic settings. The multiMarker model is applied in the motivating context of inferring apple intake in Section  \ref{sec:apples}, with the concluding Section \ref{sec:discussion} discussing the application outcomes, the multiMarker model and possible extensions. An R \citep{R2020} package, multiMarker \citep{multiMarker2020}, is freely available to facilitate widespread use of the method, and with which all results presented herein were produced.  

\section[Modelling the relationship between food intake and multiple biomarkers:
apple intake as an example]{\texorpdfstring{Modelling the relationship between food intake and multiple\\biomarkers: apple intake as an example}{Modelling the relationship between food intake and multiple biomarkers: apple intake as an example}}
\label{sec:appleIntro}
In the present work we develop a model to estimate the relationship between apple intake and $P = 4$ urinary biomarkers, identified using an untargeted metabolomics approach \citep{McNamara2020}. The four urinary biomarkers are: Xylose, Epicatechin Sulfate, $[($4-$\{$3-[2-(2,4-dihydroxyphenyl)-2-oxoethyl]-4,6-dihydroxy-2-methyoxyphenyl$\}$-2-methylbut-2-en-1-yl)oxy] sulfonic acid, and Glucodistylin. Throughout the paper we will refer to the third biomarker as $(4-3-[2-(2,4-dihydroxyphenyl)-2-oxoethyl]-DHMPMB-SA)$.

The data were collected as part of an intervention study, where a group of $32$ participants consumed different quantities of apple daily, over a three week intervention period. The intervention study was conducted as part of the A-DIET research programme (\url{www.ucdnutrimarkers.com/a-diet}), which aims to identify new metabolomic biomarkers of dietary intake. Data were available following consumption of $D = 3$ apple quantities: $50$, $100$ and $300$ grams. The participants were fed the different food quantities in each of three intervention weeks, where each intervention week was followed by a resting week. Out of the $32$ participants, $29$ consumed $50$ grams of apple, while respectively $28$ and $29$ ate $100$ and $300$ grams. Throughout the paper we treat the total of $n=86$ recorded intakes as independent observations and this assumption is assessed in Section \ref{sec:apples}. 

\begin{figure}[!h]
\begin{center}
  \includegraphics[scale=.39]{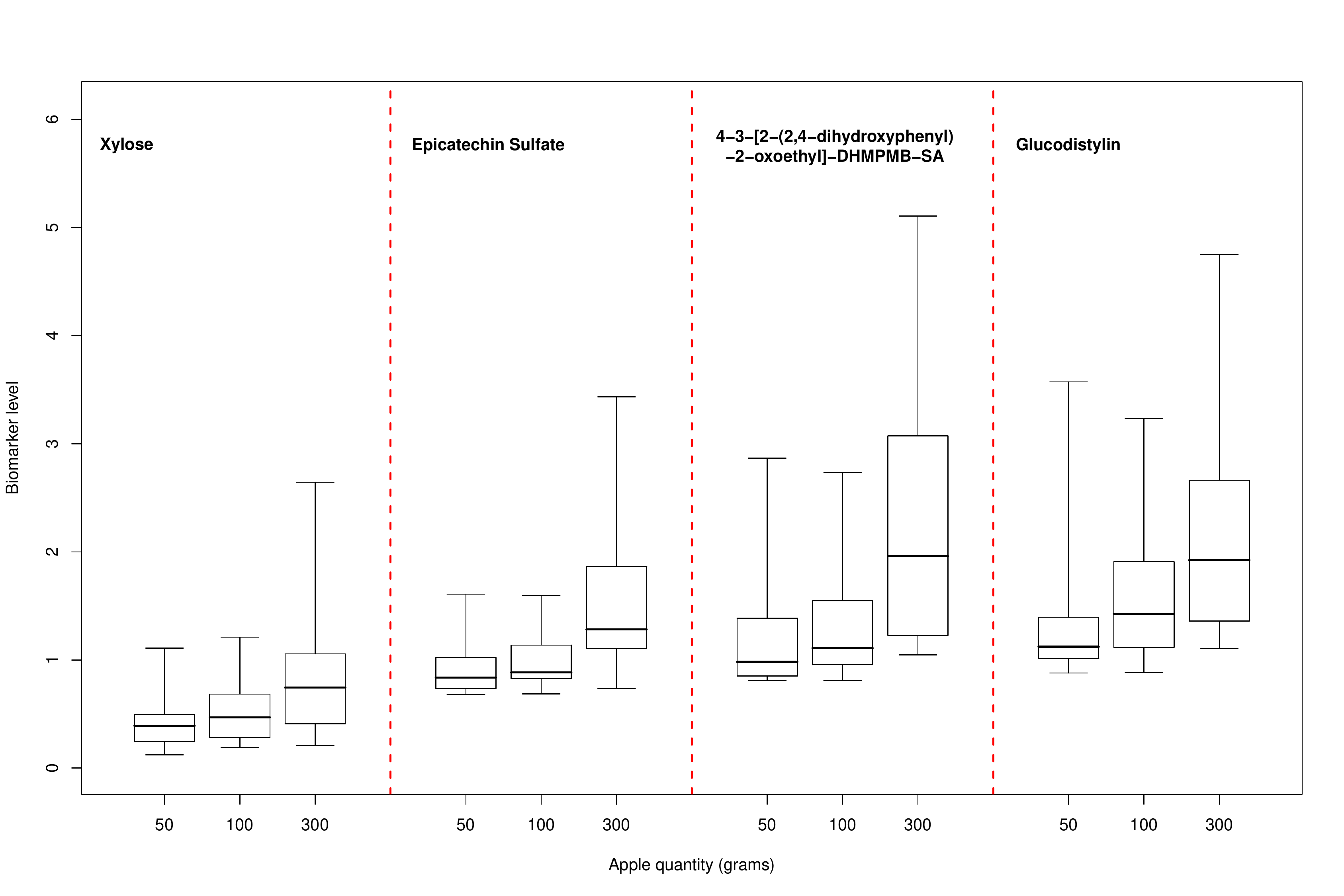}
\end{center}
    \caption{Biomarker levels following consumption of different quantities of apple. Boxplots for the $n=86$ observations on the four biomarkers, for the three apple quantities ($50$, $100$ and $300$ grams).}
    \label{fig:applesdata0}
\end{figure}

The original values for Epicatechin Sulfate, $(4-3-[2-(2,4-dihydroxyphenyl)-2-oxoethyl]-DHMPMB-SA)$ and Glucodistylin caused computational instability (most values were larger than $10^7$) and consequently were scaled (see Section 3.1 of the Supplementary Material). However, the transformation did not alter the correlations between the four biomarkers. A visualization of the data is given in  Figure \ref{fig:applesdata0}. The first three urinary biomarkers have similar median values for the first two apple quantities, and all biomarkers are highly variable for the 300 grams apple quantity. The boxplots corresponding to different apple quantities partially overlap, indicating that such quantities cannot be easily discriminated.

Given the intervention study data, we estimate the relationship between the panel of biomarkers and apple intake. We then infer the apple consumption (in grams) of new observations, for which only the metabolomic biomarker measurements are available.

\section{The multiMarker model}
\label{sec:modello}
To quantify food (in this case apple) intake from a panel of biomarkers we propose the multiMarker model which integrates factor analytic and mixture of experts models. The model estimates the relationship between the multiple biomarkers and food intake and facilitates inference on the latent intake when only biomarker data are subsequently observed.

\subsection{Model specification}
\label{sec:modello_dettagli}
Consider a biomarker matrix $Y$ of dimension $n \times P$, storing  $P$ different biomarkers measurements on a set of $n$ independent observations. 
The $D$ food quantities considered in the intervention study are denoted by $\mathbf{X}=\{X_1, \dots, X_d, \dots, X_D\}$. 
Elements in $\mathbf{X}$ are ordered such that $X_d < X_{d+1}$. 
We assume that the biomarker measurements are related to an unobserved continuous intake value, for which the food quantities are proxies, leading to the following factor analytic model:
\begin{equation}
y_{ip} = \alpha_p + \beta_p z_i +\epsilon_{ip}, \quad \forall \quad  i=1,\dots,n, \quad p = 1, \dots,P,
    \label{eq:marker_dose}
\end{equation}
where the (one dimensional) latent variable $z_i$ denotes the latent intake of observation $i$.
The $\alpha_p$ and $\beta_p$ parameters characterize, respectively, the intercept and the scaling effect for the $p^{\text{th}}$ biomarker. We assume a truncated Gaussian prior distribution, $\alpha_p \thicksim \mathcal{N}_{[0, \infty)}\bigl(\mu_{\alpha}, \sigma_{\alpha}^2 \bigr)$, for the intercept parameters. The non-negative assumption is required as biomarker levels cannot be negative. 
The $p^{\text{th}}$ scaling parameter $\beta_p$ captures the effect of an increment in consumption of a given food on the observed level of biomarker $p$. As validated biomarkers are known to be reflective of intake, we assume $\beta_p >0$, $\forall p = 1,\dots,P$ and place a truncated Gaussian prior distribution on  $\beta_p$ i.e. $ \beta_p \thicksim \mathcal{N}_{(0, \infty)}\bigl(\mu_{\beta}, \sigma_{\beta}^2 \bigr)$. 

The error term $\epsilon_{ip}$ is the variability associated with the $p^\text{th}$ biomarker. As is common in factor analytic models, we assume that  $\epsilon_{ip}\thicksim \mathcal{N}(0, \sigma_p^2)$, where $\sigma_p^2$ serves as a proxy for the precision of the $p^\text{th}$ biomarker. A precise biomarker will have a value of $\epsilon_{ip}$ close to zero but may take positive or negative values and thus $\epsilon_{ip} \in (-\infty, \infty)$. However, to guarantee non-negative biomarker values, the further assumption $y_{ip} \thicksim\mathcal{N}_{[0,\infty)}(\alpha_p + \beta_p z_i, \sigma_p^2)$ is required. Here, differently from standard factor analytic models, the scale of the latent variable plays a central role, and its accurate, quantitative recovery is a central requirement of the motivating application.

The likelihood function conditional on $\mathbf{z}=(z_1,\dots,z_n)^{T}$ is
\begin{equation}
  \ell (\mathbf{Y} | \alpha, \beta, \mathbf{z}, \Sigma) 
 = C + \sum_{p=1}^P\sum_{i=1}^n \Biggl( -\frac{1}{2} \log\bigl( 2\pi\sigma_p^2\bigr) -\frac{1}{2\sigma_p^2} \bigl( y_{ip} - \alpha_p -\beta_pz_i\bigr)^2 \Biggr)
    \label{eq:likel}
\end{equation}
where $\Sigma = diag(\sigma_1^2, \dots, \sigma_P^2)$ and $C$ is a constant. 

\subsection{Modelling the latent intakes}
\label{sec:lat_int}

Although the quantities of interest $\mathbf{z} = (z_1,\dots,z_{n})^{T}$ cannot be directly measured, food quantities from the intervention study are available and are employed to inform the distribution of the latent intakes. While the specific food quantity consumed by each intervention study observation is known, here this information is not directly modelled, but is used to inform hyperparameter settings and starting values when estimating the multiMarker model (see Supplementary Material, Section 1.1). This approach ensures that the same structural model is used when both estimating the relationship between the biomarkers and intake from the intervention study observations' data, and when inferring intake for new observations for which biomarker data only are available.

To specify a prior distribution for the latent intake, we assume that the true but latent intakes of the intervention study's observations will be distributed around the food quantities used in the study. Thus we assume a mixture of experts model with $D$ truncated Gaussian  components as the prior distribution for the latent intake
\begin{equation}
z_i \thicksim \sum_{d=1}^D \pi_{id} \mathcal{N}_{[0,\infty)}\bigl( X_d, \theta_d^2 \bigr), \quad \pi_i= \{ \pi_{i1}, \dots, \pi_{iD}\},
\label{eq:zprior}
\end{equation}
for $i=1,\dots,n$ and $d=1,\dots, D$. The use of truncated distributions follows naturally from the definition of intake, which is non-negative. 
The $d^{\text{th}}$ component in the mixture of experts model is centred at the $d^{\text{th}}$ food quantity $X_d$. The variances $\Theta=\{\theta_1^2, \dots, \theta_D^2\}$ represent intake variability around the intervention study's food quantities; in such intervention studies the range of food quantities considered is carefully chosen to realistically reflect feasible intake amounts.
Finally, the observation-specific mixture weights are assumed to be a function of the observed biomarker values, $\pi_{id} = \pi_{id}(y_i)$, ensuring that estimation of the relationship between the biomarkers and food intake, and the subsequent inference of food intake from biomarkers alone, are derived from the same model, philosophically and structurally.

As is standard in mixture models, an unobserved component allocation label $c_i$ is introduced for observation $i$ and $\mathbf{c} = \{ c_1, \ldots, c_n\}$.  These latent auxiliary variables indicate the food quantity/component to which each observation belongs, depending on the corresponding biomarker measurements. Formally, $c_i \sim \mbox{Multinomial}\{1, \pi_i = (\pi_{i1}, \ldots, \pi_{iD})\}$. For more details on $\mathbf{c}$ see Section \ref{sec:model_weights}. The mixture model in (\ref{eq:zprior}) can then be written according to its complete data representation:
\begin{equation}
z_i \mid c_i \thicksim \prod_{d=1}^D \Big\{ \mathcal{N}_{[0,\infty)} (X_d, \theta_d^2)\Big\}^{[c_{i}=d]}
\label{eq:zpriorC}
\end{equation}
where $[\cdot]$ is the Iverson bracket ($[A]$ returns $1$ if the proposition $A$ is true, and $0$ otherwise).

\subsubsection{Modelling observation-specific weights}
\label{sec:model_weights}
Given the inherent ordering of the food quantities $X_1, \dots, X_D$ in the intervention study, and hence the ordering in the mixture components, we employ an ordinal regression model with Cauchit link function to model the observation-specific weights. The Cauchit link function is more robust than the standard logit or probit link functions and thus is suited to the often highly variable biomarker data. 
Specifically, we assume:
\begin{equation}
\begin{split}
p(\pi\mid \gamma, &\eta, \mathbf{c}, \mathbf{Y}, \mathbf{X}) 
= \prod_{i =1}^n\prod_{d=1}^D \pi_{id}^{[c_i=d]}
=
\prod_{i =1}^n\prod_{d=1}^D \pi_{id}(y_i|   \gamma_d, \eta)^{[c_i=d]}\\
& = \prod_{i =1}^n \prod_{d=1}^D \Bigl[Pr(x_i \leq X_d\mid \gamma_{d}, \eta, y_i) - Pr(x_i \leq X_{d -1} \mid \gamma_{d-1}, \eta,  y_i) \Bigr]^{[c_i=d]}
\end{split}
    \label{eq:weights}
\end{equation}
where $Pr(x_i \leq X_d\mid \gamma_{d}, \eta, y_i) =  \frac{1}{\pi} \Bigl[  \arctan{  \Bigl(\frac{1}{2} (\gamma_d + \eta y_i) \Bigr)+ \frac{\pi}{2}}\Bigr]$ is the probability that observation $i$'s consumed quantity $x_i$ is less than or equal to the food quantity $X_d$, given its $P$ biomarker measurements $y_i$, expressed through a Cauchit link. 
The vector $\gamma=\{-\infty = \gamma_0, \gamma_1, \dots,\gamma_d,\dots,\gamma_D= \infty\}$ contains food quantity-specific intercepts. The $\eta$ parameter is a $P$ dimensional scaling coefficient for the biomarkers, expressing the contribution of each to the determination of the observation-specific mixture weights.

\section{An MCMC algorithm for parameter estimation}
\label{sec:MCMC}
We adopt a Bayesian approach to estimate the hierarchical model's parameters, implemented through a Metropolis within Gibbs Markov chain Monte Carlo (MCMC) algorithm. 
Hyperprior distributions are assumed on the prior parameters with the corresponding hyperparameter values fixed based on the data at hand, following an empirical Bayes approach. Hyperparameter specifications are reported in the Supplementary Material (Section 1.1). Figure \ref{fig:modelloClust} provides a graphical representation of the multiMarker model, with the steps of the MCMC algorithm detailed below.
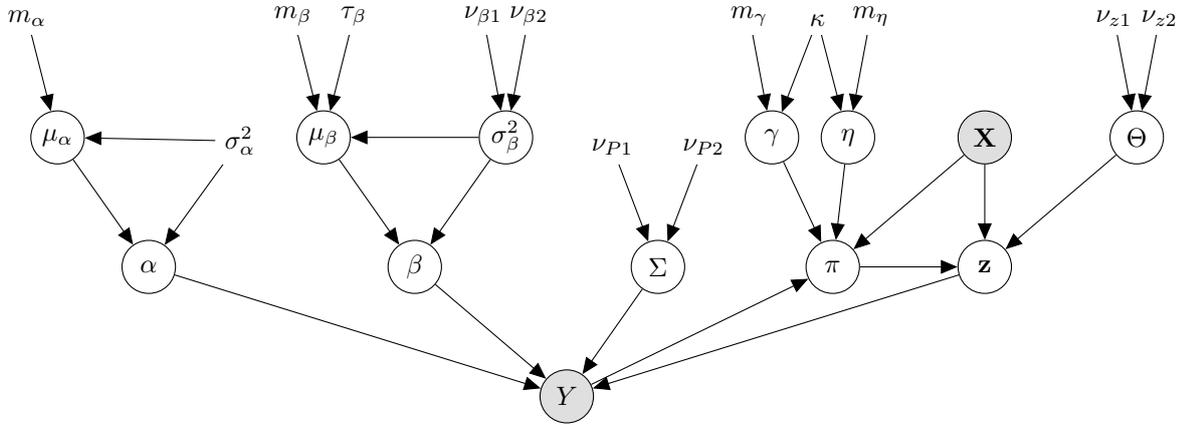
\begin{figure}[ht]
\scalebox{1}{
    \begin{tabular}{cc}
      \begin{tikzpicture}

  % Define nodes
  \node[obs]                               (y) {$Y$};
  \node[latent, above=of y, xshift=-5.5cm] (a) {$\alpha$};
  \node[latent, above=of y, xshift=-2cm] (b) {$\beta$};
  \node[latent, above=of y, xshift=5.5cm] (z) {$\mathbf{z}$};
  \node[latent, above=of y, xshift=1.2cm] (sp) {$\Sigma$};
  
  \node[latent, above=of a, xshift=-1.2cm]  (ma) {$\mu_{\alpha}$};
  \node[above=of a, xshift=1.2cm]  (sa) {$\sigma_{\alpha}^2$}; 
  \node[latent, above=of b, xshift=-1.2cm]  (mb) {$\mu_{\beta}$};
  \node[latent, above=of b, xshift=1.2cm]  (sb) {$\sigma_{\beta}^2$}; 
 
    \node[latent, above=of z, xshift=2cm]  (sz) {$\Theta$}; 
 % \node[latent, above=of sz, xshift=-1cm]  (tauz) {$\tau_z$};

  \node[latent,above=of y, xshift=3.5cm]  (D) {$ \pi$};

  \node[above=of sp, xshift=-.6cm]  (t1) {$\nu_{P1}$};
  \node[above=of sp, xshift=.6cm]  (t2) {$\nu_{P2}$}; 
  \node[obs,above=of z]  (xd) {$\mathbf{X}$};

 \node[above=of ma, xshift=-.4cm]  (mma) {$m_{\alpha}$};
 \node[above=of mb, xshift=-.4cm]  (mmb) {$m_{\beta}$};
 \node[above=of mb, xshift=.4cm]  (taub) {$\tau_{\beta}$};
  \node[above=of sb, xshift=-.3cm]  (nu1b) {$\nu_{\beta 1}$};
  \node[above=of sb, xshift=.3cm]  (nu2b) {$\nu_{\beta 2}$};
 \node[above=of sz, xshift=-.3cm]  (nuz1) {$\nu_{z1}$};
  \node[above=of sz, xshift=.3cm]  (nuz2) {$\nu_{z2}$};
\node[latent, above=of D, xshift=-.8cm]  (Gammastima) {$\gamma$};
 \node[latent, above=of D, xshift=.2cm]  (Etastima) {$\eta$};
 \node[ above=of Etastima, xshift=.3cm]  (etam) {$m_{\eta}$};
\node[ above=of Gammastima, xshift=-.3cm]  (gammam) {$m_{\gamma}$};
\node[ above=of Gammastima, xshift=.6cm]  (Kappa) {$\kappa$};
  % Connect the nodes

  % Connect the nodes
  \edge {a} {y} ; %
  \edge {b} {y} ; %
  \edge {z} {y} ; %
  \edge {sp} {y} ; %
  \edge {D} {z} ; %
  \edge {t1} {sp} ; %
  \edge {t2} {sp} ; %
  \edge {ma} {a} ; %
  \edge {mb} {b} ; %
  \edge {sa} {a} ; %
  \edge {sb} {b} ; %
  \edge {sa} {ma} ; %
  \edge {sb} {mb} ; %
   \edge {sz} {z} ; %
 % \edge {xd} {tauz} ; %
\edge  {xd}{z};
\edge  {xd}{D};
%\edge  {tauz}{sz}

  \edge {mma} {ma} ; %
\edge{Gammastima}{D};
\edge{Kappa}{Gammastima};
\edge{Kappa}{Etastima};
\edge{etam}{Etastima};
\edge{gammam}{Gammastima};
\edge{Etastima}{D};
  \edge {mmb} {mb} ; %
  \edge {taub} {mb} ; %
  
\edge{y}{D};
  \edge {nu1b} {sb} ; %
  \edge {nu2b} {sb} ; %
  \edge {nuz1} {sz} ; %
 \edge {nuz2} {sz} ; %
\end{tikzpicture} 
    \end{tabular}
    }
   \caption{Hierarchical structure of the multiMarker model. Shaded circles represent the data. Parameters and latent variables are indicated with transparent circles and hyperparameters are indicated with no circles.}
\label{fig:modelloClust}
\end{figure}
\subsection{Biomarker-specific regression parameters}
For the $p^{\text{th}}$ biomarker, the truncated Gaussian prior distributions of both $\alpha_p$ and $\beta_p$  are combined with the likelihood function leading to the full conditional distributions of $\alpha_p$ and $\beta_p$:
\[
\alpha_p \thicksim \mathcal{N}_{[0, \infty)} \bigl( \mu_{\alpha_p}^*, \sigma_{\alpha_p}^{2*}\bigr); \qquad \beta_p \thicksim \mathcal{N}_{(0, \infty)} \bigl( \mu_{\beta_p}^*, \sigma_{\beta_p}^{2*}\bigr)
\]
where
\[
 \sigma_{\alpha_p}^{2*} = \frac{\sigma_p^2 \sigma_{\alpha}^2}{n\sigma_{\alpha}^2 + \sigma_p^2};  \qquad \mu_{\alpha_p}^* =  \sigma_{\alpha_p}^{2*}\Biggl[\frac{\sum_{i=1}^n \bigl( y_{ip} -\beta_pz_i \bigr)}{\sigma_p^2} +\frac{\mu_{\alpha}}{\sigma_{\alpha}^{2}} \Biggr]; \]
 \[
 \sigma_{\beta_p}^{2*} = \frac{\sigma_p^2 \sigma_{\beta}^2}{\sigma_{\beta}^2 \sum_{i=1}^n z_i^2 + \sigma_p^2}; \qquad \mu_{\beta_p}^* =  \sigma_{\beta_p}^{2*}\Biggl[\frac{\sum_{i=1}^n z_i\bigl( y_{ip} -\alpha_p\bigr)}{\sigma_p^2} +\frac{\mu_{\beta}}{\sigma_{\beta}^{2}} \Biggr]
\]
\subsection{Error term's variance parameter}
Assuming an inverse gamma prior distribution for the $p^{\text{th}}$ variance term $\sigma_p^2$, with shape parameter $\nu_{P1}$ and scale parameter $\nu_{P2}$, leads to the following full conditional distribution:
\[
\sigma_p^2 \thicksim Inv\Gamma \Biggl(\nu_{P1}^{*}= \frac{n}{2} +\nu_{P1}, \quad\nu_{P2}^{*}=\quad \frac{1}{2} \sum_{i=1}^n (y_{ip} -\alpha_p -\beta_p z_i)^2 + \nu_{P2}\Biggr)
\]

\subsection{Nuisance parameters}
To allow for the uncertainty in the parameters of the prior distributions of $\alpha_p$ and $\beta_p$, hyperprior distributions are specified for $\mu_{\alpha}$, $\mu_{\beta}$ and $\sigma_{\beta}^2$. Specifically, $\mu_{\alpha} \thicksim \mathcal{N}_{[0,\infty)} \bigl( m_{\alpha}, \tau_{\alpha}\sigma_{\alpha}^2 \bigr)$, $\mu_{\beta} \thicksim \mathcal{N}_{[0,\infty)} \bigl( m_{\beta}, \tau_{\beta}\sigma_{\beta}^2 \bigr)$, and $ \sigma_{\beta}^2 \thicksim Inv\Gamma \bigl(\nu_{\beta 1},\nu_{\beta 2} \bigr)$, leading to the following full conditionals:
\[
\mu_{g}  \thicksim \mathcal{N}_{[0,\infty)}  \bigl( \mu_g^*, \sigma_g^{2*} \bigr), \quad
 \sigma_{\beta}^2  \thicksim Inv \Gamma \bigl(\nu_{\beta1}^*, \nu_{\beta2}^* \bigr), 
\]
with
\[
 \sigma_g^{2*} = \frac{\tau_{g} \sigma_{g}^2}{ \tau_{g} P +1 }, \quad \mu_g^* = \sigma_g^{2*} \Bigl[ \frac{\tau_{g} \sum_{p=1}^P g_p +m_g}{\tau_g\sigma_g^2}\Bigr],
\]
\[
\nu_{\beta1}^* = \frac{P+1+2\nu_{\beta1}}{2}, \quad \nu_{\beta2}^* = \nu_{\beta2} +\frac{\tau_\beta \sum_{p=1}^P (\beta_p -\mu_{\beta})^2 +(\mu_{\beta} -m_{\beta})^2}{2 \tau_\beta},
\]
where $g = \alpha, \beta$. The set of hyperparameters $\bigl(m_{\alpha},m_{\beta}, \tau_{\alpha},\tau_{\beta}, \nu_{\beta 1},\nu_{\beta 2} \bigr)$ are fixed in advance, some of which are informed by the observation-specific food quantity information from the intervention study. Practical guidelines are outlined in the Supplementary Material (Section 1.1). Finally, we fix the value of $\sigma_{\alpha}^2=1$, for identifiability.

\subsection{Observation-specific weight parameters}
\label{sec:4.4}
The prior distributions for the $\gamma=\{\gamma_0, \gamma_1, \dots, \gamma_d\}$ parameters represent the constrained characteristic of ordinal data, that is: $X_1 < \dots < X_d < \dots < X_D$. These constraints correspond to the following in terms of model parameters: $\gamma_0 < \dots < \gamma_d < \dots < \gamma_D$. Thus, the prior distribution for $\gamma_d$ is taken to be $\gamma_d\thicksim \mathcal{N}_{(m_{\gamma_{d-1}}, m_{\gamma_{d+1}})}(m_{\gamma_d}, \kappa)$, for $ d=1,\dots,D-1$.
Biomarkers' intercepts' $\eta=\{\eta_1, \dots, \eta_P\}$ prior distribution is $\eta_p \thicksim \mathcal{N}(m_{\eta_p}, \kappa)$, for $ p=1,\dots,P$.

Both $\gamma$ and $\eta$ parameters are updated via a Metropolis Hastings step using random walk proposal distributions. Hyperparameters $m_{\gamma}$ and $m_{\eta}$, where $m_{\gamma} = \{m_{\gamma_{0}}, m_{\gamma_{1}}, \dots, m_{\gamma_{D-1}}\}$ and $m_{\eta} = \{ m_{\eta_{1}}, \dots, m_{\eta_{P}} \}$, are set via their corresponding estimates from the ordinal regression of the known observation-specific component allocation labels on biomarker measurements from the intervention study. More details on the hyperparameters, and on $\gamma$ and $\eta$ initialization are deferred to the Supplementary Material (Section 1.2).

Given current $\mathbf{z}$, $\theta_d^2$, $\gamma$ and $\eta$ values, weights are updated using (\ref{eq:weights}), together with the unobserved observation-specific component allocation labels: 
$c_i = \arg \max_d \frac{\pi_{id}\mathcal{N}_{[0,\infty)} (z_i \mid X_d, \theta_d^2 )}{\sum_{d=1}^D\pi_{id}\mathcal{N}_{[0,\infty)} (z_i\mid X_d, \theta_d^2)}
$.

\subsection{Latent intakes}
\label{sec:lat_int_fc}
As is common in latent variable models, following (\ref{eq:zpriorC}), we sample the $i^{\text{th}}$ observation's latent intake value from $z_i \mid c_i = d,\dots \thicksim \mathcal{N}_{[0,\infty)} \bigl( \mu_{id}^*, \theta_{id}^{2*}\bigr)$, with
\[
\theta_{id}^{2*} = \Biggl( \sum_{p=1}^P \frac{\beta_p^2 \theta_d^2 +\sigma_p^2/P}{ \theta_d^2 \sigma_p^2} \Biggr)^{-1}, \qquad
 \mu_{id}^* = \sigma_{id}^{2*} \Biggl[\sum_{p=1}^P \frac{\beta_p\bigl( y_{ip} -\alpha_p\bigr)}{\sigma_p^2} + \frac{X_{d}}{\theta_{d}^{2}}  \Biggr].
\]
Further, assuming an inverse gamma prior distribution on the components' variance parameters, $\theta_d^2 \thicksim Inv \Gamma \bigl( \nu_{z1}, \nu_{z2}\bigr)$, the full conditional distribution is $\theta_d^2 \thicksim Inv \Gamma \bigl( \nu_{z1}^*, \nu_{z2}^*\bigr)$, with
\[
\nu_{z1}^*  = \frac{n_d }{2} + \nu_{z1}, \quad \nu_{z2}^* = \nu_{z2} + 
\frac{\sum_{i =1}^n [c_i=d](z_i -x_d)^2  }{2},
\]
where $n_d = \sum_{i=1}^n [c_i=d]$. 

\subsection{Latent intakes' posterior predictive distribution}
The model defined by (\ref{eq:likel}) and (\ref{eq:zpriorC}) facilitates estimation of the relationship between biomarkers and latent intake, taking advantage of the information on food quantities used in an intervention study. We are then interested in inferring latent intake values having observed only biomarker measurements for a new group of $n^*$ observations. Such inference is available through the latent variable's posterior predictive distribution. 
Denoting by $z_j^{*}$ the $j^{\text{th}}$ new observation's latent intake, and by $y_j^{*}=(y_{j1}^{*},.\dots,y_{jP}^{*})^{T}$ their corresponding biomarker measurements, following 
(\ref{eq:likel}), (\ref{eq:zpriorC}) and (\ref{eq:weights}), the $j^{\text{th}}$ latent intake's sampling distribution is:
\begin{equation}
\begin{split}
p(z_j^*\mid y_j^*,\Omega) &= \mathcal{N}_{[0,\infty)}(\mu_{zj}, \sigma_{zj}^2) \sum_{d=1}^D \pi_{jd} \mathcal{N}_{[0,\infty)}(X_d, \theta_d^2)\\
& =  \sum_{d=1}^D \pi_{jd}(y_j^*| \gamma_d, \eta) \mathcal{N}_{[0,\infty)}\Bigl[\frac{\mu_{zj}\theta_d^2 + X_d\sigma_{zj}^2}{\sigma_{zj}^2 + \theta_d^2} , \bigl( \frac{1}{\theta_d^2}+\frac{1}{\sigma_{zj}^2} \bigr)^{-1}\Bigr]
\end{split}
\label{eq:zsamplingPredW}
\end{equation}
where for brevity $\sigma_{zj}^2 = \bigl( \sum_{p=1}^P\frac{\beta_p^2}{\sigma_p^2} \bigr)^{-1}$, $\mu_{zj} = \sigma_{zj}^2  \bigl( \sum_{p=1}^P   \frac{\beta_p(y_{jp}^{*} - \alpha_p)}{\sigma_p^2 }\bigr)$  and $\Omega= \{ \alpha, \beta, \Sigma, \mathbf{X}, \Theta, \eta, \gamma \}$.

As in Section \ref{sec:lat_int}, let $c_j^*$ be the unobserved $j^{\text{th}}$ observation-specific component allocation label. Hence (\ref{eq:zsamplingPredW}) can be written in its complete data representation:
\begin{equation}
p(z_j^*\mid y_j^*, c_j^*,\Omega) = \prod_{d=1}^D \Bigl[  \mathcal{N}_{[0,\infty)}\Bigl(\frac{\mu_{zj}\theta_d^2 + X_d\sigma_{zj}^2}{\sigma_{zj}^2 + \theta_d^2} , \bigl( \frac{1}{\theta_d^2}+\frac{1}{\sigma_{zj}^2} \bigr)^{-1}\Bigr)\Bigr]^{[c_j^*=d]}
\label{eq:zsamplingPredW1}
\end{equation}
Thus, the posterior predictive distribution of latent intake $z_j^*$, given the observed biomarker data $y_j^*$ is:
\begin{equation}
p(z_j^{*}\mid y_j^*, c_j^*) \propto 
\int p(z_j^{*} \mid y_j^*,  c_j^*,\Omega) p(\Omega \mid \mathbf{Y})  \,d\Omega
    \label{eq:posteriorpred}
\end{equation}
Note that the conditional distribution in equation \ref{eq:posteriorpred} depends on both the intervention study's biomarker data $\mathbf{Y}$ and new observation $j$'s biomarker measurements $y_j^{*}$. Hence, intuitively, to infer latent intake values for $n^*$ new observations for which biomarker data only are available, one needs to have first estimated the multiMarker model on some intervention study data. Moreover, the $P$ biomarkers measured for the $n^*$ new observations must be the same as the ones used in the intervention study.  

Given $P$ biomarker measurements $y_j^*$, latent intake values $z_j^{*}$ can be sampled from the posterior predictive (\ref{eq:posteriorpred}) in two steps. 
First, relevant model parameters, $\Omega$, are sampled from their estimated full conditional and proposal distributions, given observed $y_j^*$ values and current $z_j^*$, $c_j^*$ values, for $j=1\dots,n^*$. Then, given the updated parameter values $\Omega$, observation-specific weights $\pi_{jd}$ are computed together with the unobserved component-allocation labels
$c_j^* = \arg \max_d \frac{\pi_{jd}\mathcal{N}_{[0,\infty)} (z_j^* \mid X_d, \theta_d^2 )}{\sum_{d=1}^D\pi_{jd}\mathcal{N}_{[0,\infty)} (z_j^*\mid X_d, \theta_d^2)}$, for $d=1\dots, D$ and $j=1,\dots,n^*$. Finally, the latent intake value $z_j^{*}$ is sampled from (\ref{eq:zsamplingPredW1}), conditioning on the current $c_j^*$ value and model parameters $\Omega$.
The choice of sampling latent intakes from (\ref{eq:zsamplingPredW1}) rather than from (\ref{eq:zsamplingPredW}) is driven by the fact that the latter implies averaging across the whole mixture distribution to draw $z_j^{*}$, leading to inferred intake values close to the food quantities' sample mean $\bar{X}_d = \frac{1}{D}\sum_{d} X_d$. This leads to poorly inferred $z_j^{*}$ values, particularly for observations that truly belong to components located near small or large food quantities. Thus (\ref{eq:zsamplingPredW1}), which conditions on the current component allocation labels, is employed as the sampling distribution for the latent intakes in the MCMC algorithm.

\section{Simulation studies}
\label{sec:sim}
Several simulation scenarios have been constructed to explore the performance of the proposed approach. The term ``training data'' refers to simulated intervention study data employed to estimate the multiMarker model while ``test data'' refers to simulated biomarker only data, from which intake is to be inferred. To depict common experimental settings, the number of biomarkers considered is $P=4$. To represent a variety of real-world scenarios, different food quantity values and sample sizes have been considered. Training datasets' sample sizes are $n=\{30, 60, 99, 150\}$, representing a low sample size scenario to a larger sample size one. The test datasets have $n^*=\floor{0.4\times n}$ observations.

\begin{figure}
    \centering
    \subfloat[][\emph{Scenario 1.}]
{\includegraphics[width=.52\textwidth]{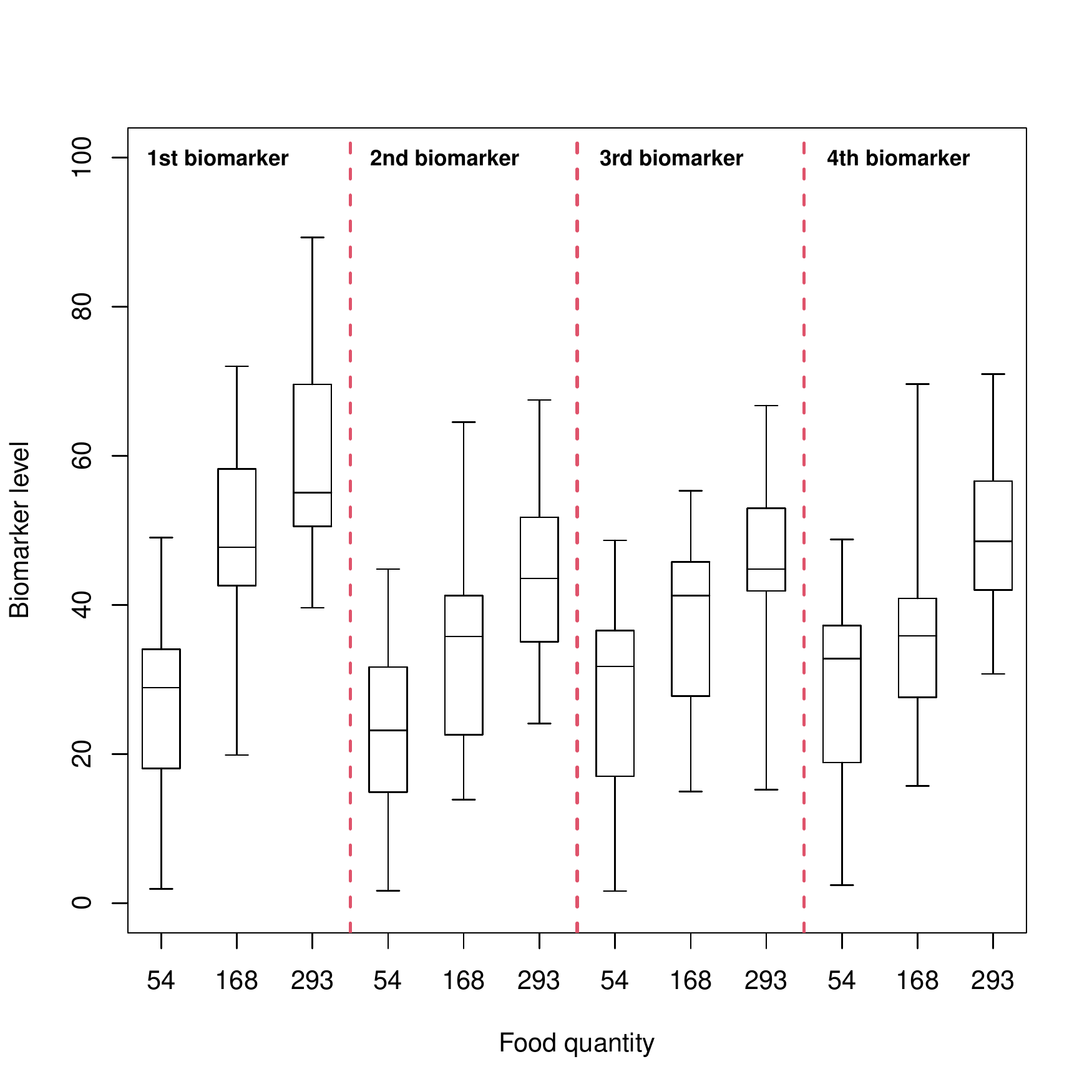}}
\subfloat[][\emph{Scenario 3.}]
{\includegraphics[width=.52\textwidth]{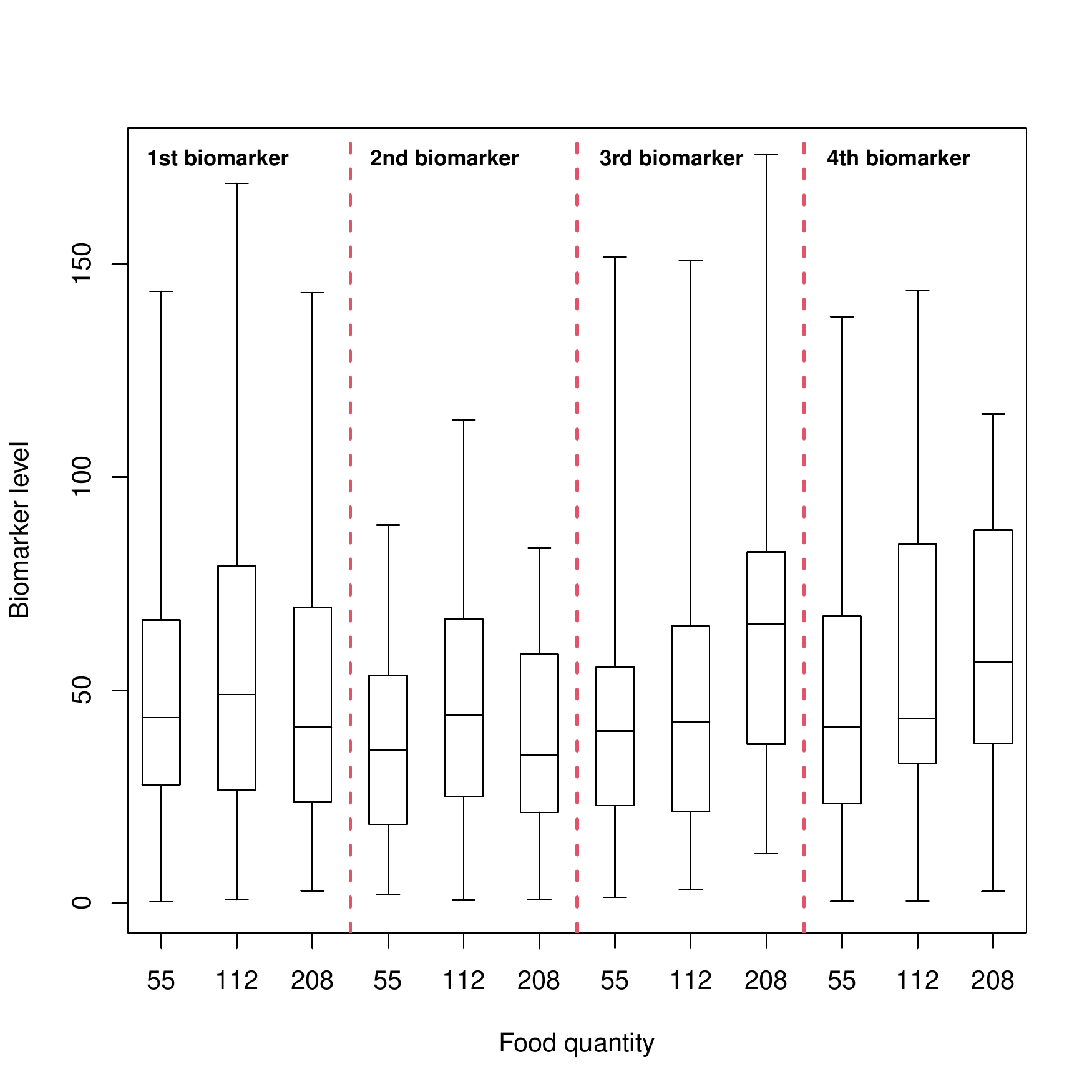}}
    \caption{Simulation study I. Examples of simulated biomarker data ($P=4,D=3$) under (a) the small (Scenario 1) and (b) large (Scenario 3) biomarkers' variability scenarios. Here, $n=99$ with (a) ``stable increments'' and (b) ``increasing increments'' .}
    \label{fig:exampledata}
\end{figure}

Three different specifications are considered for the $\alpha_p$ and $\beta_p$ parameters, to represent different types of biomarker measures. Also, across the $P$ biomarkers, three different specifications are considered for $\sigma_p^2$, to explore the impact of increasing biomarker variability: a small (\emph{scenario 1})(S.1), mixed (\emph{scenario 2})(S.2) or large (\emph{scenario 3})(S.3) range of variance values. Finally, the mixture components' parameters $\{\mathbf{X},\theta_1^2,\dots,\theta_D^2\}$ employed reflect real-world intervention study scenarios: food quantities with stable increments (``stable increments''), food quantities with increasing increments (``increasing increments''), and food quantities with decreasing increments (``decreasing increments''). Two settings are implemented for the components variances $\theta_d^2$, corresponding to low and high variability.
Observation-specific component allocation labels $\mathbf{c}$ are sampled at random from the set $\{1, \dots, D\}$, leading to similar, yet not equal, proportions of observations being assigned to the food quantities in the training datasets.
Combinations of these settings (20 simulated datasets for each) are used to investigate different aspects of performance in three simulation studies detailed below. Figure \ref{fig:exampledata} illustrates two simulated training datasets with low (\emph{scenario 1}) and high (\emph{scenario 3}) biomarker variability levels.

For the MCMC algorithm (Section \ref{sec:MCMC}), $30000$ iterations were run, both when estimating the model from a training dataset and when inferring intake from a test set's biomarker data only, with the first $6000$ iterations being discarded. Further details on the simulation settings, hyperparameter specifications and initializations are deferred to the Supplementary Material (Section 2).

In the absence of comparable approaches for intake quantification in the presence of multiple biomarker data, we compare our results with those obtained using Bayesian linear regression (BLR) and partial least squares (PLS) regression. In BLR, the food quantities are regressed on a linear combination of the biomarkers in the training dataset with the resulting parameters used to infer intake in the test dataset. In PLS, the food quantities are regressed on a dimensionally reduced representation of the biomarker data in the training dataset.

\subsection{Simulation study I: intake quantification under varying biomarker variability}
\label{sec:sim1}
Simulation study I assesses performance under increasing noise levels in the $P$ biomarkers. In the case where $D = 3$, we compare results from the three biomarker variability scenarios. Table \ref{tab:sim3_abs_err} (Simulation Study I columns) reports the median absolute errors between true and estimated intakes from the training data and true and inferred intakes from the test data.

Overall, for estimated intakes, absolute error values are quite low, with median values ranging from $3$g in scenario 1 (S.1)(small biomarker variability) to $9$g in scenario 3 (S.3)(large biomarker variability). Notably, the within-scenario error variability is low, and in all cases is commensurate with the median error. Moreover, the absolute errors are relatively stable across increasing noise levels $\sigma_p^2$, suggesting robustness of the approach to biomarker variability. Notably the relevant parameters for the multiMarker model have been recovered within a $95\%$ credible interval (see Section 2.4 of the Supplementary Material). Table \ref{tab:sim3_abs_err} also shows that the estimated absolute errors under BLR and PLS are higher and more uncertain than those obtained using the multiMarker approach. Also, the BLR and PLS errors increase with increasing biomarker variability $\sigma_p^2$.

In terms of inferring intake from biomarker data alone, the multiMarker approach performs well, but with larger errors in scenario 3 (S.3); however, the data simulated under scenario 3 are extreme and unlikely to be seen in real applications (see Figure \ref{fig:exampledata}). The BLR errors for these test datasets are much larger and more variable. Under PLS, the errors are comparable with the proposed approach, however, as is borne out in the application (see Section \ref{sec:apples}), error values under PLS tend to be relatively low as the inferred intakes tend to be close to the food quantities' sample mean. While such mean-tendency yields good results in terms of median absolute error values, it corresponds to a lack of precision in the inferred intakes, with more extreme intakes being inferred in the middle of the intake range.

Finally, while a small number of food quantities (e.g. $D = 3$) is practical and cost efficient, a larger $D$ could increase the coverage of the realistic intake range. Thus, we have investigated the impact of a larger $D$ value ($D = 6$) on estimation of the multiMarker model and on inferring intake from biomarker data alone. The results and a comparison with the findings from simulation study I are reported in the Supplementary Material (Section 2.1). In general, the performance of multiMarker in both cases ($D = 3$ or $D = 6$) is comparable, suggesting that the benefit of employing a larger number of food quantities is minimal, and that modelling the latent intake range with a mixture distribution is a flexible and cost efficient approach. 

\begin{table}[ht]
\caption{Simulation studies' results. Median ($95\%$ CI width) absolute error values (in grams) computed between true and estimated (E) or inferred (I) latent intakes. The values are reported for the three simulation studies (I,II,III) and the three biomarkers' variability scenarios (S.1,S.2,S.3). Results for the multiMarker (MM) model are reported, as well as those from Bayesian linear regression (BLR) and PLS (PLS) regression. }
\begin{center}
\label{tab:sim3_abs_err}
\scalebox{0.8}[0.83]{
  \begin{tabular}{|cccccccccccccccc|}
  \toprule
  &&  & & \multicolumn{11}{c}{Simulation Study} & \\
 &&  & & \multicolumn{3}{c}{I} &&  \multicolumn{3}{c}{II} & &  \multicolumn{3}{c}{III}  & \\
 &Model &  & & S.1 & S.2 & S.3 & & S.1  & S.2 & S.3 & &   S.1 & S.2 & S.3   & \\
\midrule
  \footnotesize
  & &  & &  &  &  &  &  & & & &  & & &\\
 & \multirow{2}{*}{MM} & E & & 3(7) & 4(8) & 6(18) & & 3(8) & 4(8) & 7(62) & &  6(35) & 7(44) & 11(62)  &\\
  & & I & & 4(9) & 4(33) & 9(59) & & 5(25) & 7(22) & 22(77) & &  8(49) & 9(56) & 27(74) &\\
  & &  & &  &  & &  &  &  & & &   & & &\\
  & \multirow{2}{*}{BLR} & E & & 77(200) & 62(136)& 62(136)& & 61(181) & 113(222) & 80(118) & & 66(227) & 98(269) & 87(197) &\\
  & & I & & 76(216) & 111(263) & 64(137) & & 62(184) & 112(224) & 88(108) & &  67(231) & 112(298) & 101(138) &\\
   & &  & &  &  &  &  &  &&  & &  & & &\\
  & \multirow{2}{*}{PLS} & E & & 10(24) & 20(39) & 35(60) && 9(30) & 22(33) & 39(56) &&  26(72) & 43(80) & 64(65)  &\\
 & & I & & 10(26) & 21(41) & 37(69) & & 10(37) & 23(43) & 41(67) & &  31(77) & 46(87) & 70(80) &\\
 \bottomrule
  \end{tabular}
  }
  \end{center}
\end{table}

\subsection{Simulation study II: intake quantification under discrepancies between training and test data generation}
\label{sec:sim2}

Due to the controlled environment, observations in an intervention study will have intake values which are similar to the administered food quantities. Thus the associated latent intakes will exhibit low variability around the intake mixture components' means (i.e. low $\Theta$ values). However, this is unlikely to be the case when new biomarker measurements for the $n^*$ test observations are considered. Hence we explore the performance of multiMarker when the variability of the intakes in the test dataset is larger than that of the intakes in the training data. 

We simulated data as in simulation study I, but where $\theta_d^2 \:  \forall d = 1, \ldots, D$ are sampled from inverse gamma distributions with expected values of $6$ and $12$ in training and test sets respectively (in the small variances setting), and with expected values of $12$ and $24$ in training and test sets respectively (in the large variances setting). Table \ref{tab:sim3_abs_err} (Simulation Study II columns) presents the absolute errors between estimated and inferred latent intakes and the true values. Here, median error values are low and similar to those computed under simulation study I, for scenarios 1 (S.1) and 2 (S.2). Scenario 3 (S.3) presents a relatively larger median error value, and associated $95\%$ CI width, between inferred and true latent intakes ($22$). However, the multiMarker approach still performs well, given the range of the simulated intakes and the performance of the competitors under the same scenario. This is due to the flexibility of the model for the latent intakes, and multiMarker's factor analytic aspect, that allow refinement of the inferred latent intakes from biomarker data only to values that potentially differ from those of the intervention study. The BLR and PLS results overall yield similar conclusions to those resulting from simulation study I.

Three additional settings were explored in which there were discrepancies between how training and test data were generated (see Supplementary Material, Section 2.2). In the first additional setting different food quantity values, sampled in comparable and realistic ranges, are considered to generate training and test datasets. Cases in which the number of food quantities in the test data is $D^* \neq D$ with either $D^* = D+1$ or $D^*= D-1$ were considered.  In the second additional setting, latent intakes in the test data were generated from a diffuse uniform distribution on $(0, 350)$, to assess robustness to deviations from the mixture of Gaussians assumption in (\ref{eq:zprior}). The third additional setting explored the impact of differently-sized mixture components in the test data. Overall, results (Table 1 in the Supplementary Material) are similar to those obtained under simulation studies I and II (see Table \ref{tab:sim3_abs_err}), with the exception of larger median absolute errors under variability scenario 3 in the additional settings. However
the median error values (and corresponding $95\%$ CIs) are smaller than those of PLS and BLR.

\subsection{Simulation study III: intake quantification under model misspecification}
\label{sec:sim3}

The multiMarker model assumes a linear relationship between latent intake and observed biomarkers. To assess performance of multiMarker in the presence of model misspecification, we simulated data according to the settings of simulation study I, but using a non-linear relationship between intake and biomarkers:
\begin{equation}
    y_{ip} = \alpha_p + \beta_p z_i^2 +\epsilon_{ip}, \quad \forall \quad  i=1,\dots,n, \quad p = 1, \dots,P
    \label{eq:marker_dose2}
\end{equation}
Here the scaling coefficients $\beta_p$ have been re-scaled to $0.1\%$ of the values used in simulation study I, to obtain similar simulated biomarker data ranges.

Table \ref{tab:sim3_abs_err} (Simulation Study III columns) reports the absolute errors between estimated and inferred latent intakes and the true values. Here, error values are low and similar to those computed under simulation study I.
Indeed, median error values are always lower than $27$g in the test data, indicating that true intakes can still be recovered quite well, even when the underlying model is misspecified. This is due to the role of the mixture distribution on the latent intake, which anchors the intakes around the intervention study's food quantities. However, model parameters are hardly ever recovered and often overestimated, indicating that the model misspecification is absorbed by their estimates. The price of biased parameters estimates is a relatively low one to pay in this context, as our goal is to obtain reliable inference on intake.
Further, the range of errors' confidence intervals is larger in simulation study III than in simulation study I, but comparable to that of simulation study II. Finally, comparison with the BLR and PLS results yields similar conclusions to those presented in simulation studies I and II.

\section{Application of multiMarker to biomarkers associated with apple intake}
\label{sec:apples}

The motivation for multiMarker is estimation of the relationship between four biomarkers and apple intake from an intervention study, with a view to subsequently inferring intake when only the four biomarkers are available. MultiMarker was employed to estimate the relationship between the biomarkers and apple intake using $30000$ MCMC iterations, discarding the first $6000$, on a Intel(R) Core(TM) i7-8565U@1.80GHz computer. Model hyperparameters were fixed as outlined in the Supplementary Material (Section 1.1).

To assess the ability of multiMarker to infer apple intake, we use leave-one-out cross validation. Thus, $n = 86$ multiMarker models were fitted, each with a different set of $(n-1)$ observations. In each case, the intake for the left out observation was then inferred using only its biomarker data. The MCMC algorithms were run several times and demonstrated no substantial difference in the parameter estimates and inferred latent intakes. Details of expected sample size values are available in the Supplementary Material (Section 3.3). The computation time was approximately $3.51$ minutes per estimated model, and $2.67$ seconds per data set when inferring intake. 

\begin{figure}[hpt]
    \centering
\includegraphics[scale=.74]{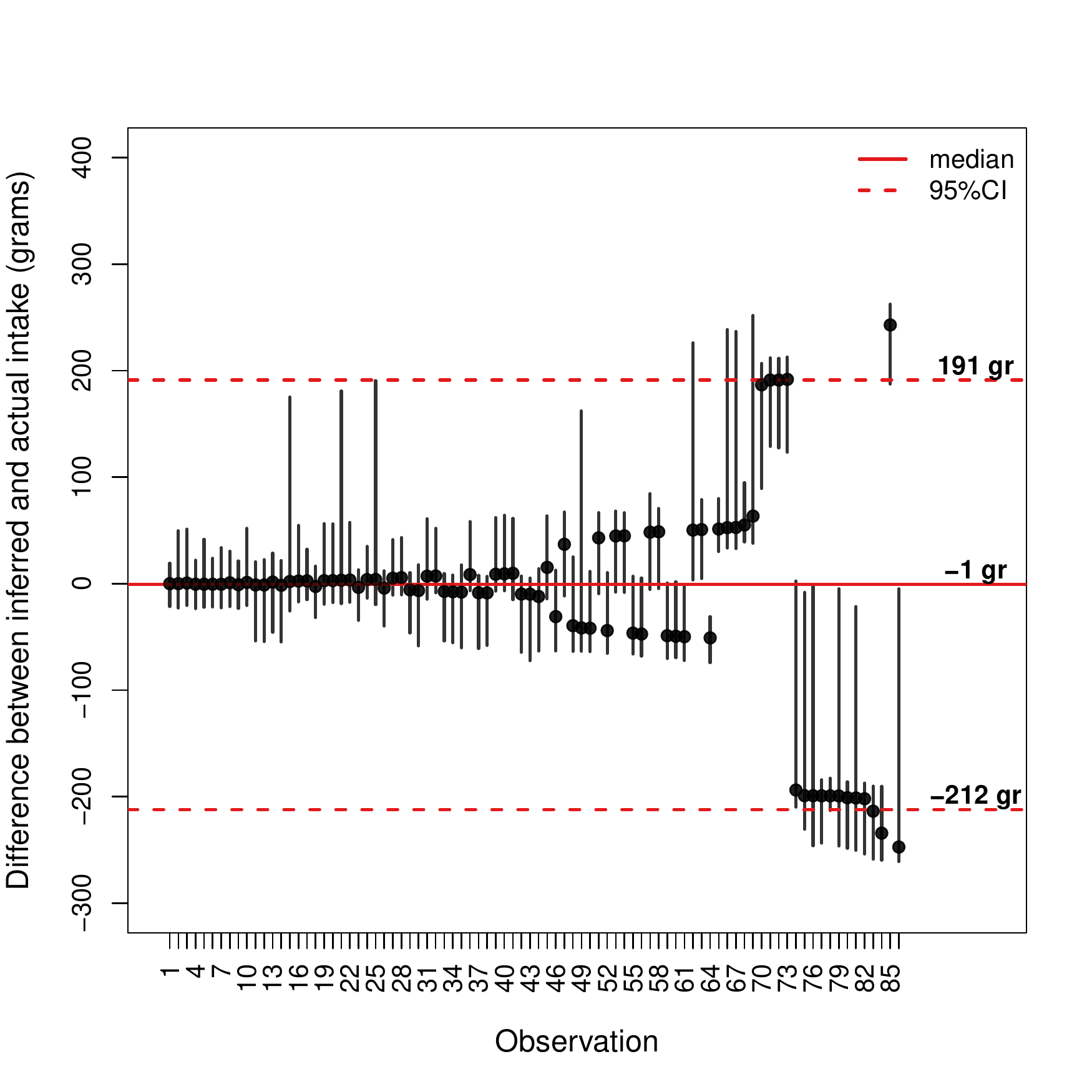}
 \caption{Performance of the multiMarker model when inferring apple intake (in grams) from four biomarkers. Each dot represents the difference between posterior median predicted intake and known consumed apple quantity, with vertical lines illustrating the $95\%$ credible interval. Horizontal lines represent overall median and $95\%$ credible intervals for the differences, with the corresponding numerical values reported in bold.}
    \label{fig:applesdata1}
\end{figure}

Results from the leave-one-out cross validation procedure are shown in Figure \ref{fig:applesdata1}. Differences between the known consumed apple quantities and median inferred intake values are reported, together with corresponding $95\%$ credible intervals. Overall, there is good agreement between consumed apple quantities and inferred intakes. Indeed, median inferred values are concentrated around the true quantities and generally the latent intake posterior predictive distributions include the true apple quantity values. Only a few observations have median inferred intakes far from their true values. In general, such ``miss-inferred'' intakes have larger associated uncertainty, presented through their wider $95\%$ credible intervals. The median absolute difference between inferred intake and true consumed food quantity is $15.29$ grams for the $50$ grams apple quantity group, $35.22$ grams for the $100$ grams quantity and $9.19$ grams for the $300$ grams quantity. These errors correspond to, respectively, $31\%$, $35\%$ and $3\%$ of the true apple quantity.

\begin{figure}[hpt]
   \centering
     \subfloat[][]
{\includegraphics[scale=.335]{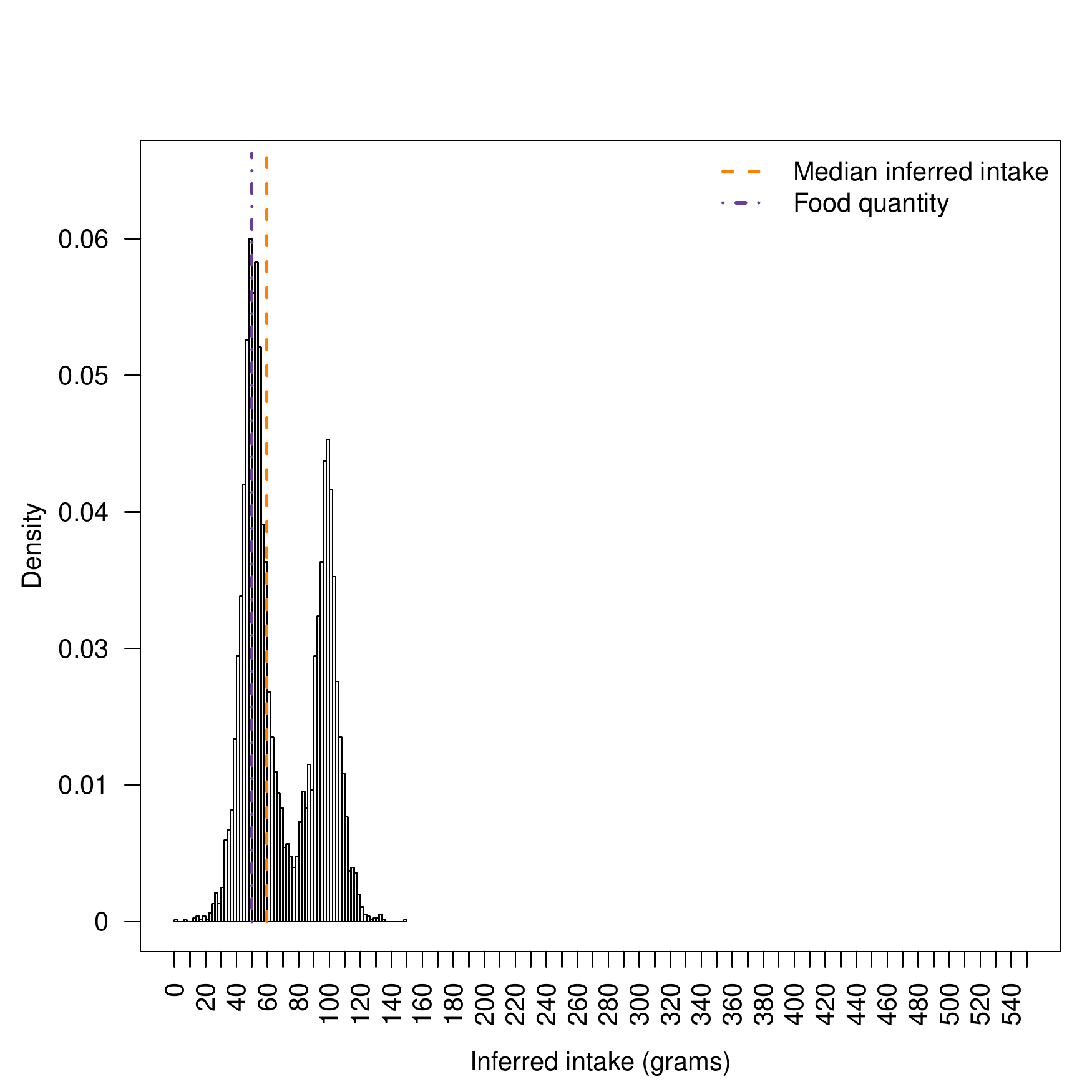}}
 \subfloat[][]
{\includegraphics[scale=.335]{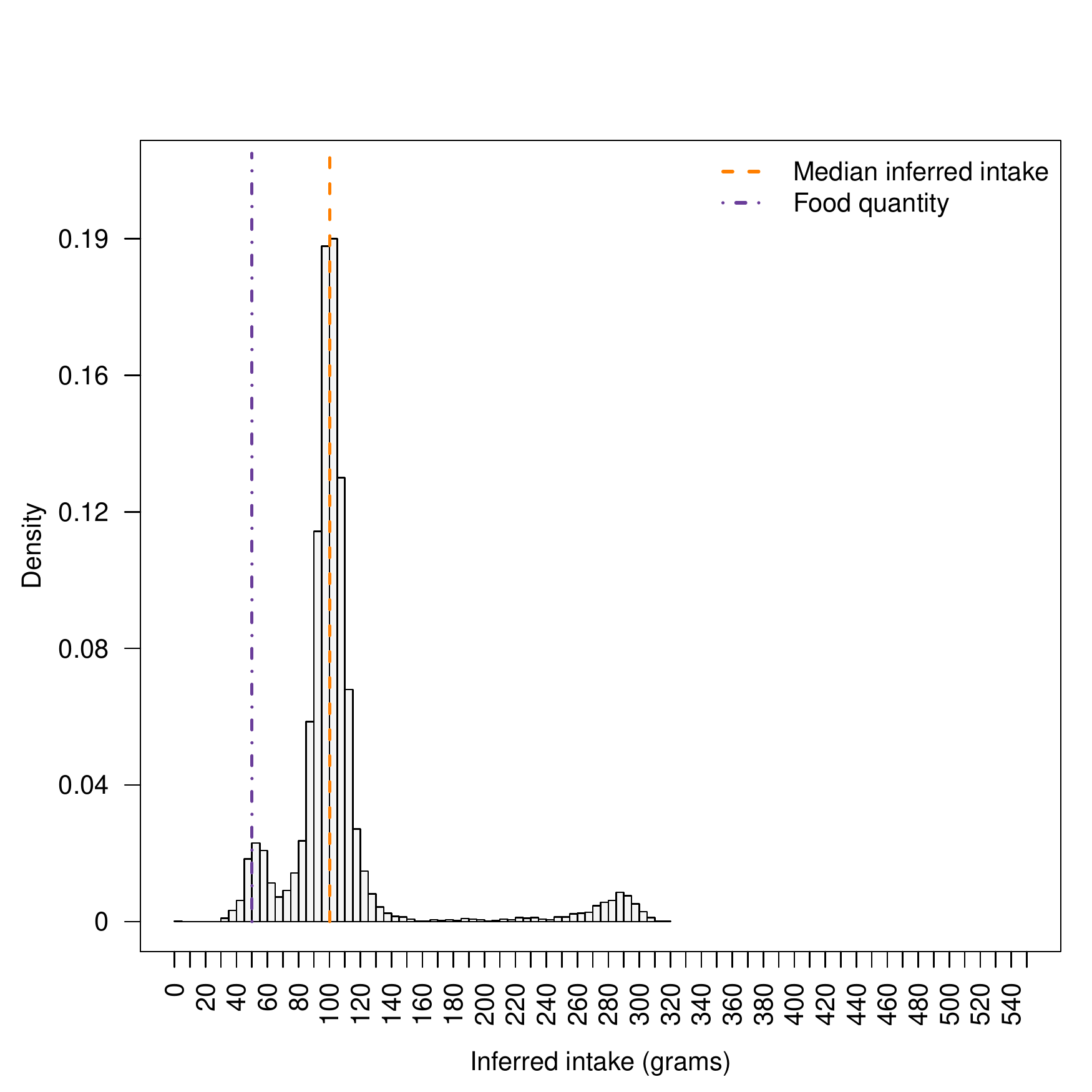}}\quad
     \subfloat[][]
{\includegraphics[scale=.335]{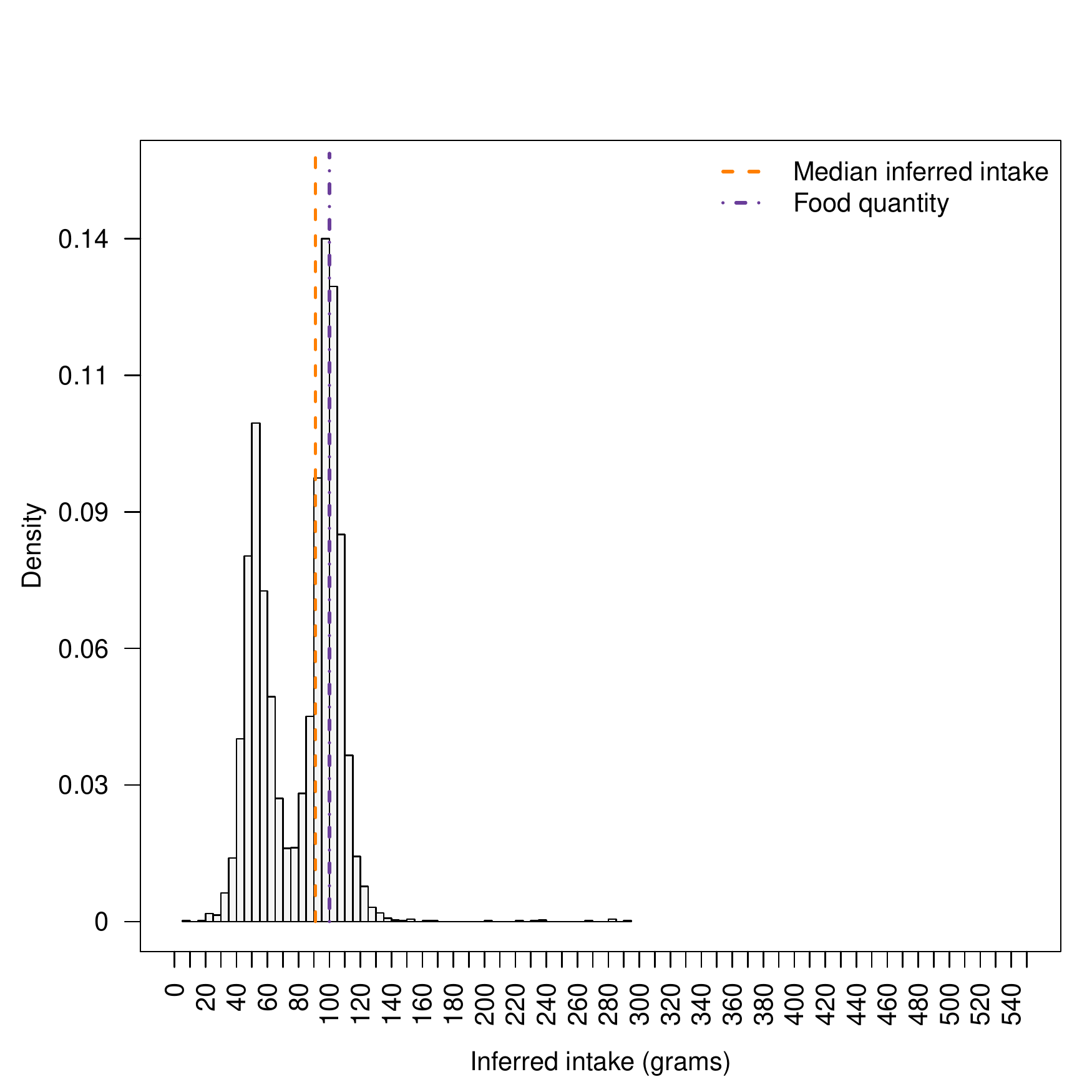}}
\subfloat[][]
{\includegraphics[scale=.335]{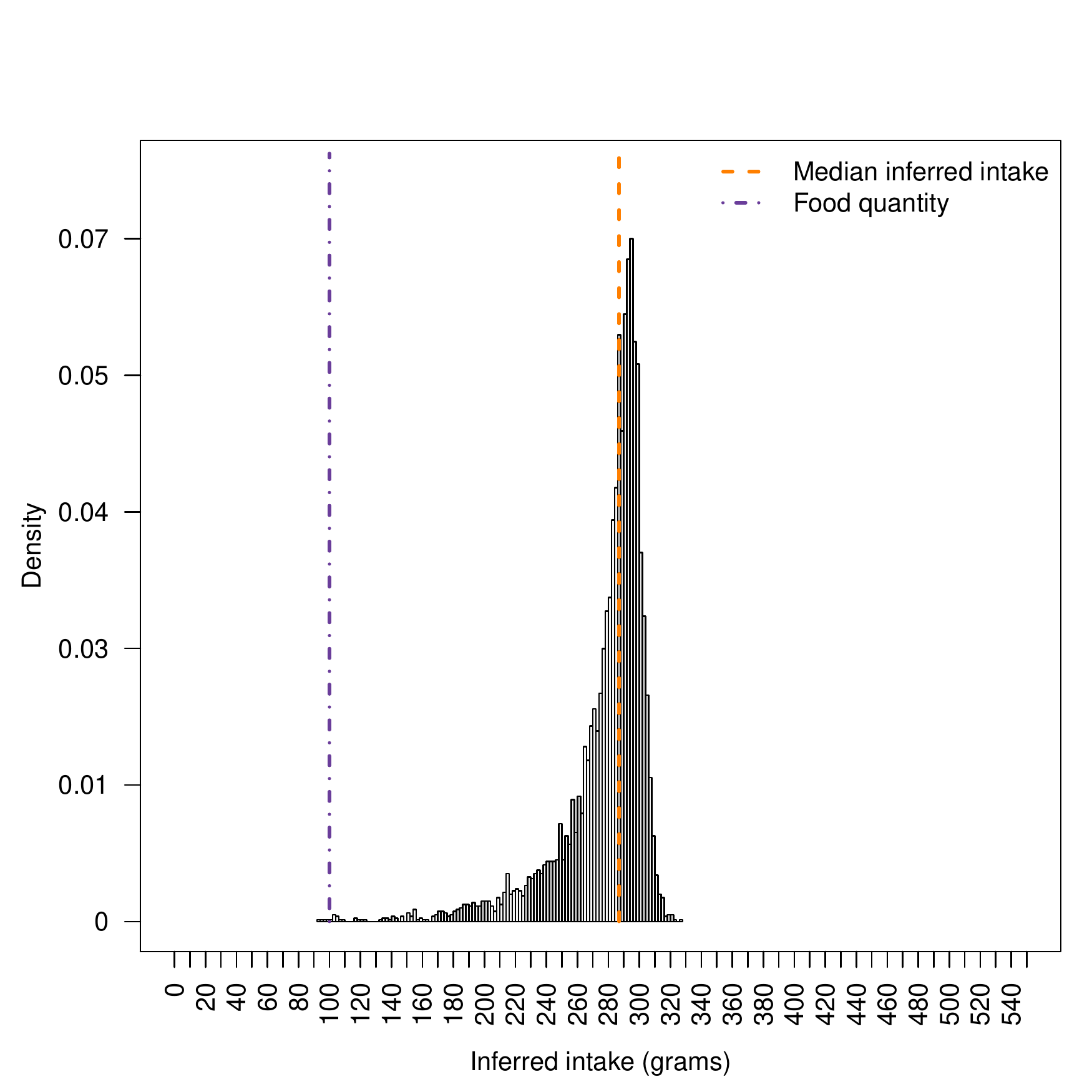}}\quad
    \subfloat[][]
{\includegraphics[scale=.335]{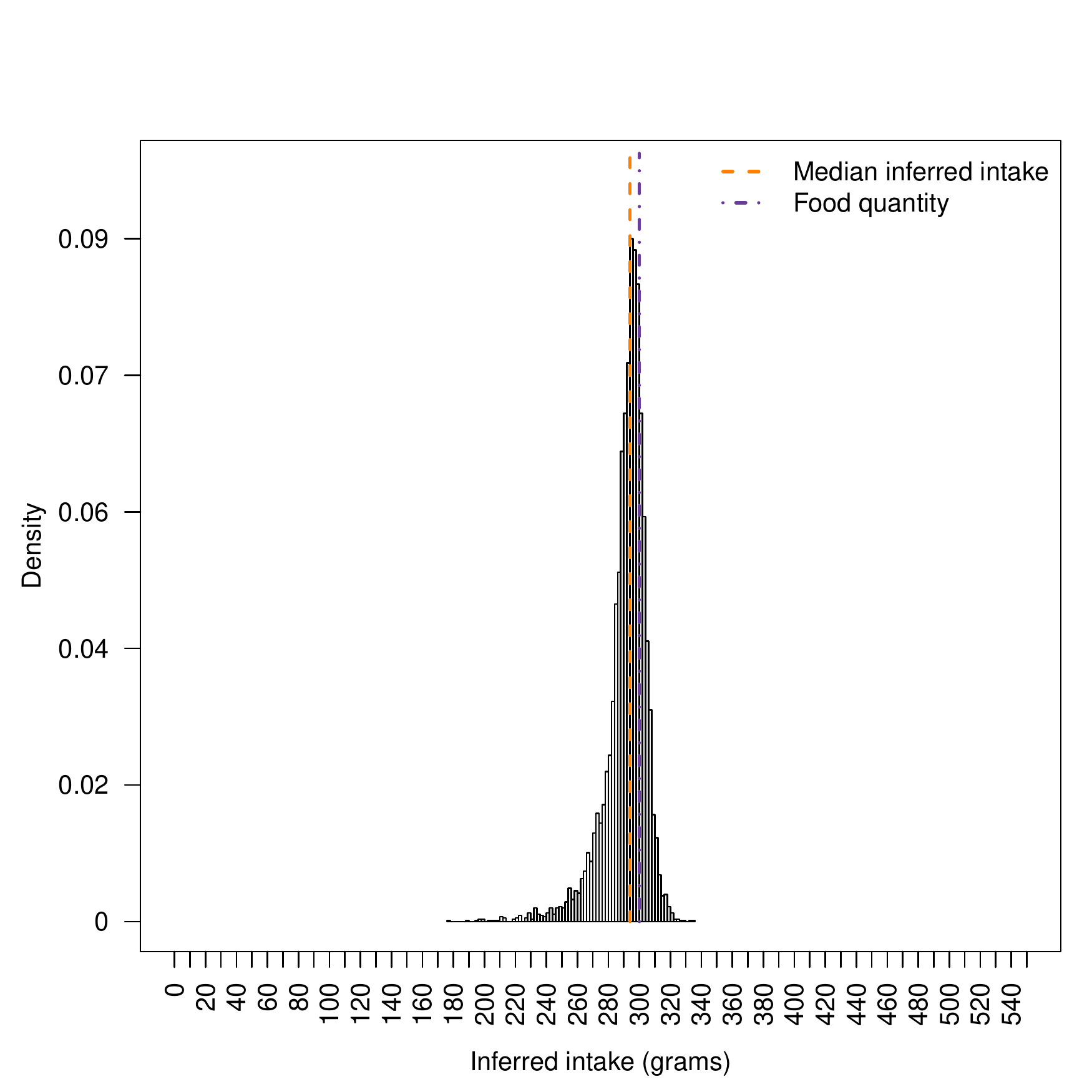}}
 \subfloat[][]
{\includegraphics[scale=.335]{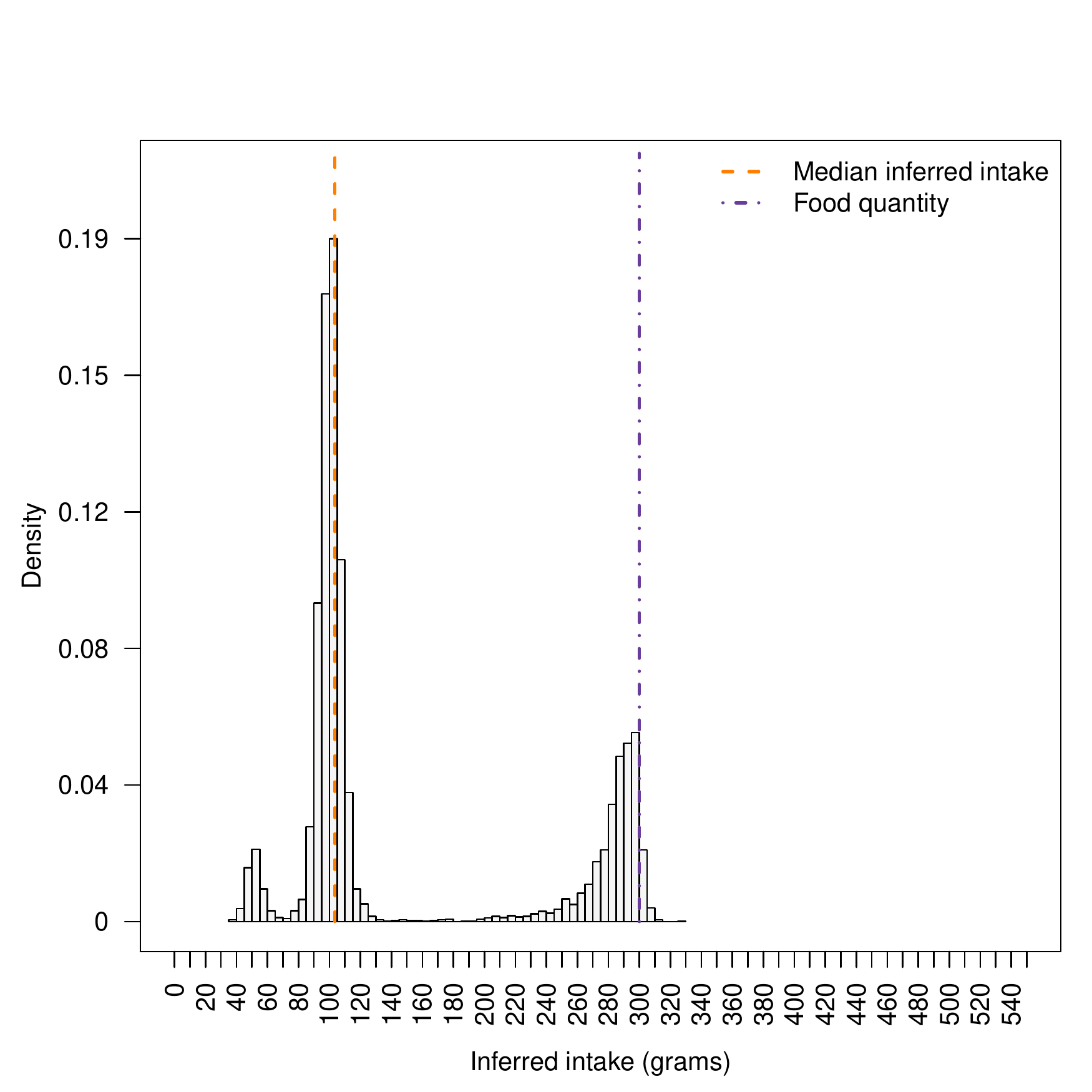}}
 \caption{Posterior predictive distributions of apple intake (in grams) for 6 observations. Orange and purple lines denote median inferred intake and known apple quantity, respectively. }
    \label{fig:appledata2}
\end{figure}

To further visualize the results, Figure \ref{fig:appledata2} presents inferred posterior predictive intake distributions, for 6 observations, two for each apple quantity. Plots on the left side of Figure \ref{fig:appledata2} correspond to accurately inferred intakes, while those on the right are poorly inferred.  When intakes are correctly inferred, the range of the posterior predictive distribution does not incorporate apple quantities far from the truth (e.g. observations $13$, $32$ and $82$). For some of these observations, when the corresponding true intake was either $50$ or $100$ grams, the inferred distribution is bimodal (e.g. observations $13$, $32$). However, for all of them the associated $95\%$ credible interval is always relatively narrow and the mode closer to the true intake value is consistently higher than the other one.  Among ``miss-inferred'' intakes, those whose true quantity was $300$ grams tend to have notably underestimated inferred intakes (e.g. observation $73$). This is due to the ordinal approach in the modelling of the observation-specific weights, which is more robust to overestimation than to underestimation. Similar posterior predictive distributions to those presented in Figure \ref{fig:appledata2} were obtained for the remaining observations. 

For comparative purposes, BLR and PLS regression were also used to infer apple intake via leave-one-out cross validation. Similar to the observed performance in the simulation studies, both methods proved unable to precisely infer intake, with most predictions around the average of the three apple quantities ($\approx 150$ grams). Furthermore, $95\%$ credible intervals obtained using BLR are more than twice the size of those obtained under multiMarker, indicating low reliability of the inferred values. Further, the uncertainty quantification under PLS regression was not meaningful, as the constructed $95\%$ confidence intervals via cross validation yielded very narrow ranges. Further details on these results are deferred to the Supplementary Material (Section 3.2).

To assess whether treating the $n = 86$ observations as independent was appropriate, given that they resulted from a set of 32 intervention study participants, we estimated the multiMarker model using slightly modified versions of the dataset. Specifically, we constructed modified versions of the original dataset such that the same participant appeared only once i.e. each participant is present in only one apple quantity, which was selected at random. This procedure was repeated $100$ times, and each time the multiMarker model was fitted to the modified dataset. Table \ref{tab:apple_par} presents the posterior median parameter estimates obtained from fitting the model to the original data and to the modified datasets. For the $\alpha_p$, $\beta_p$, $\sigma_p$ parameters,
dimensions 1 to 4 correspond to the four biomarkers, while for the $\theta_d$ parameter the dimensions correspond to the three apple quantities. The median estimates for the first three parameters ($\alpha_p$, $\beta_p$, $\sigma_p$) are similar under the original and modified data although the similarity is weaker for the variances $\sigma_p$. Also, posterior medians of these parameters from the original data are always contained in the corresponding $95\%$ credible intervals inferred from the modified datasets. These results suggest the inference is robust to our assumption of independence between the $n = 86$ observations. As for the $\theta_d$ parameters, the estimated component variances are much larger in the original data than of the modified data. This finding is in line with our expectations, as the $\theta_d$ parameters are a function of the sample size (see Section \ref{sec:lat_int_fc}). Indeed, as discussed in simulation study II, component variances are nuisance parameters, as their role is instrumental to adapting the inferred latent intake distribution to the data analysed, but their value is of no interest.

\begin{table}[!h]
\begin{center}
   \caption{Posterior median ($95\%$ CI width) parameter values inferred from the original apple data, and average posterior median (average $95\%$ CI width) parameter values inferred from the modified apple datasets.}
   \label{tab:apple_par}
   \scalebox{0.9}[0.95]{
 \begin{tabular}{|cccccccc|}
  \toprule
   & & & \multicolumn{4}{c}{Parameter} & \\
  Data & Dimension && $\alpha_p$ & $\beta_p$ & $\sigma_p$ & $\theta_d$ &   \\
\midrule
\multirow{4}{*}{Original} & 1 && 0.206 (0.267) & 0.003 (0.002) & 0.353 (0.117) & 5.546 (6.514) &   \\
 & 2 && 0.489 (0.215)  & 0.005 (0.002) & 0.271 (0.104) & 7.546 (14.038) &   \\
 & 3 && 0.612 (0.514) & 0.007 (0.004) & 0.677 (0.234) & 98.989 (71.595) &   \\
& 4 && 0.614 (0.323) & 0.008 (0.004) & 0.382 (0.174) & - &   \\
& && &  &  & &   \\
\multirow{4}{*}{Modified} & 1 && 0.214 (0.433) & 0.003 (0.003) & 0.597 (0.157) &  1.806 (1.176) &   \\
 & 2 && 0.515 (0.416) & 0.005 (0.004) & 0.600 (0.264) &  2.032 (1.773) &   \\
 & 3 &&  0.517 (0.755) & 0.009 (0.007)  & 0.746 (0.258) & 8.407 (10.358) &   \\
& 4 &&  0.662 (0.635) & 0.008 (0.006) & 0.676 (0.226) & - &   \\
\bottomrule
  \end{tabular}
  }
  \end{center}
\end{table}

\section{Discussion}
\label{sec:discussion}
Motivated by the need to estimate the relationship between metabolomic biomarkers and food intake to allow inference on intake from biomarker data alone, we have introduced the general and flexible multiMarker model. MultiMarker builds upon two classical regression models, multiple linear regression and ordinal regression, combining them in the wider frameworks of factor analysis and mixture of experts models. The multiMarker model facilitates estimation of the relationship between multiple biomarkers and food intake from intervention studies. Further, it allows one to subsequently infer intake in a quantitative manner when only biomarker data are available. The multiMarker model advances on current approaches, which focus on inferring categorical levels of intake, by providing more detailed inference. Moreover, as multiMarker is embedded in a Bayesian framework, uncertainty quantification is readily available, which is an important requirement in the motivating application domain.

The multiMarker model and inferential algorithm were assessed in an extensive simulation study, where a large variety of biomarker data have been generated to check model performances under, among other things, different levels of biomarker or intake variability. Furthermore, performances under model misspecification have been verified. In all of the cases the framework performed well, as intake values and their range were recovered with low error, even in a model misspecification context.

The multiMarker model was successfully employed to estimate the relationship between biomarkers and apple intake and to subsequently infer apple intake from biomarker data alone, leveraging data collected under the motivating A-DIET research programme.  Leave-one-out cross validation results showed that generally inferred intakes were concentrated around the true apple quantities. Comparison with existing regression models showed that multiMarker was able not only to provide useful uncertainty quantification, but also much more reliable inference on apple intake.

The proposed framework allows quantification of any unobserved quantity of interest for which prior information on its scale is observed (here, food quantities) and for which proxies can be measured (here, biomarkers). Furthermore, the observed variables' scales are not relevant to recover that of the latent variable.  Such a feature is appealing for researchers, as biomarkers for the same intake could correspond to quite different measurement ranges, due to their sources (blood, urine, etc), the instrument used for collection and so on. 

Although multiMarker has been motivated by the specifications of the A-DIET study, the model is general and designed to be accessible to non-statisticians. In multiMarker, the extra layer of complexity introduced by the latent variable and associated model parameters bestows the advantage of inferring refined, quantitative intakes and allows the model to adapt to different scenarios in the data. Modelling the weights of the mixture distribution using a mixture of experts framework guides quantitative inference, informing the latter on which part of the intake range is to be considered when inferring intake. Unlike previous approaches, we explicitly model the ordinal nature of the food quantities. However, multiMarker could easily be modified to suit different application contexts, and different specifications could be employed to represent different data settings. For example, should the ordinal regression feature be inappropriate for a particular setting, multinomial regression can easily be incorporated as a substitute. Further, some of the application-driven restrictions imposed here, for example bounded prior distributions for some parameters, could be easily lifted to adapt the proposed model to different contexts without resulting in any additional complications from either a model definition or estimation point of view.

As a key requirement of the motivating application was to quantify uncertainty, particularly when inferring intakes from biomarker data alone, inference was carried out via a Metropolis within Gibbs MCMC algorithm. Alternative inferential procedures could easily be considered and may be particularly fruitful in alternative specifications of multiMarker that use different distributional assumptions, leading to more complex and computationally expensive posterior distributions. The MCMC approach may also have limitations should the number of biomarkers or observations increase. A natural alternative in such a setting is a variational inference approach \citep{Blei2017}, however it can also be sensitive to initialization and only guarantees convergence to a local optimum. Further, uncertainty quantification is not automatic under a variational framework, although \cite{chen18} demonstrate that a bootstrap procedure is still valid, albeit for variational parameters which may or may not coincide with the true parameters of interest.

Although designed for a particular problem, the quantification of apple intake from a panel of urinary biomarkers, the multiMarker approach has general applicability outside of nutrition. Indeed, the model could have applicability in any scenario where multiple outcomes are associated with an unobserved variable of interest, such as in toxicology or social science studies.

Possible extensions to the multiMarker model are many and varied, including explicitly modelling the repeated biomarker measurements in the same group of participants. Individual heterogeneity could be accounted for by introducing random effects in the model (see for example \cite{Muthen89}). Further, the introduction of subject-specific covariates in the latent intake mixture weights, to allow a more flexible model fit, would be feasible. Thus there is much potential in the use of latent variable models to infer food intake.

{\bf Acknowledgments.} Supported by a research grant from the European Research Council (ERC)(647783). 

This publication has emanated from research conducted with the financial support of Science Foundation Ireland under Grant number SFI/12/RC/2289$\_$P2. For the purpose of Open Access, the author has applied a CC BY public copyright licence to any Author Accepted Manuscript version arising from this submission’

We would like to acknowledge the help of the following people with respect to the Apple Biomarker data: Aoife E. McNamara, Cassandra Collins, Pedapati S. C. Sri Harsha and Diana Gonzalez-Pena.

\bibliographystyle{agsm}
\bibliography{bibl}
\vspace{1cm}
\pagebreak
\appendix

\section{Posterior distribution and MCMC algorithm}
Considering the multiMarker model’s likelihood, prior and hyperprior distributions, the posterior distribution is:
\begin{equation}
\begin{split}
    P(\Omega \mid \mathbf{Y},  \mathbf{X})  = &
    L(\mathbf{Y} \mid \alpha, \beta, \mathbf{z}, \Sigma) p(\alpha \mid \mu_{\alpha}, \sigma_{\alpha}^2) p(\mu_{\alpha} \mid m_{\alpha}, \tau_{\alpha}, \sigma_{\alpha}^2)p(\beta \mid \mu_{\beta}, \sigma_{\beta}^2) p(\mu_{\beta} \mid m_{\beta}, \tau_{\beta},\sigma_{\beta}^2) \\
    & p(\sigma_{\beta}^2\mid \nu_{\beta1}, \nu_{\beta2}) p(\Sigma\mid \nu_{p1}, \nu_{p2}) p(\mathbf{z} \mid \pi, \mathbf{X}, \Theta)p(\Theta\mid \nu_{z1}, \nu_{z2}) \\
    & p(\pi \mid \gamma, \eta, \mathbf{Y}, \mathbf{c}, \mathbf{X}) p(\gamma\mid m_{\gamma}, \kappa) p(\eta\mid m_{\eta}, \kappa) \\
\end{split}    
\end{equation}
where for brevity $\Omega$ denotes the set of model parameters.

\subsection{Hyperparameter settings }
Given an observed dataset, hyperparameter values are fixed automatically according to the following procedures. The overall means $m_{\alpha}$ and $m_{\beta}$ are fixed, respectively, as the estimated intercept and slope coefficient of the multiple linear regression defined using biomarkers as response variable and food quantity values as predictor.
Variances' hyperparameters are fixed differently according to the parameter to which they refer: $(\nu_{\beta 1},\nu_{\beta 2})=(2,3 )$ and $(\nu_{p 1},\nu_{p 2})=(1,3 )$.
The $(\nu_{z 1},\nu_{z 2})$ latent intake variances' hyperparameters are $D$-dimensional vectors, with $\nu_{z1d}= \frac{D -d+1}{2}$, and $\nu_{z 2d}= n $, $\forall d$. 
Regarding the $\alpha$ and $\beta$ vector of parameters, their values are initialized solving the following system of equations:
\[
\sum_{(x_i = X_d)} (y_{ip}) \approx \alpha_p + \beta_p x_d \quad \text{for} \quad p=1, \dots, P \quad \text{and} \quad d=1,\dots,D. 
\]
Biomarkers' error variances are initialized exploiting the definition of estimated error variances under the factor analytic model, adjusted for the extra variability brought in by the latent intakes prior distribution: $\Sigma = \widehat{V(\mathbf{Y})} - \frac{1}{D}\beta \beta^{T}  $. Last, mixture components' variance parameters are initialized with the following values:
\[
\sigma_d^2 = \frac{1}{P}\sum_{p=1}^P \frac{ \widehat{V(Y_{pd})} -\sigma_{\alpha}^2 - \sigma_p^2 }{\sigma_{\beta}^2}
\]
where $\widehat{V(Y_{pd})}= \hat{var}\bigl(\mathds{1}(x_i = X_d) y_{ip}\bigr)$.
Last, denoting with $\{x_1, \dots, x_i, \dots, x_n\}$ the food quantities known to be consumed by the $n$ observations in an intervention study, latent intake values are initialized as $z_i\thicksim \mathcal{N}_{(0, \infty)}(x_i, \theta_d)$, where $\theta_d^2 = 5^2$, $\forall d$, is the starting value for the components' variance parameters.

\subsection{Components' weights}
The prior distributions for the $\gamma = \{ \gamma_0, \gamma_1, \dots, \gamma_D\}$ weights' parameters are defined to represent the constrained characteristic of ordinal data, that is: $X_1 < \dots < X_d < \dots < X_D$. These constraints correspond to the following in terms of model parameters: $\gamma_0 < \dots < \gamma_d < \dots < \gamma_D$. A natural choice for the corresponding prior distributions is the following:
\[
\gamma_d \thicksim \mathcal{N}_{(m_{\gamma_{d-1}}, m_{\gamma_{d+1}}) } \bigl( m_{\gamma_d}, \kappa \bigr), \quad \text{for } \quad d=1, \dots D-1
\]
Further, biomarkers' intercepts $\eta = \{\eta_1, \dots, \eta_P\}$ have been given the following prior distribution: $\eta_p\thicksim \mathcal{N}(m_{\eta_p}, \kappa)$, $p=1,\dots,P$.
Hyperparameters $\{m_{\eta_p}\}_{p=1}^P$ and $\{m_{\gamma_d}\}_{d=1}^{D-1}$ are set via their corresponding estimates from the ordinal regression of the known observation-specific component allocation labels on biomarker measurements from the intervention study. Such estimates are obtained with the \emph{ordinalNet} package, available on CRAN (\url{https://CRAN.R-project.org/package=ordinalNet}).
Both $\eta$ and $\gamma$ parameters are initialized with their hyperameter values ($\{m_{\eta_p}\}_{p=1}^P$, $\{m_{\gamma_d}\}_{d=1}^{D-1}$) and are updated with a random walk Metropolis Hastings step inside the MCMC algorithm. Variance hyperparameter $\kappa$ is fixed to some value $\kappa<=2$. Small variations around this threshold have been tested and did not produce any substantial difference in the $\eta$ and $\gamma$  parameters estimates. 
\subsection{Latent intakes posterior predictive distribution}
In Section 4.2 we have introduced a sampling distribution for the latent intakes, to be used when deriving the latent intake posterior predictive distribution. This distribution was derived as the product of two terms, the first being the log-likelihood of the model, expressed as a function of the latent intakes:
\[
\begin{split}
  \ell (y_j^{*} \mid \alpha, \beta, z_j^{*}, \Sigma)  & =   p (z_j^{*} \mid \alpha, \beta, y_j^{*}, \Sigma) \propto \sum_{p=1}^P p(z_{j}^{*}\mid \alpha_p, \beta_p, y_j^{*}, \sigma_p^2)\\
  & = \sum_{p=1}^P \Biggl( -\frac{1}{2} \log\bigl( 2\pi\sigma_p^2\bigr) -\frac{1}{2\sigma_p^2} \biggl( z_{j}^{*}\beta_p -\bigl( y_{jp}^{*} - \alpha_p \bigr) \biggr)^2 \Biggr)\\
  & = \sum_{p=1}^P \Biggl( -\frac{1}{2} \log\bigl( 2\pi\sigma_p^2\bigr) -\frac{\beta_p^2}{2\sigma_p^2} \biggl( z_{j}^{*} -\bigl( \frac{y_{jp}^{*} - \alpha_p}{\beta_p} \bigr) \biggr)^2 \Biggr)\\
  & \propto \sum_{p=1}^P \Biggl( -\frac{\beta_p^2}{2\sigma_p^2} \biggl( z_{j}^{*} -\bigl( \frac{y_{jp}^{*} - \alpha_p}{\beta_p} \bigr) \biggr)^2 \Biggr)\\
  & \propto -\frac{1}{2} z_j^{*2} \biggl( \sum_{p=1}^P \frac{\beta_p^2}{\sigma_p^2}\biggr) +z_j^{*} \biggl( \sum_{p=1}^P
  \frac{ \beta_p(y_{jp}^{*} - \alpha_p)}{ \sigma_p^2}\biggr)
  = - \frac{1}{2\sigma_z^2}z_j^{*2} +  \frac{\mu_z}{\sigma_z^2}z_j^{*},
\end{split}
\]
The second term is the mixture distribution presented in Section 3.2. The product of these two terms can be expressed as a mixture distribution:
\begin{equation}
\begin{split}
  p(z_j^{*} \mid \Omega) & = \mathcal{N}_{[0, \infty]} \bigl( \mu_z, \sigma_z^2 \bigr) \sum_{d=1}^D \pi_{jd} \mathcal{N}_{[0,\infty]}\bigl( X_d, \theta_d^2 \bigr) 
   = \sum_{d=1}^D \pi_{jd} \mathcal{N}_{[0, \infty]} \bigl( \mu_z, \sigma_z^2 \bigr) \mathcal{N}_{[0,\infty]}\bigl( X_d, \theta_d^2 \bigr)  \\
  & =  
  \sum_{d=1}^D \pi_{jd} \mathcal{N}_{[0, \infty]} \biggl( \frac{\mu_z\theta_d^2 + X_d\sigma_z^2}{\theta_d^2 +\sigma_z^2}, \bigl( \frac{1}{\theta_d^2 } + \frac{1}{\sigma_z^2} \bigr)^{-1} \biggr) =  \sum_{d=1}^D \pi_{jd} \mathcal{N}_{[0, \infty]} (\mu_{zd}, \sigma_{zd}^2) \\
  \end{split}
\end{equation}
where $\Omega = \{ \mu_{z1}, \dots, \mu_{zD}, \sigma_{zq}^2, \dots , \sigma_{zD}^2\}$.

\section{Additional simulation study details}
\label{sec:sim_add}
In the simulation studies described in the paper, biomarkers' intercepts and scale coefficients $(\alpha,\beta)$ are sampled from their prior distributions with hyperparameters:
\[\{\mu_{\alpha},\mu_{\beta}, \sigma_{\alpha^2}, \sigma_{\beta}^2\}= \{(1, 0.01, 1,0.01 ), (20, 0.1, 4, 0.1 ),(100, 1, 14, 1 )\}\]
These hyperparameter specifications correspond to small, medium and large range biomarker values, and are used to represent different types of biomarker measures (as for example measurements coming from different instruments or measurements non normalised by osmolality). Regarding the error standard deviations terms $\sigma_p^2$, these values are sampled from inverse gamma distributions with expected values dependent on the biomarkers' range considered: $(1^2,3^2,15^2)$ for small variances and $(3^2,20^2,100^2)$ for large ones, respectively in the small, mixed and large biomarkers' range frameworks. 
Values for $X_d$ are sampled from $D$ zero-truncated Gaussian distributions with means $\mu_{X_d}$ ranging in between $30$ and $320$, with $\mu_{X_d}  < \mu_{X_{d+1}}$. Three different settings are explored to represent ``food quantities with stable increments'', ``food quantities with increasing increments'', and ``food quantities with decreasing increments''. In the first setting (``stable increments''), means $\mu_{X_d}$ are equispaced, that is $d(\mu_{X_{d-1}},\mu_{X_d}) = d(\mu_{X_{d}},\mu_{X_{d+1}}) $. Instead, in the second and third settings we have that $d(\mu_{X_{d-1}},\mu_{X_d}) < d(\mu_{X_{d}},\mu_{X_{d+1}}) $ and $d(\mu_{X_{d-1}},\mu_{X_d}) > d(\mu_{X_{d}},\mu_{X_{d+1}}) $, respectively. In simulation studies I and III, values for $\theta_d^2$ are sampled from inverse gamma distributions with either expected value of $8^2$ (small variances) or of $16^2$ (large variances).
In simulation study II, values for $\theta_d^2$ are sampled from Inverse Gamma distributions with
\begin{itemize}
    \item Small variances setting: expected value of $6^2$ in the intervention study data and of $12^2$ in the biomarker only data;
    \item Large variances setting: expected value of $12^2$ in the intervention study data and of $24^2$ in the biomarker only data.
\end{itemize}

\subsection{Simulation study I: comparison between $D=3$ and $D=6$}
Table \ref{tab:sim3_abs_err12} (first column block) reports the absolute errors computed between estimated and inferred latent intakes and the truth when $D = 6$. As observed in Simulation Study I in the paper, the errors under BLR are large and uncertain when compared to the multiMarker and PLS approaches which perform similarly when the number of food quantities is large. The multiMarker approach performs similarly when $D = 3$ or $D = 6$, suggesting that the benefit of employing of a larger number of food quantities is minimal. 

\subsection{Simulation study II: discrepancies between training and test data generation}
\subsubsection{Varying $\mathbf{X}$ values}
This simulation study considers the case in which different food quantity values, sampled in comparable ranges, are used to generate train and test data. In particular, latent intakes used to generate train data have been simulated from a mixture distribution with $D=3$ components, food quantity vector $\mathbf{X}$, and a vector $\Theta$ of component variances. Instead, to generate test data, we either sample latent intake values from:
 \begin{itemize}
    \item a new vector of food quantities $\mathbf{X}^*$, of length $D^*=D$, whose range is similar to that of $\mathbf{X}$,
    \item a new vector of food quantities $\mathbf{X}^*$, of length $D^*=D-1$, whose range is similar to that of $\mathbf{X}$,
   \item a new vector of food quantities $\mathbf{X}^*$, of length $D^*=D+1$, whose range is similar to that of $\mathbf{X}$.
\end{itemize}
As in Section \ref{sec:sim_add}, food quantity values $\mathbf{X}^*$ are sampled from $D^*$ zero-truncated Gaussian distributions with means $\mu_{X_d^*}$ ranging in between 30 and 320, with $\mu_{X_d^*}  < \mu_{X^*_{d+1}}$, and such that $X_d \neq X_d^*$, $\forall d, d^*$. Further, $\mathbf{X}^*$ have been simulated from all three increments settings: ``food quantities with stable increments'', ``food quantities with increasing increments'', and ``food quantities with decreasing increments''.

For comparability purposes, all other parameters and hyperparameters settings are set as in Simulation study I (Section 5.1 of the paper). Table \ref{tab:sim3_abs_err12} (second column block) reports the absolute errors computed between estimated and inferred latent intakes and simulated values under ``varying $\mathbf{X}$'' scenarios. In general, our approach is able to recover quite well ``true'' latent intakes under this type of mis-specification of the latent intake distribution. Indeed, absolute error values computed between true and estimated intakes under this setting are comparable with those of Simulation Study I (Section 5.1), that is to the case where  the latent intake distribution is correctly specified.

\subsubsection{Diffuse intake distribution in the test data}
This simulation study considers the case in which intakes in the test data are generated from a single diffuse distribution, to assess the impact of deviations from the mixture of Gaussians assumption (see Section 3.2 of the paper) on performance. Here, latent intakes used to generate train data are simulated from a mixture distribution with $D=3$ components (see Section \ref{sec:sim_add}), while latent intakes in the test data are sampled from a Uniform distribution in $(0,350)$.  For all other parameters and hyperparameter settings, see Section 5.1 of the paper. Table \ref{tab:sim3_abs_err12} (third column block ``Uniform intakes'') reports the absolute errors between estimated and inferred latent intakes and simulated values. In general, absolute error values are in line with those obtained with other settings (see Table 1 of the paper), indicating that intakes sampled from a single diffuse distribution (here, a uniform distribution) can still be well recovered using a mixture of Gaussians construct.

\subsubsection{Unbalanced components in the test data}
This simulation study considers the case in which observations in the test data have intake values which are concentrated in a specific sub-region of the intake range explored in the intervention study. 
This corresponds to assuming $n_d^* >> n_k^*$, for $k \in \{1, \dots,D\}, k \neq d$, where $n_d^*$ is the number of observations in the $d^\text{th}$ component, and $n_k^*$ the number of observations allocated in a component different from $d$. 
To assess multiMarker's performance under this setting, intakes are simulated intakes in the train data from a ``balanced'' mixture distribution with $D=3$ components (see Section 5.1 of the paper and Section \ref{sec:sim_add}). However, intakes in the test data have been generated as follows:
\begin{enumerate}
    \item A component index $d$ is sampled at random from $\{1, \dots, D\} = \{1, 2, 3\}$.
    \item Given $d$, $n_d^*$ observations are simulated from the $d^\text{th}$ component, where $n_d^*$ is sampled at random in the range $(0.7n^*, 0.8n^*)$.
    \item $n_{-d}^* = \{n_k^*, n_l^*\}$ observations, where $l,k$ are the indexes for the two remaining components,  are simulated from the corresponding components. The $n_{-d}^*$ values are sampled at random in $(0, n^* - n_d^*)$, and normalized so that $(n_k^*+n_l^*) = (n^* - n_d^*)$.
\end{enumerate}
All other parameters and hyperparameters are specified as in Section 5.1 of the paper. Table \ref{tab:sim3_abs_err12} (fourth column block ``Unbalanced components'') reports the absolute errors computed between estimated and inferred latent intakes and simulated values. Results are in agreement with those obtained in all previous scenarios (see Table 1 of the paper) and the other column blocks in Table \ref{tab:sim3_abs_err12} for a comparison. This suggests that the ability to recover intakes in the test data is not impacted by how the intakes are spread across the intake range.\\

\begin{table}[!h]
\caption{Simulation studies' results. Median ($95\%$ CI width) absolute error values (in grams) computed between true and estimated (E) or inferred (I) latent intakes. The values are reported for the cases $D=6$, ``Varying $\mathbf{X}$'', ``Uniform intakes'' and ``Unbalanced components'', and for the three biomarkers' variability scenarios (S.1,S.2,S.3), introduced in Section 5 of the manuscript. Results for the multiMarker (MM) model are reported, as well as those from Bayesian linear regression (BLR) and PLS (PLS) regression. }
\begin{center}
\label{tab:sim3_abs_err12}
\scalebox{0.7}[0.7]{
  \begin{tabular}{|cccccccccccccccccccc|}
  \toprule
  &&  & & \multicolumn{3}{c}{$D=6$} & &  \multicolumn{3}{c}{Varying $\mathbf{X}$}  & & \multicolumn{3}{c}{Uniform intakes} & & \multicolumn{3}{c}{Unbalanced components}  & \\
 &Model &  & & S.1 & S.2 & S.3 & &   S.1 & S.2 & S.3   & & S.1 & S.2 & S.3 & & S.1 & S.2 & S.3 &\\
\midrule
  \footnotesize
  & &  & &  &  &  & &  & & & &&&& &&&&\\
 & \multirow{2}{*}{MM} & E & & 4(11) & 5(16) & 14(15)  & & 4(9) & 6(52) & 12(15)   & & 4(7) & 5(8) & 10(48)  & & 4(8) & 5(8) & 9(38)  &\\
  & & I & & 4(15) & 6(47) & 33(59) & & 9(47) & 10(51) & 40(62)  && 10(21) & 22(38) & 37(74)  && 5(12) & 19(23) & 27(66) &\\
  & &  & &  &  &  & &  & & & &&&& &&&&\\
  & \multirow{2}{*}{BLR} & E & &  73(207) & 94(234) & 69(138) & & 66(145) & 81(60) & 76(70)  && 93(263) & 114(231) & 72(109)   && 108(206) & 142(228) & 98(117) &\\
  & & I & & 86(172) & 126(282) & 72(145)   & & 68(150) & 87(114) & 81(296)  && 88(256) & 110(245) & 96(187)  & & 96(187) & 128(275) & 96(169)  & \\
   & &  & &  &  &  & &  & & & &&&& &&&&\\
  & \multirow{2}{*}{PLS} & E & & 8(19) & 21(46) & 34(49) & & 19(60) & 46(70) & 14(15)  && 10(15) & 23(30) & 43(61)  && 10(25) & 20(31) & 38(65)  &\\
 & & I & & 8(26) & 22(51) & 38(58)  & &  10(65) & 19(67) & 49(83)   && 11(20) & 25(45) & 47(74)   & & 11(29) & 22(44) & 39(81)  &\\
 \bottomrule
  \end{tabular}
  }
  \end{center}
\end{table}

\subsection{Simulation study: examples of simulated data}
\begin{figure}[!h]
  \centering
     \subfloat[][\emph{Simulation study II. Examples of simulated latent intakes in train and test data.}]
{\includegraphics[height=10cm, width=13cm]{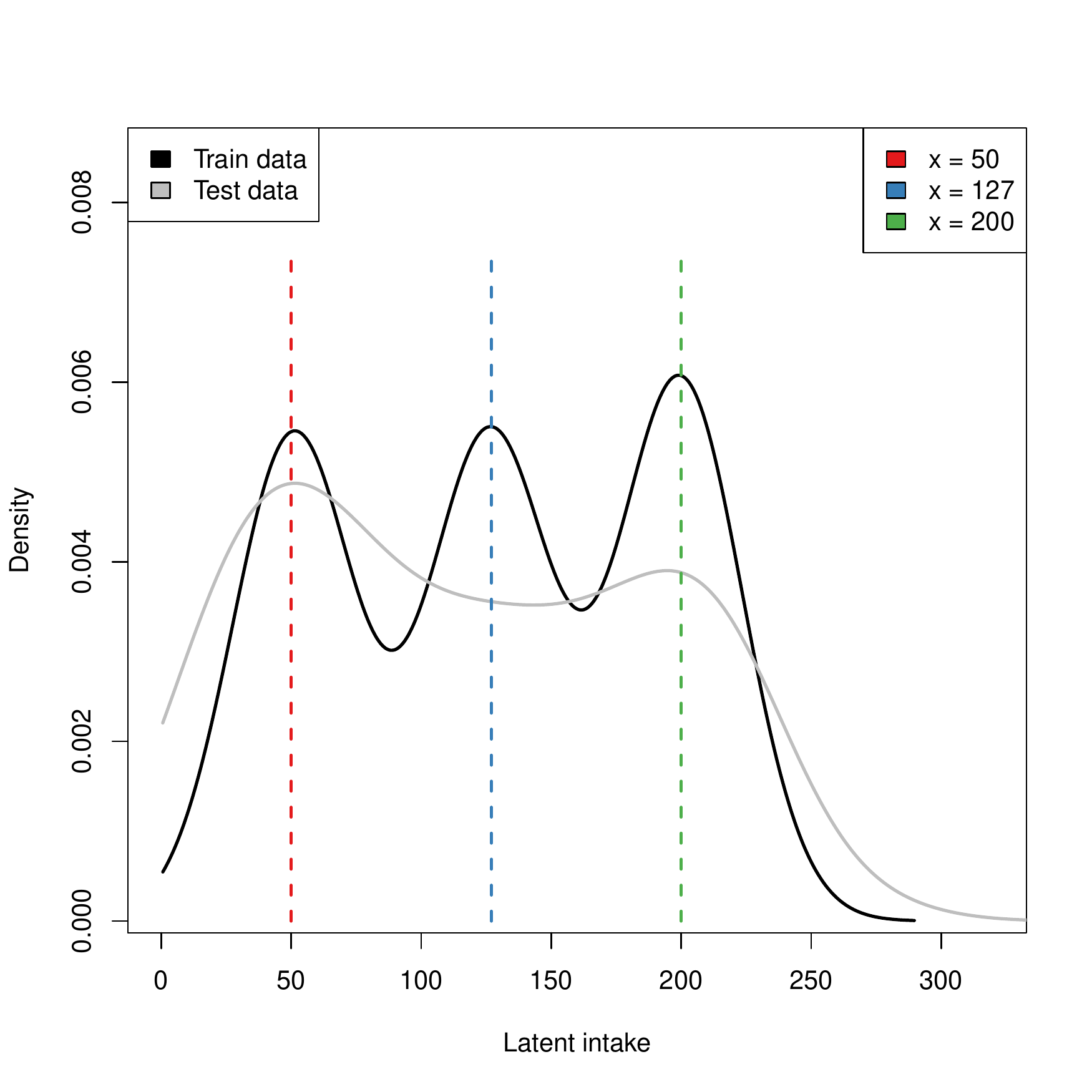}}\\[-7ex]  
\subfloat[][\emph{Simulation studies I and III. Examples of simulated biomarker data. }]
{\includegraphics[height=10cm, width=13cm]{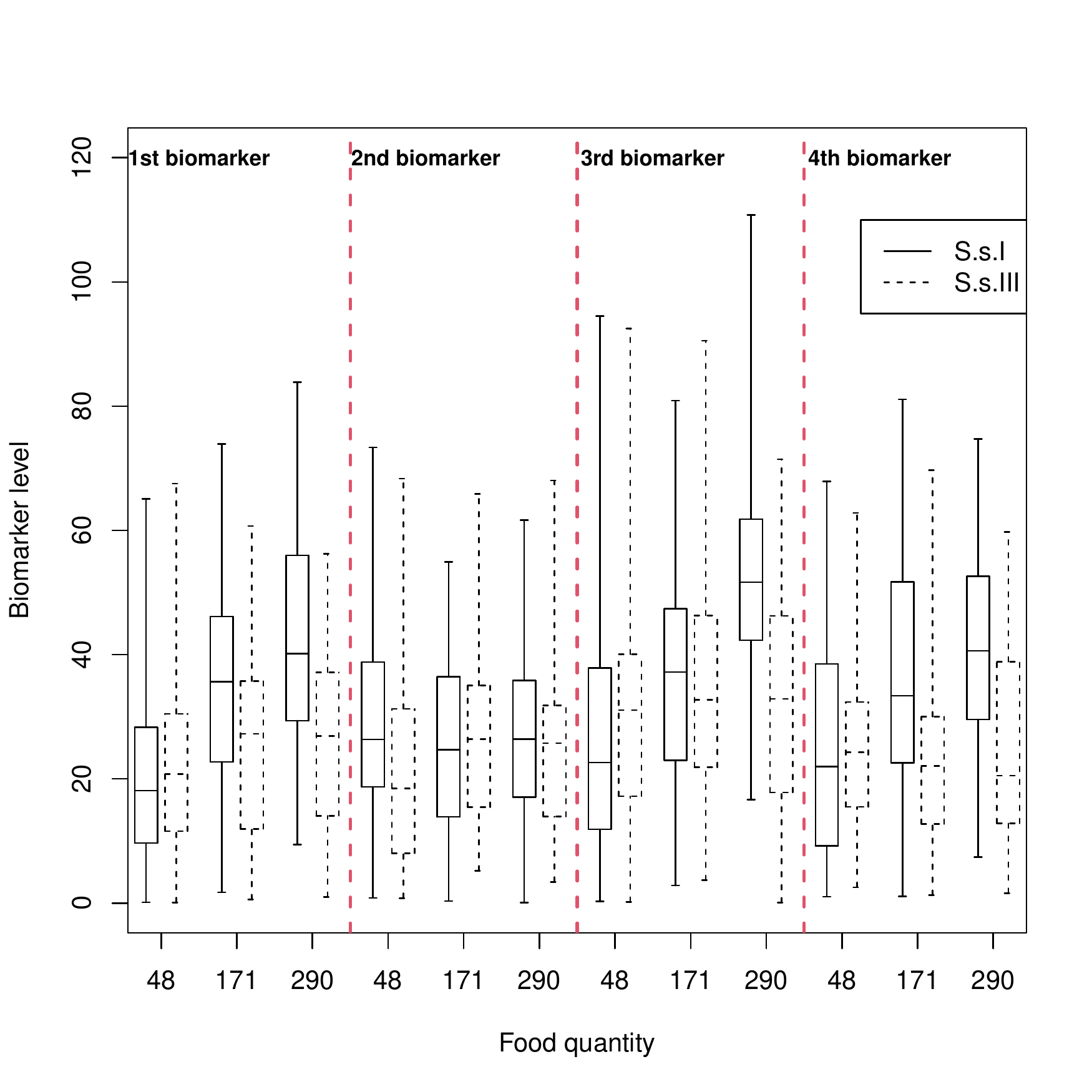}}\\[-1ex] 
\caption{Simulation study. Example of simulated latent intakes (study II) (a) and biomarker data (studies I, III) (b).}
    \label{fig:exampledata_sim}
\end{figure}

Figure \ref{fig:exampledata_sim} shows an example of simulated latent intakes in train and test data, under Simulation study II. In Figure \ref{fig:exampledata_sim} (a), an example of simulated latent intakes in simulation study II is reported, with intakes from the train data and test data depicted in black and grey, respectively. Simulated food quantity values are also reported, to show how in the test data increased $\theta_d^2$ values, $d=1,\dots,D$, lead to a flatter latent intake density. Indeed, latent intakes are no longer clearly concentrated around the food quantities, but are more widely spread across the intake range (roughly $0$ to $310$). Figure \ref{fig:exampledata_sim} (b) reports a comparison between simulated biomarker data under simulation studies I and III. Simulated biomarkers are in the medium range scenario, with food quantities presenting ``stable increments'' and under variability scenario S.1. 

\subsection{Estimated model parameters}
\begin{figure}[!h]
    \centering
     \subfloat[][\emph{Small biomarker data range.}]
{\includegraphics[scale=.55]{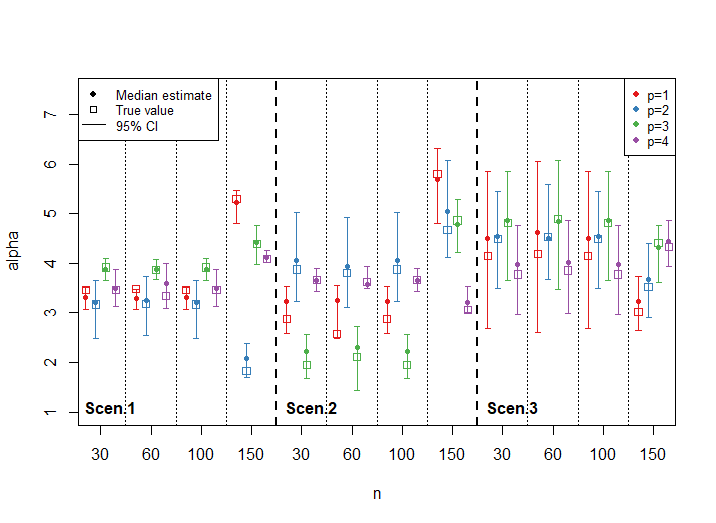}}\\[-7ex]  
 \subfloat[][\emph{Medium biomarker data range.}]
{\includegraphics[scale=.55]{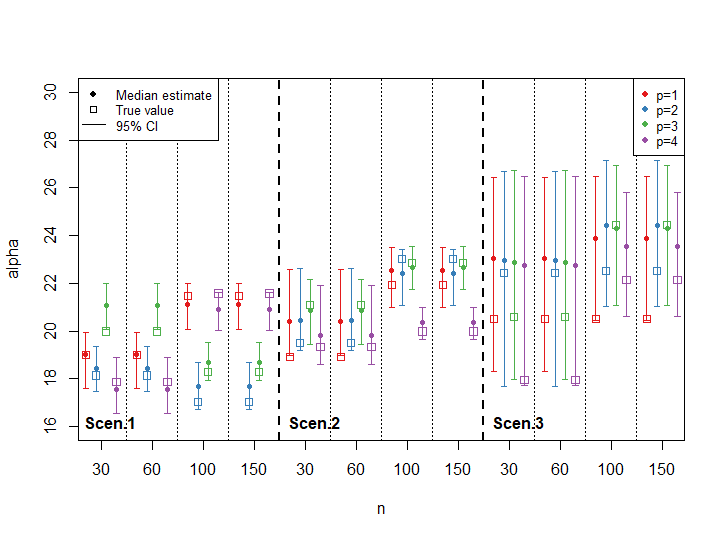}}\\[-7ex]  
\subfloat[][\emph{Large biomarker data range.}]
{\includegraphics[scale=.55]{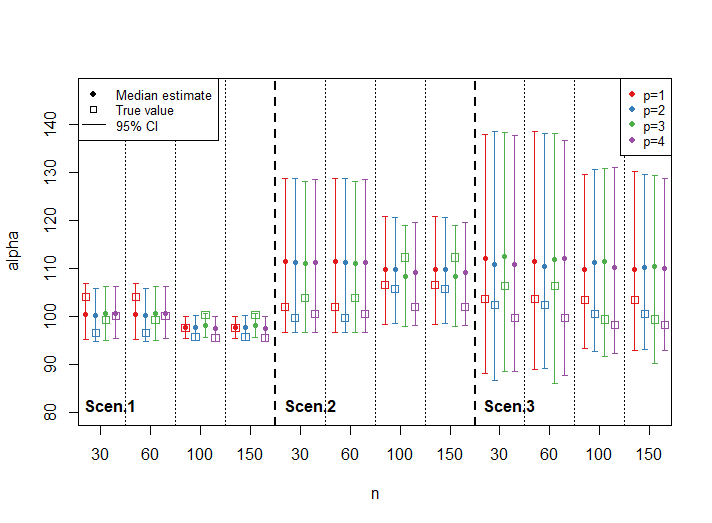}}\\[-1ex]  
 \caption{Intercept parameters ($\alpha_1, \dots, \alpha_P$). Dots, lines and squares represent, respectively, median estimates, $95\%$ credible interval estimates and true values. Four different colors represent parameter referred to different biomarkers. Results are reported split by sample size and variability scenario. }
 \label{fig:estalpha}
\end{figure}

\begin{figure}[!h]
    \centering
     \subfloat[][\emph{Small biomarker data range.}]
{\includegraphics[scale=.55]{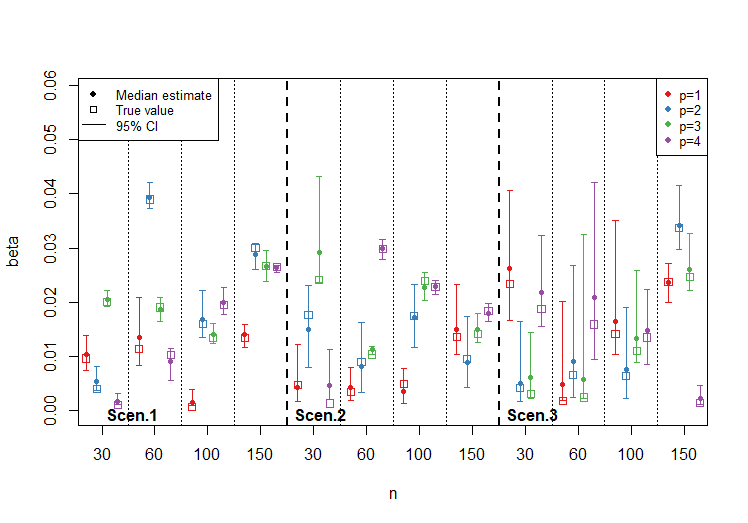}}\\[-7ex] 
 \subfloat[][\emph{Medium biomarker data range.}]
{\includegraphics[scale=.55]{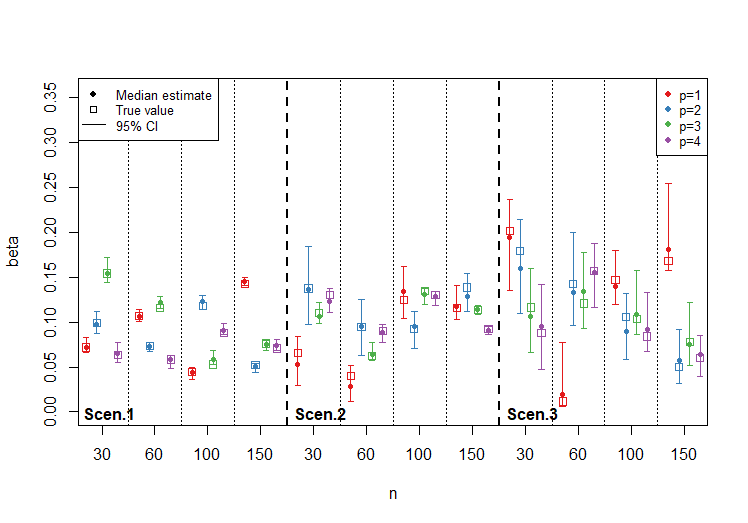}}\\[-7ex] 
\subfloat[][\emph{Large biomarker data range.}]
{\includegraphics[scale=.55]{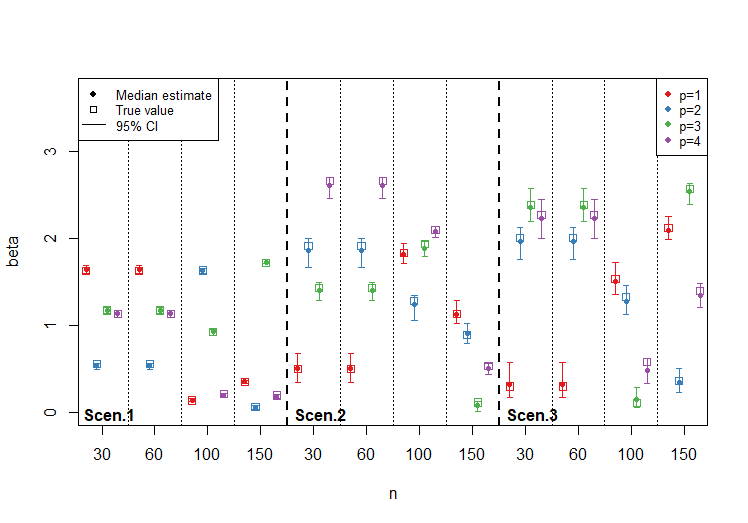}}\\[-1ex] 
 \caption{Scale coefficient parameters ($\beta_1, \dots, \beta_P$). Dots, lines and squares represent, respectively, median estimates, $95\%$ credible interval estimates and true values. Four different colors represent parameter referred to different biomarkers. Results are reported split by sample size and variability scenario.}
 \label{fig:estbeta}
\end{figure}

\begin{figure}[!h]
    \centering
     \subfloat[][\emph{Small biomarker data range.}]
{\includegraphics[scale=.55]{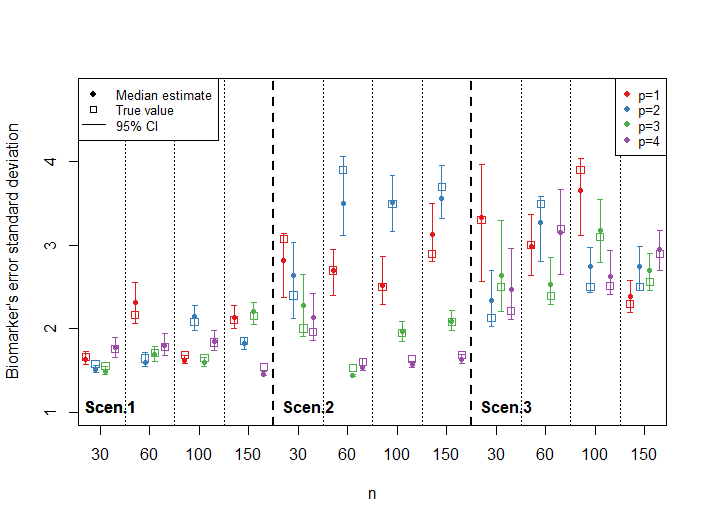}}\\[-7ex] 
 \subfloat[][\emph{Medium biomarker data range.}]
{\includegraphics[scale=.55]{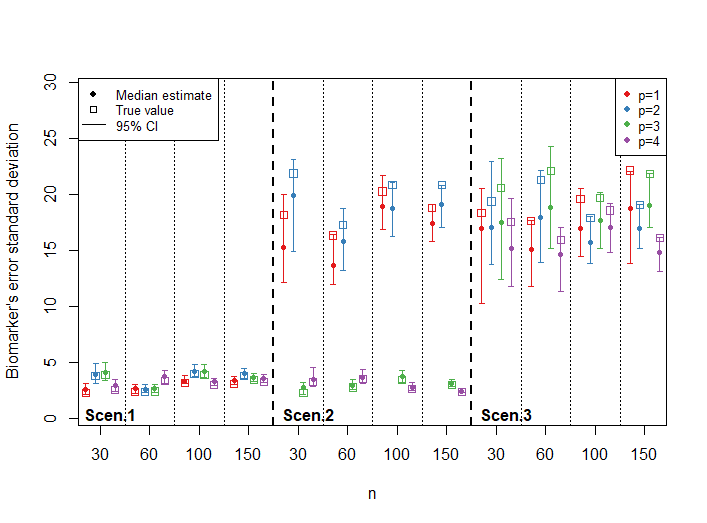}}\\[-7ex] 
\subfloat[][\emph{Large biomarker data range.}]
{\includegraphics[scale=.55]{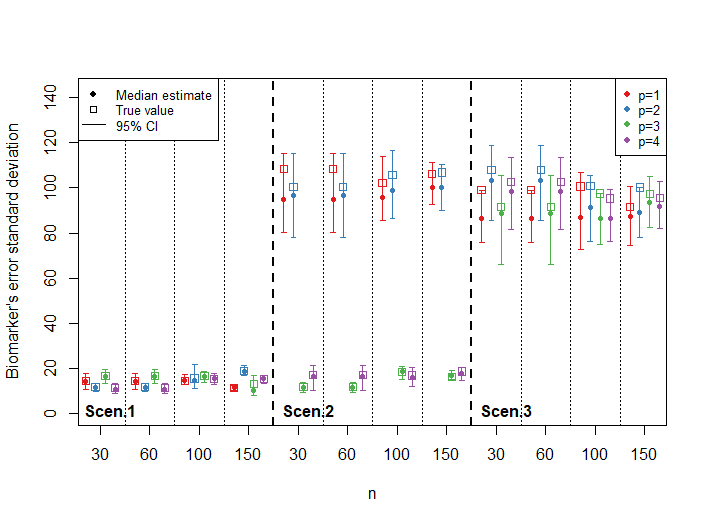}}\\[-1ex] 
 \caption{Biomarker's error standard deviations ($\sigma_1, \dots, \sigma_P$). Dots, lines and squares represent, respectively, median estimates, $95\%$ credible interval estimates and true values. Four different colors represent parameter referred to different biomarkers. Results are reported split by sample size and variability scenario.}
 \label{fig:estvp}
\end{figure}

Figures \ref{fig:estalpha}, \ref{fig:estbeta} and \ref{fig:estvp} show the median and $95\%$ credible interval estimates, respectively for the $\alpha_p$, $\beta_p$ and $\sigma_p$ parameters, $\forall p$, under Simulation study I. 
Medians and $95\%$ credible intervals are computed using the subsets of simulated intervention study datasets (and corresponding parameter estimates) that shared the same true simulated value of the parameter of interest. 
True simulated parameter values are reported for a comparison, and results are reported separately for the different error variability scenarios and biomarker data ranges.

\section{Apple intake data}
\subsection{Data scaling}
 The original values for Epicatechin Sulfate, $(4-3-[2-(2,4-dihydroxyphenyl)-2-oxoethyl]-DHMPMB-SA)$ and Glucodistylin biomarkers caused computational instability (most values were larger than $10^7$) and consequently were scaled. Given a biomarker $p$, the corresponding scaled values $\tilde{y}_{ip}$ are computed as follows: 
\[\tilde{y}_{ip} = \frac{y_{ip} - \bar{y_{p}}}{sd(y_{p})}  + 2  \bigg| \min_{i=1,\dots,n} \Bigl(\frac{y_{ip}- \bar{y_{p}}}{sd(y_{p})} \Bigr)\bigg|
\]
where $y_{ip}$ are the original measurements, $i=1,\dots,n$. The original measurements mean and standard deviation are denoted with $\bar{y_{p}}$ and $sd(y_{p})$, respectively. The transformation did not alter the correlations between the four biomarkers.
\subsection{Comparison to BLR and PLS regression}
\begin{figure}[!h]
    \centering
   \subfloat[][]
{\includegraphics[scale=.61]{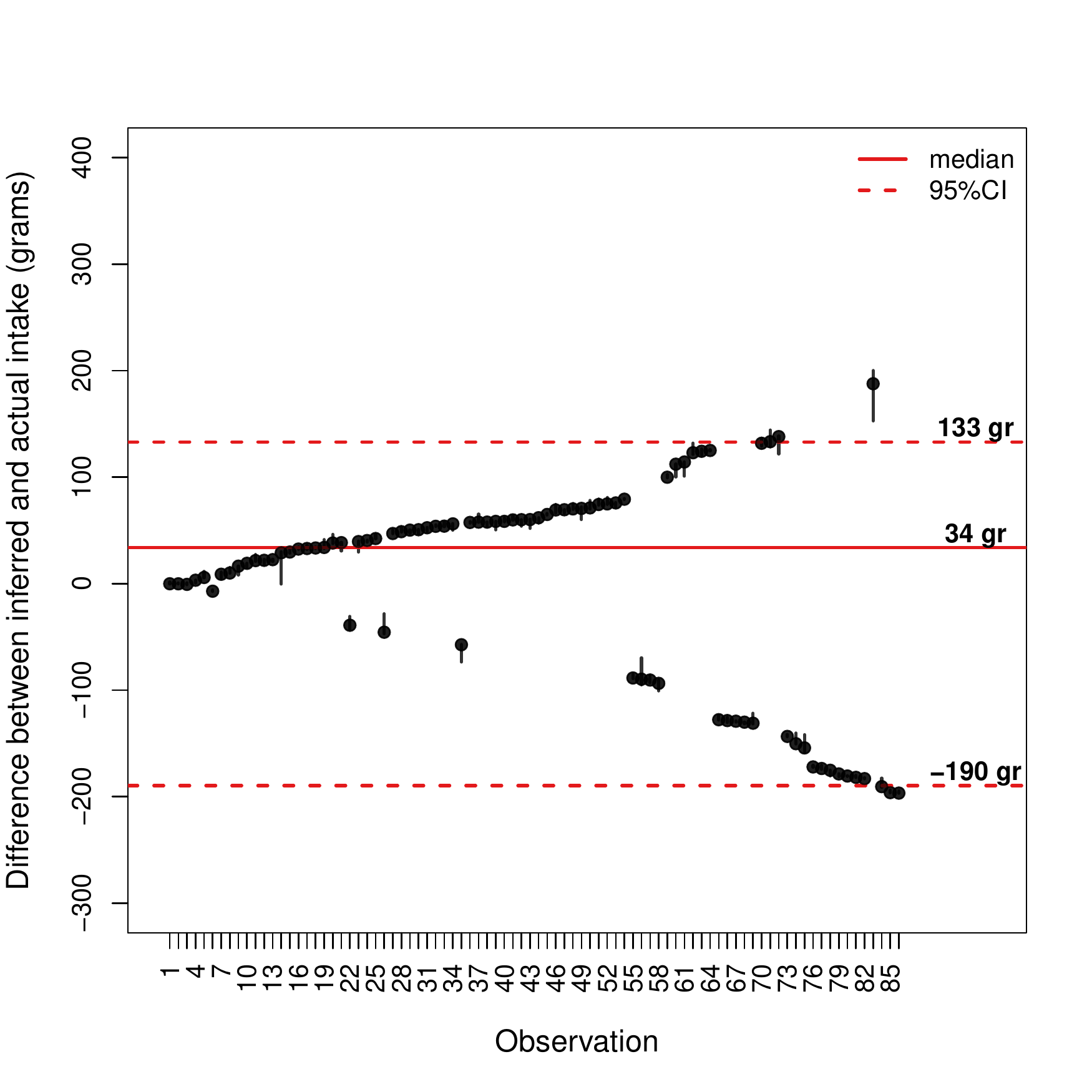}}\\[-7ex] 
 \subfloat[][]
 {\includegraphics[scale=.61]{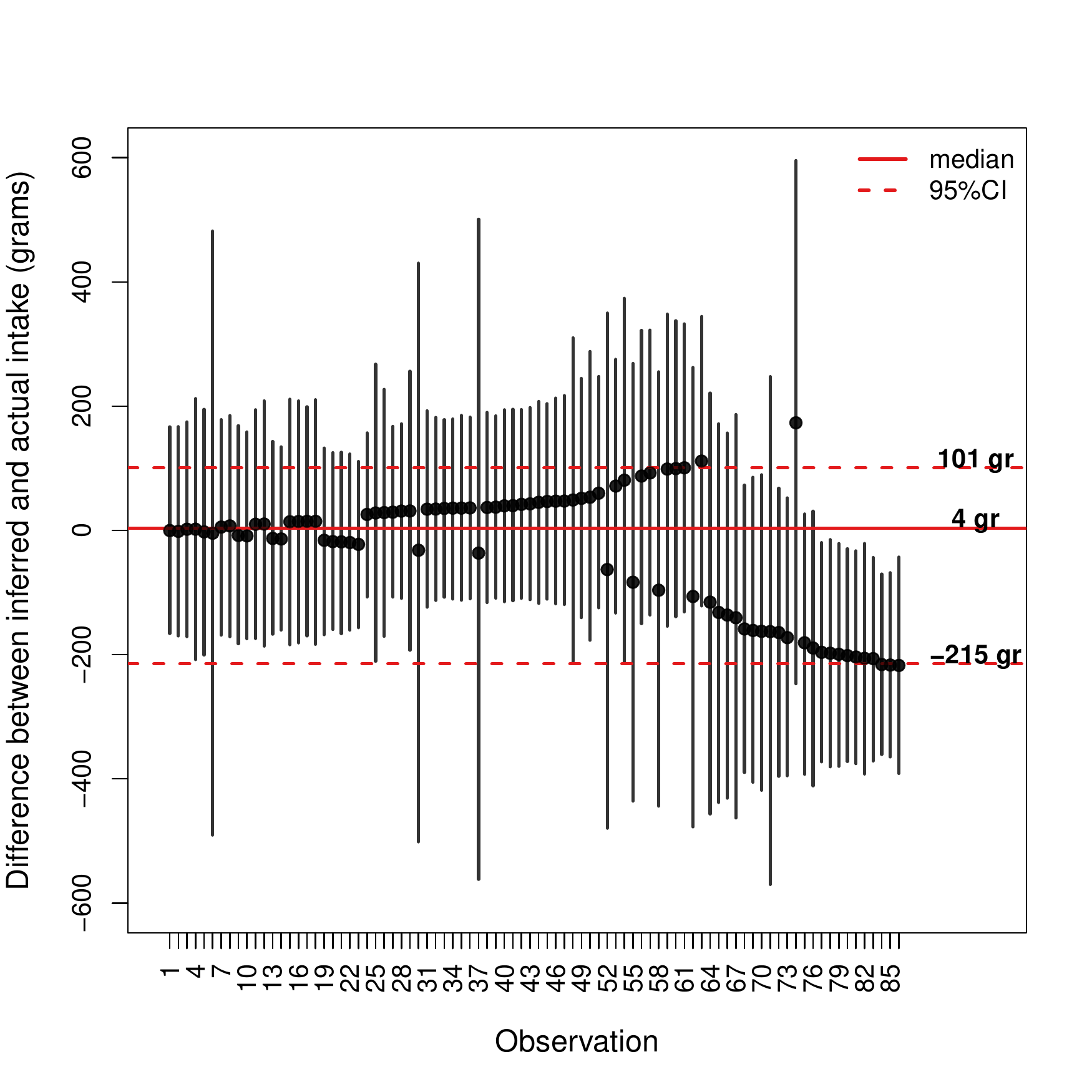}}\\[-1ex] 
 \caption{Inferred intakes (in grams), obtained using leave one out cross validation and (a) fitting the PLS regression model, and (b) fitting the Bayesian linear regression model. These plots are analogous to that in Figure 4 of the paper. }
    \label{fig:applesdata1-extra}
\end{figure}

Figure \ref{fig:applesdata1-extra} reports plots that are analogous to those in Figure 4 of the paper, obtained using either BLR or PLS regression, for a comparison. 

\subsection{MCMC diagnostics}
\begin{figure}[!h]
    \centering
     \subfloat[][\emph{Xylose.}]
{\includegraphics[width=7cm,height=7cm,]{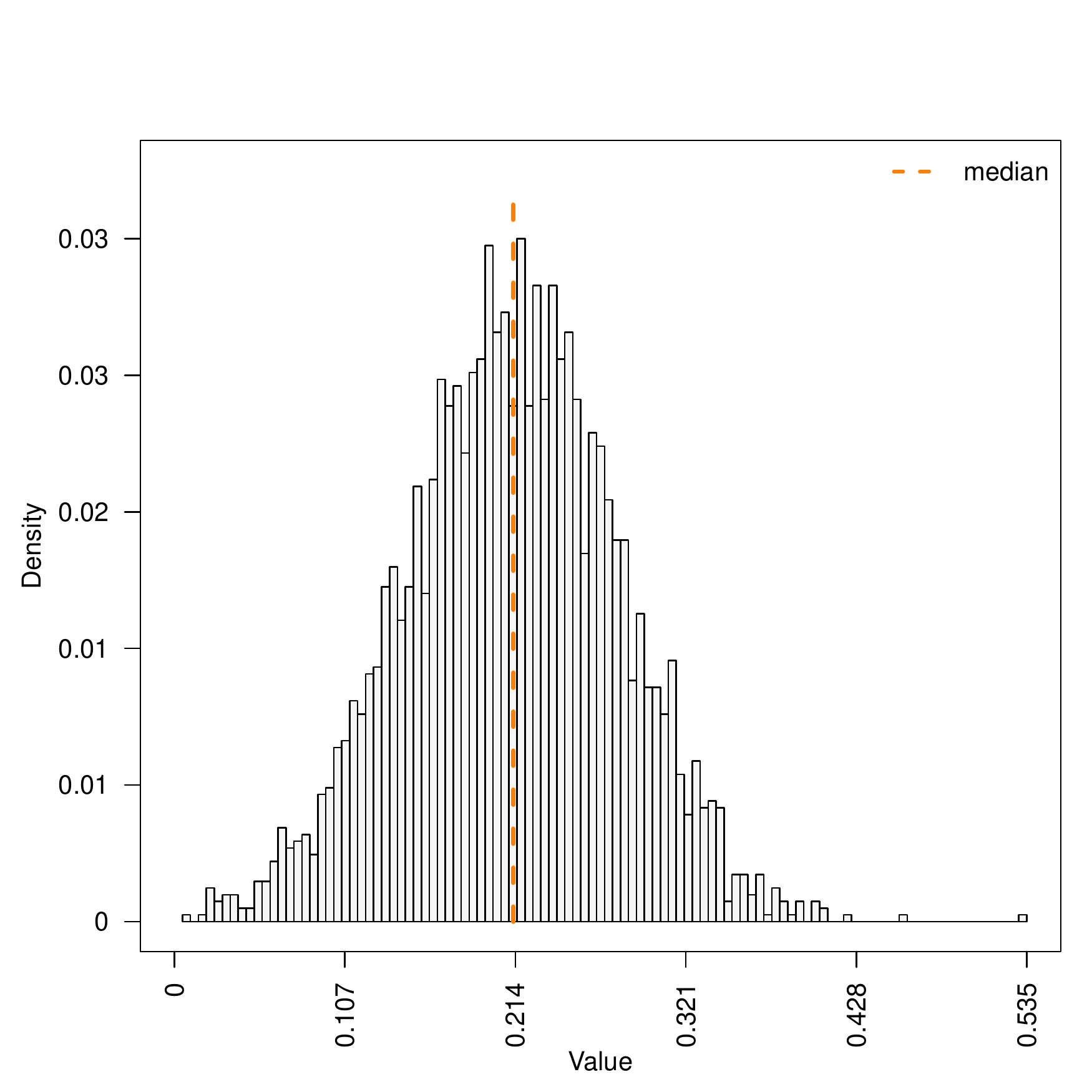}}\quad
 \subfloat[][ \emph{Epicatechin Sulfate}.]
{\includegraphics[width=7cm,height=7cm,]{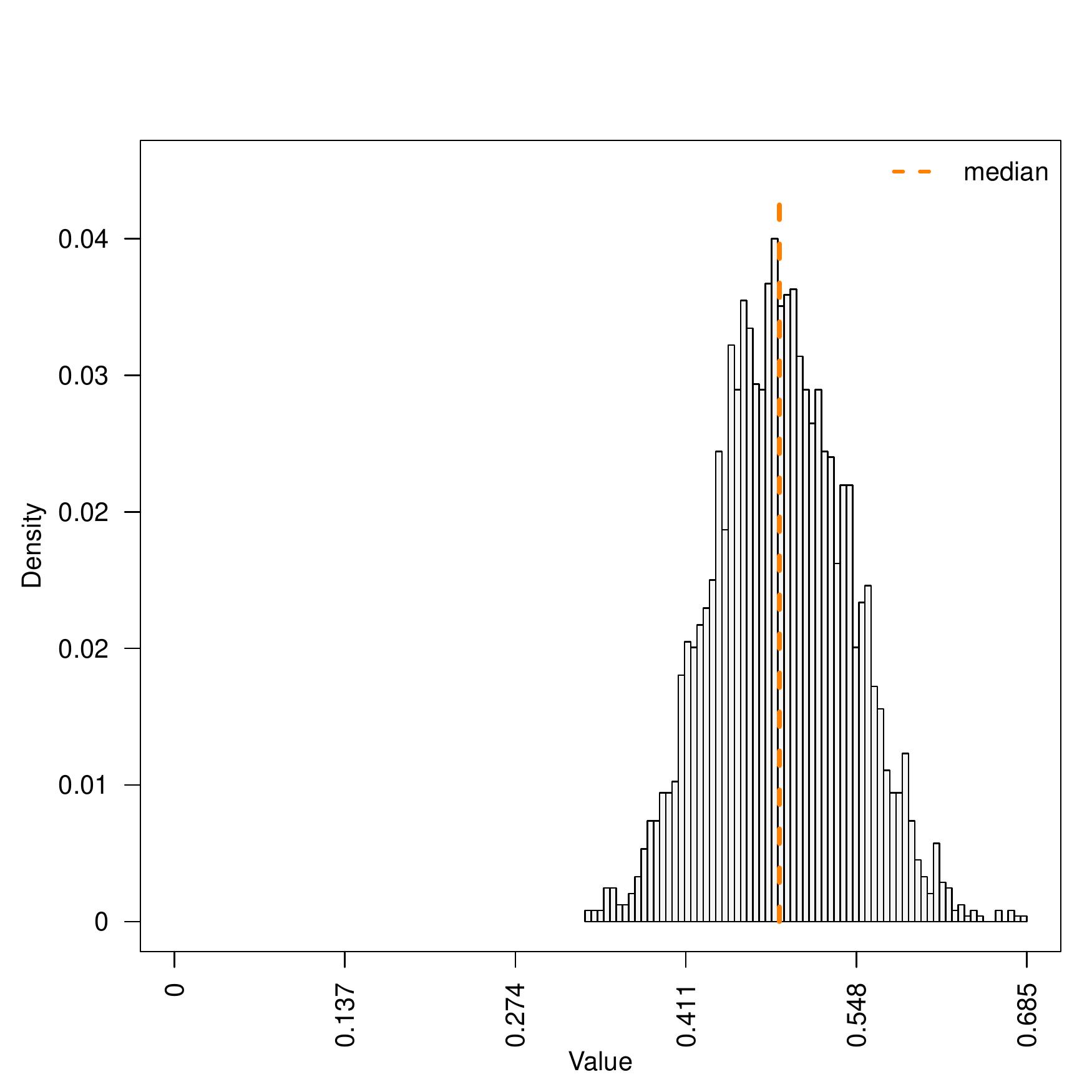}}
\quad 
\subfloat[][\emph{$(4-3-[2-(2,4-dihydroxyphenyl)-2-oxoethyl]-DHMPMB-SA)$}.]
{\includegraphics[width=7cm,height=7cm,]{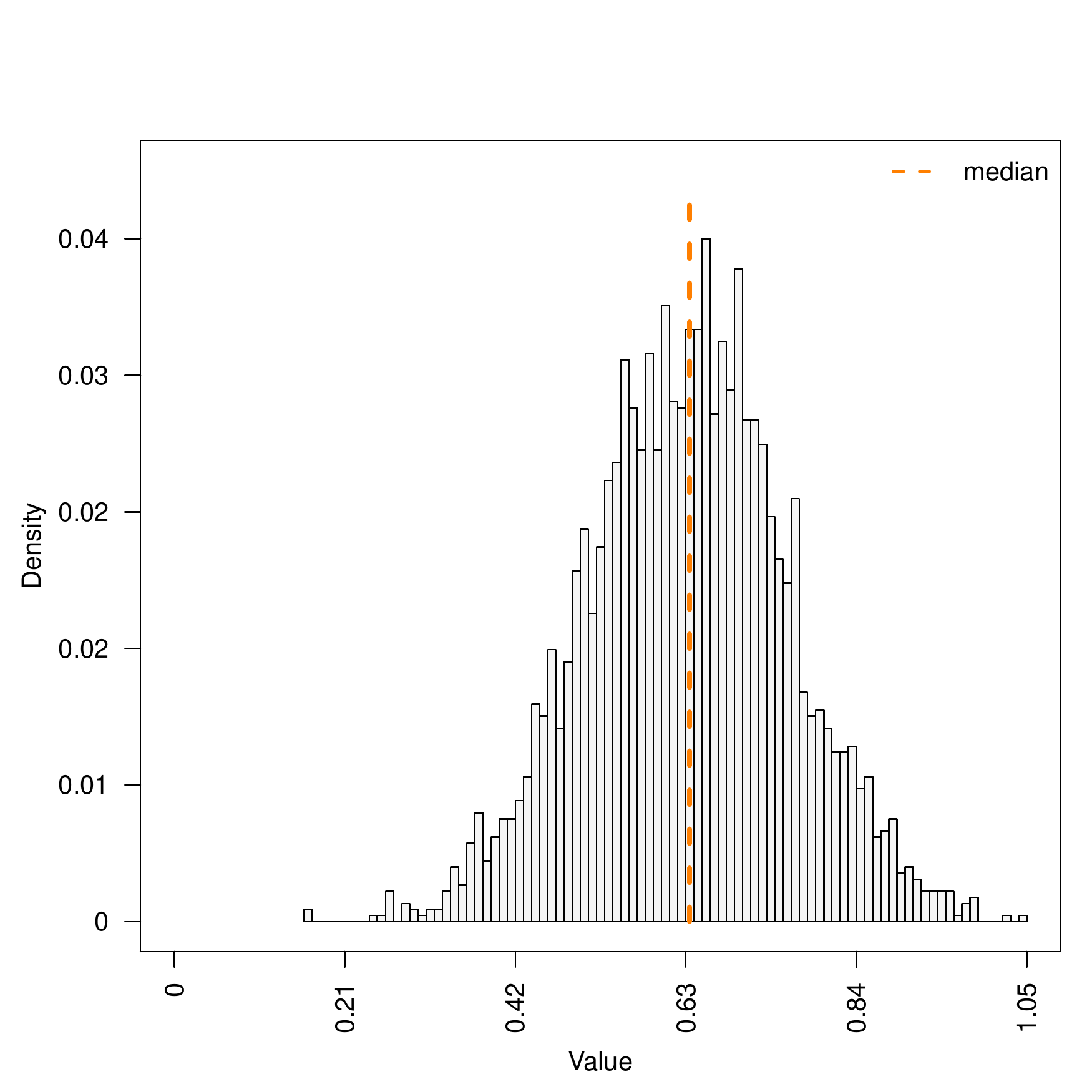}}\quad
 \subfloat[][\emph{Glucodistylin}.]
 {\includegraphics[width=7cm,height=7cm,]{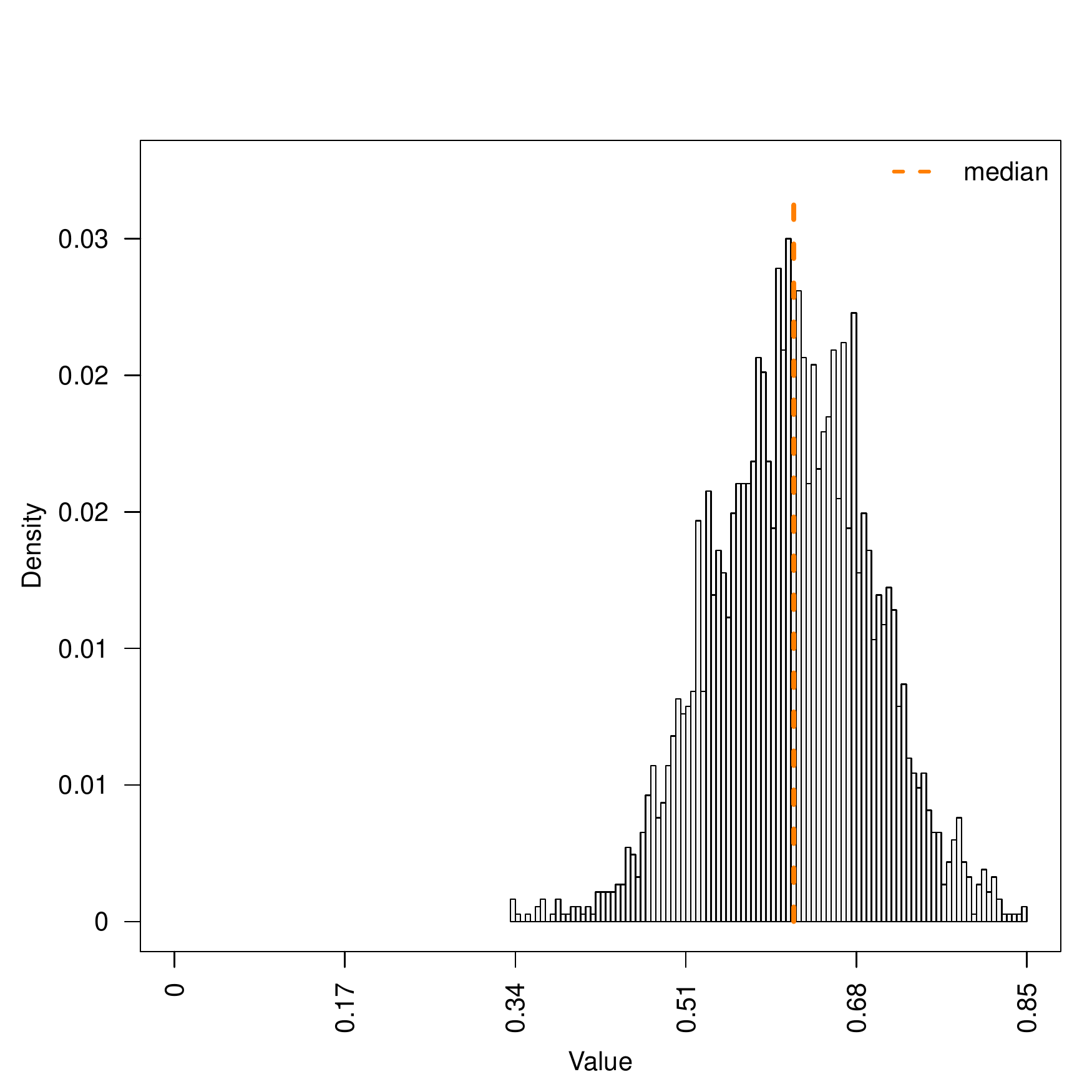}}
 \caption{Apple data. Estimated posterior distributions for the $\alpha_p$ parameters, $p=1,\dots,P$.}
  \label{fig:applesdata1-extra1}
\end{figure}

\begin{figure}[!h]
    \centering
     \subfloat[][\emph{Xylose.}]
{\includegraphics[width=7cm,height=7cm,]{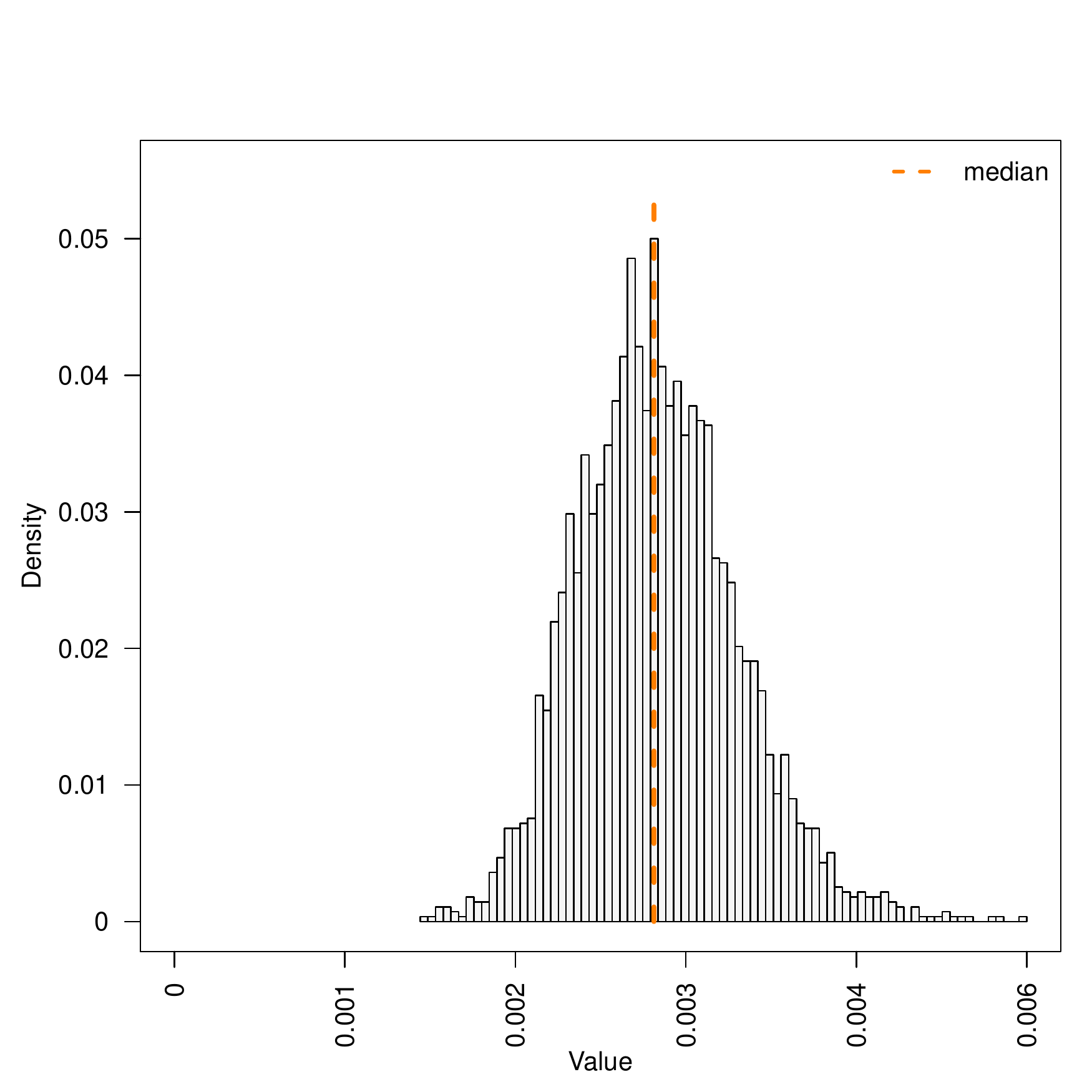}}\quad
 \subfloat[][ \emph{Epicatechin Sulfate}.]
{\includegraphics[width=7cm,height=7cm,]{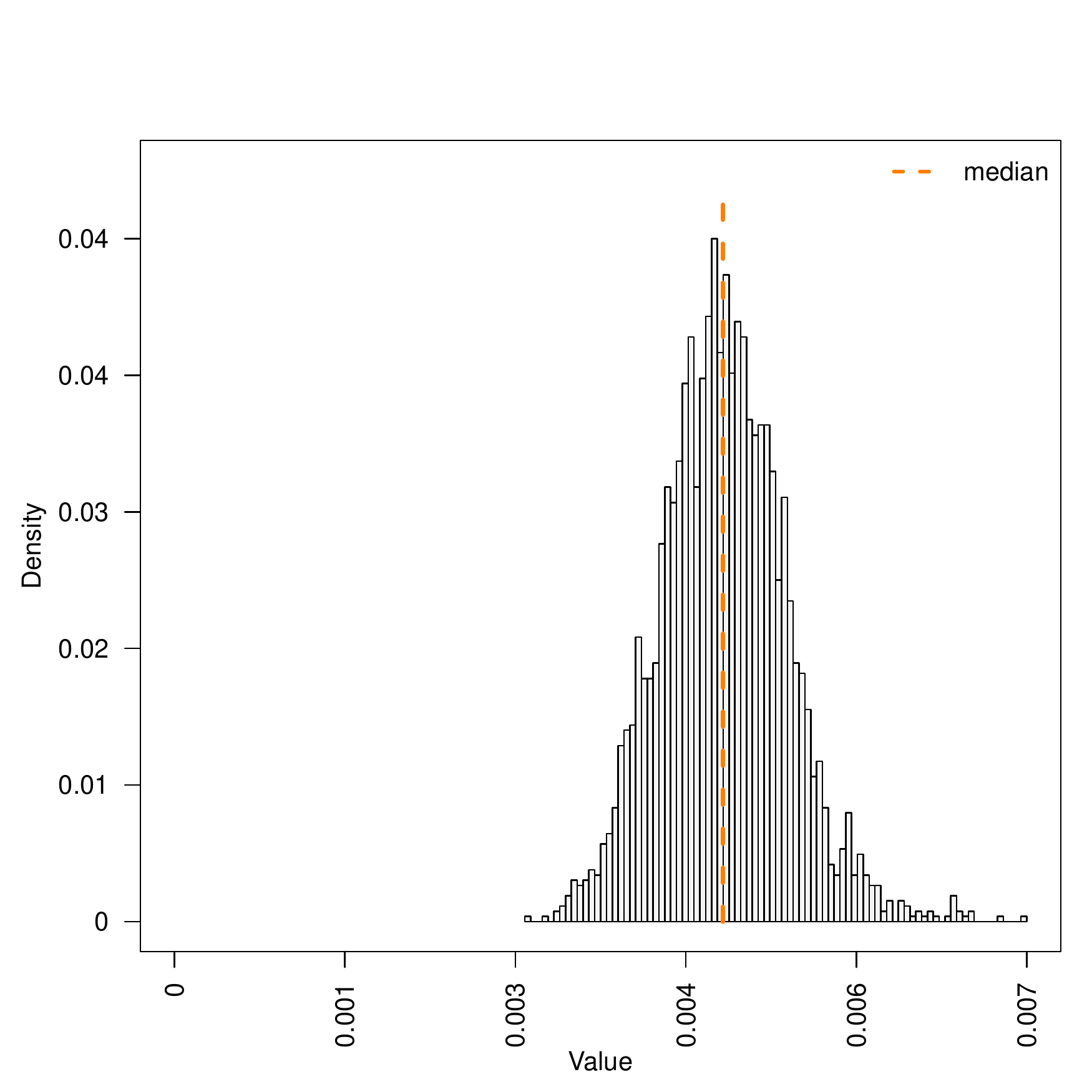}}
\quad 
\subfloat[][\emph{$(4-3-[2-(2,4-dihydroxyphenyl)-2-oxoethyl]-DHMPMB-SA)$}.]
{\includegraphics[width=7cm,height=7cm,]{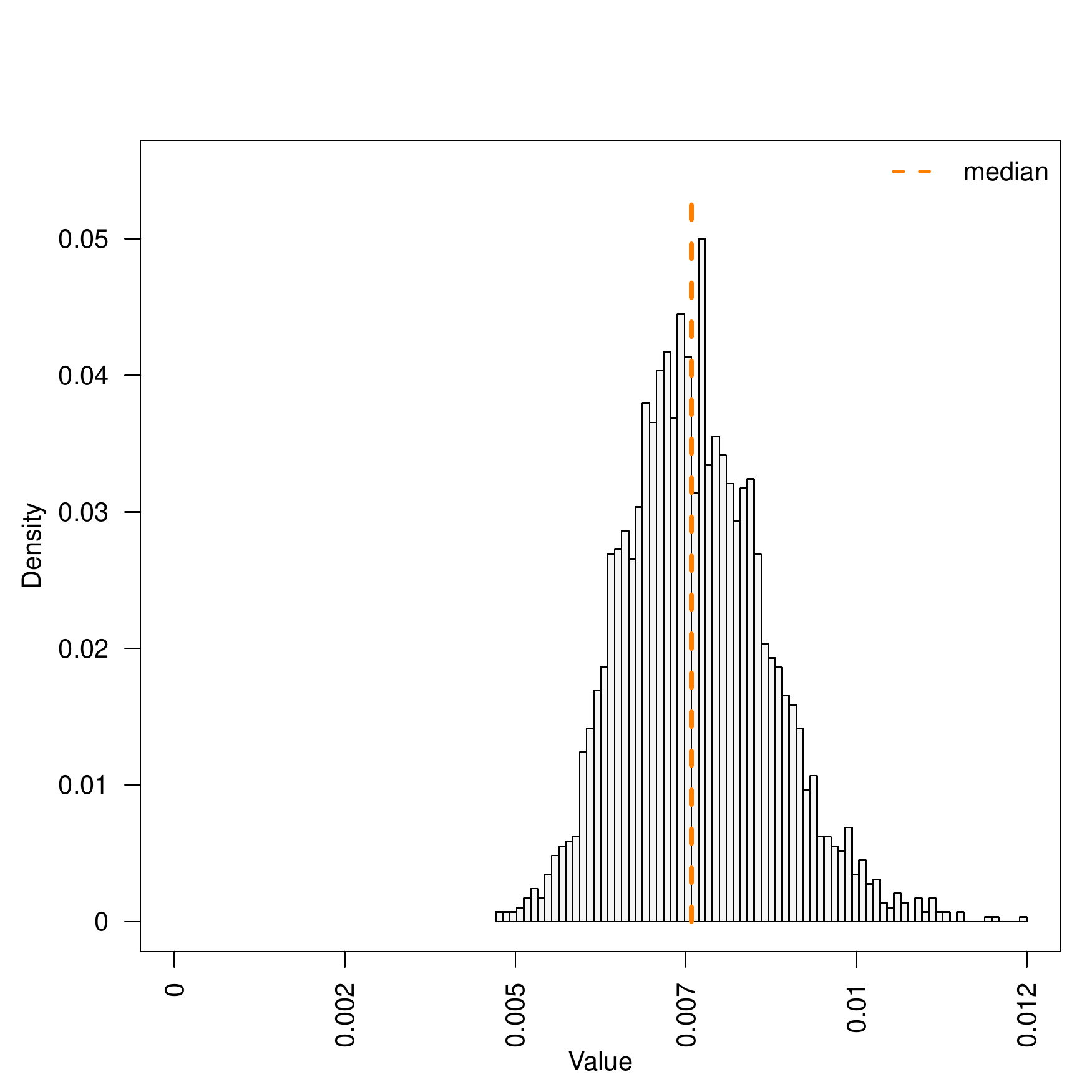}}\quad
 \subfloat[][\emph{Glucodistylin}.]
 {\includegraphics[width=7cm,height=7cm,]{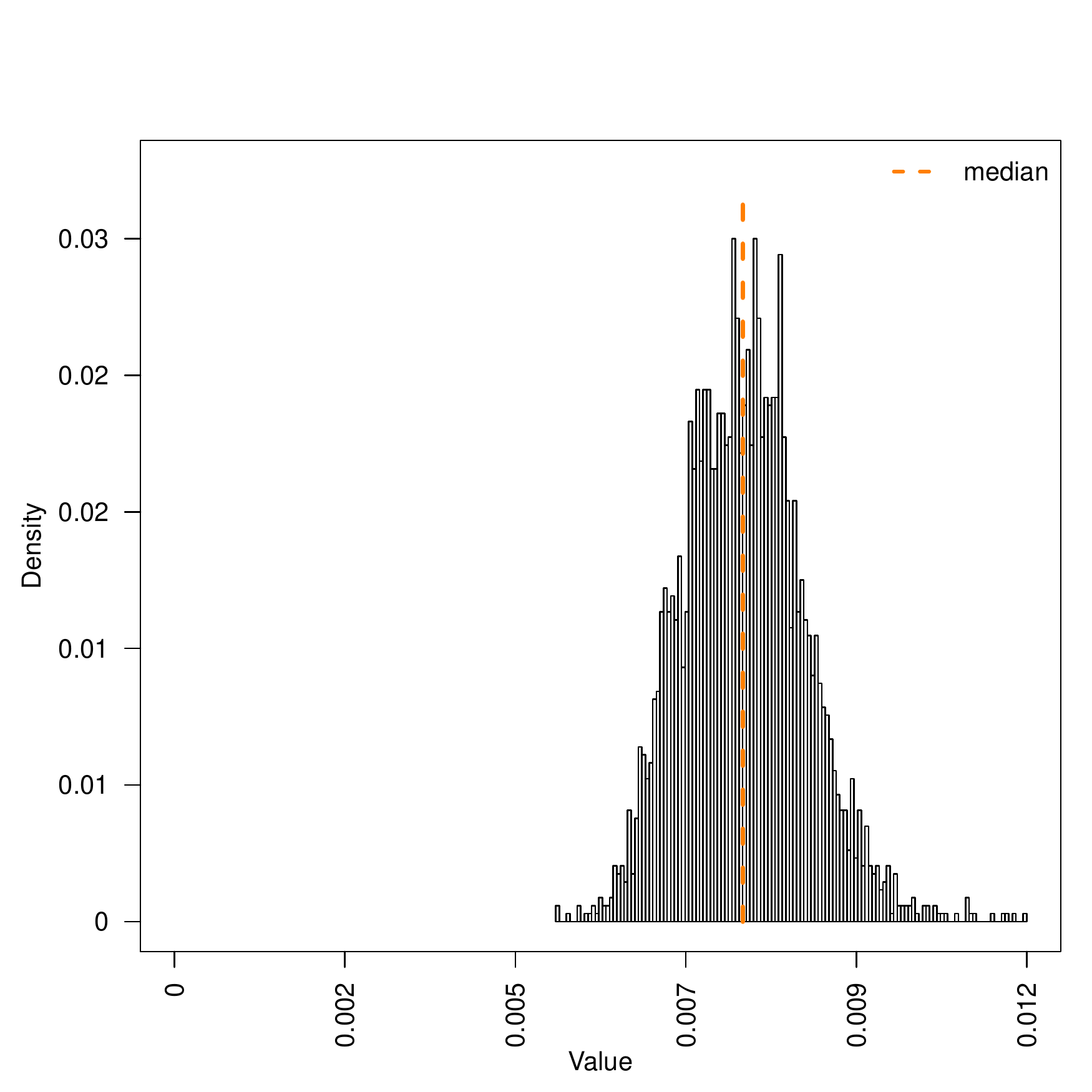}}
 \caption{Apple data. Estimated posterior distributions for the $\beta_p$ parameters, $p=1,\dots,P$.}
  \label{fig:applesdata1-extra2}
\end{figure}

\begin{figure}[!h]
    \centering
     \subfloat[][\emph{Xylose}.]
{\includegraphics[width=7cm,height=7cm,]{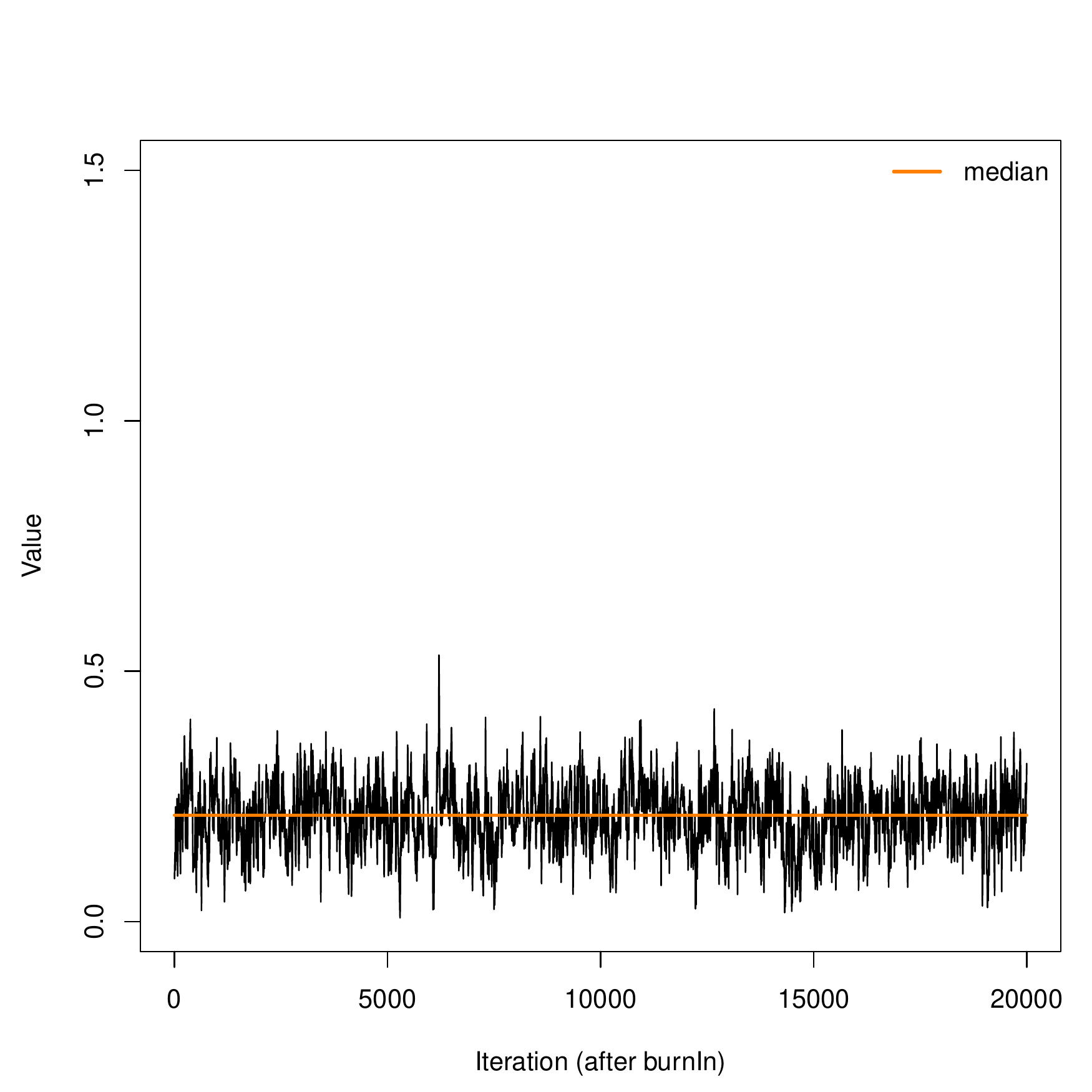}}\quad
 \subfloat[][\emph{Epicatechin Sulfate}.]
{\includegraphics[width=7cm,height=7cm,]{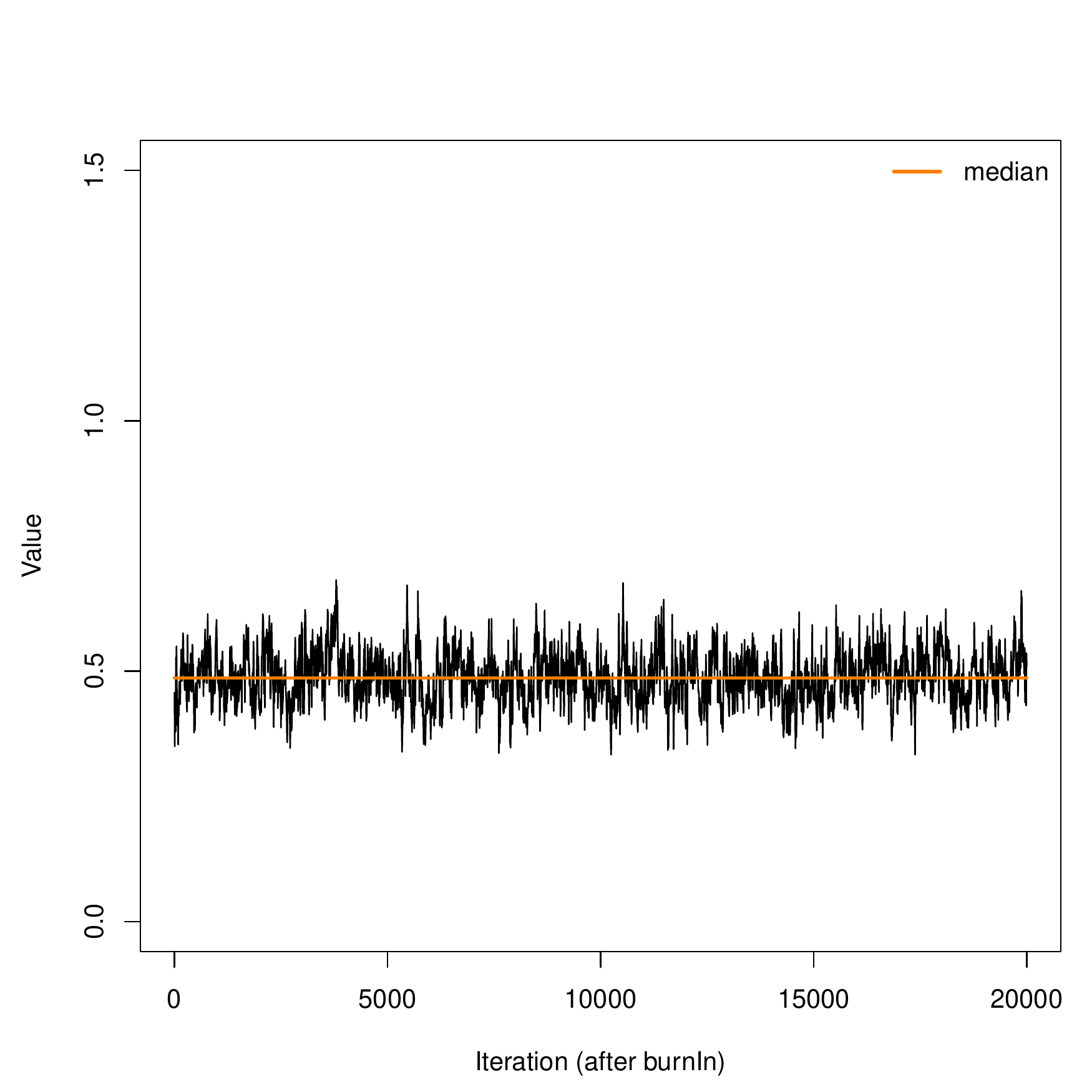}}
\quad 
\subfloat[][\emph{$(4-3-[2-(2,4-dihydroxyphenyl)-2-oxoethyl]-DHMPMB-SA)$}.]
{\includegraphics[width=7cm,height=7cm,]{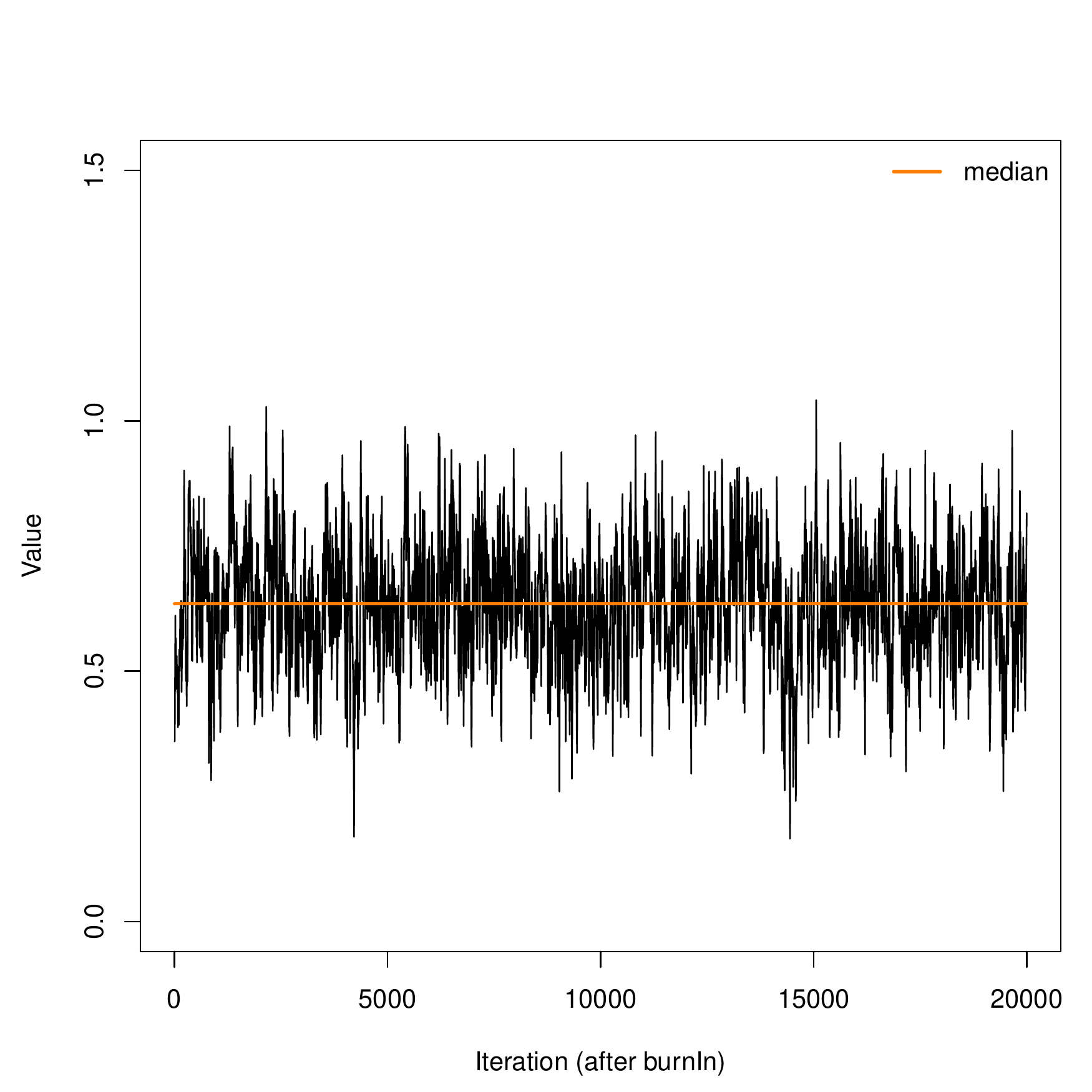}}\quad
 \subfloat[][\emph{Glucodistylin}.]
 {\includegraphics[width=7cm,height=7cm,]{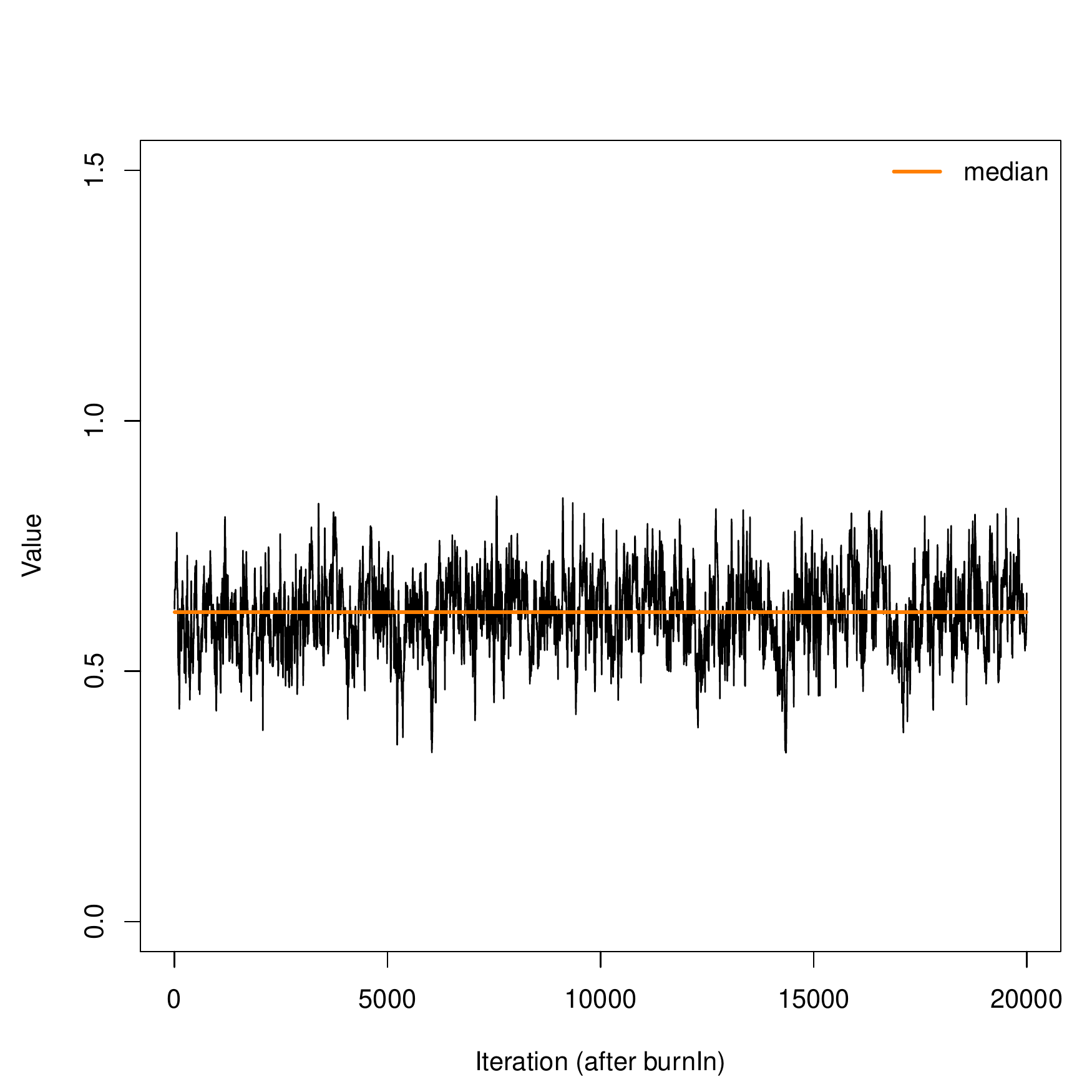}}
 \caption{Apple data. Trace plots for the estimated $\alpha_p$ parameters, $p=1,\dots,P$.}
  \label{fig:applesdata1-extra3}
\end{figure}

\begin{figure}[!h]
    \centering
     \subfloat[][\emph{Xylose.}]
{\includegraphics[width=7cm,height=7cm,]{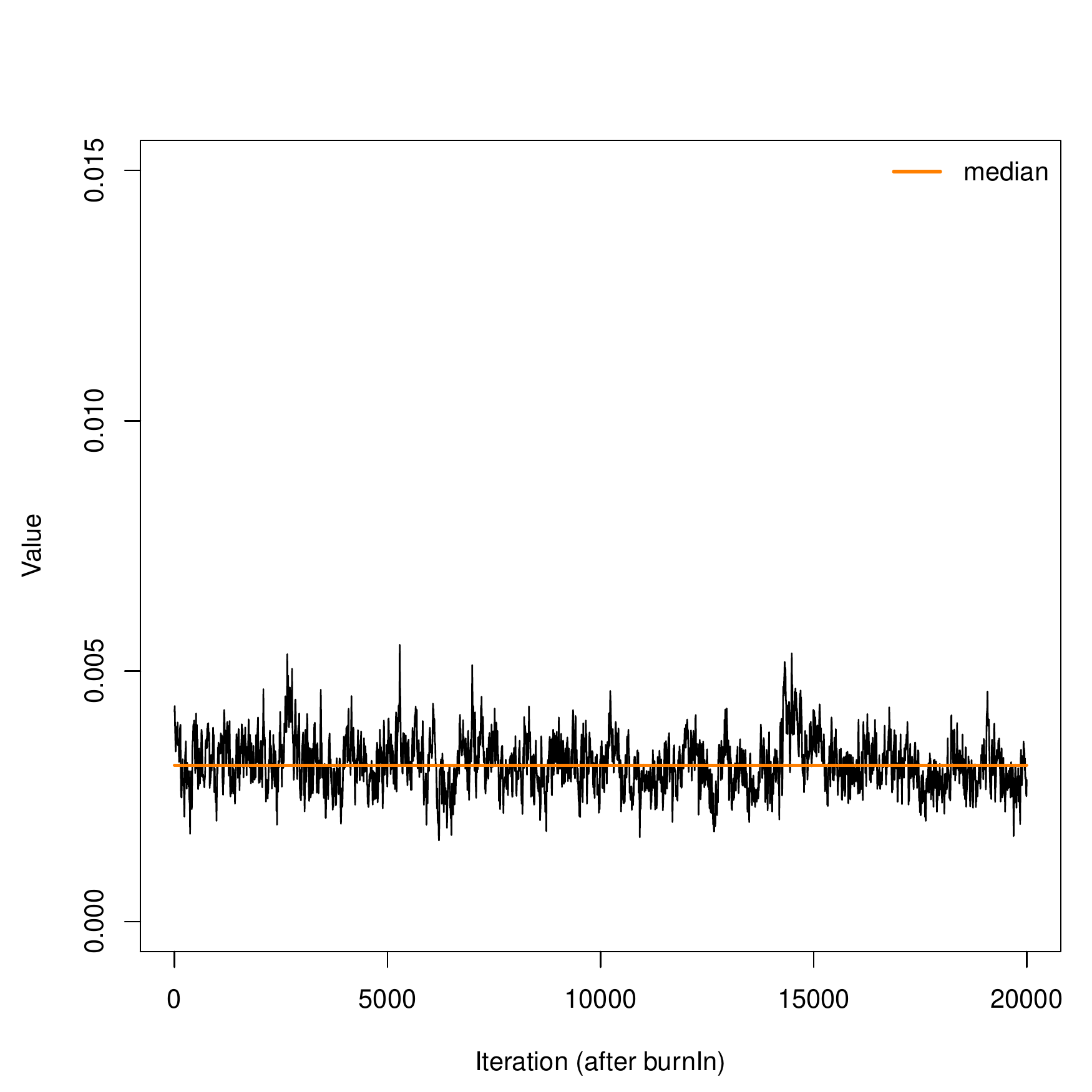}}\quad
 \subfloat[][\emph{Epicatechin Sulfate}.]
{\includegraphics[width=7cm,height=7cm,]{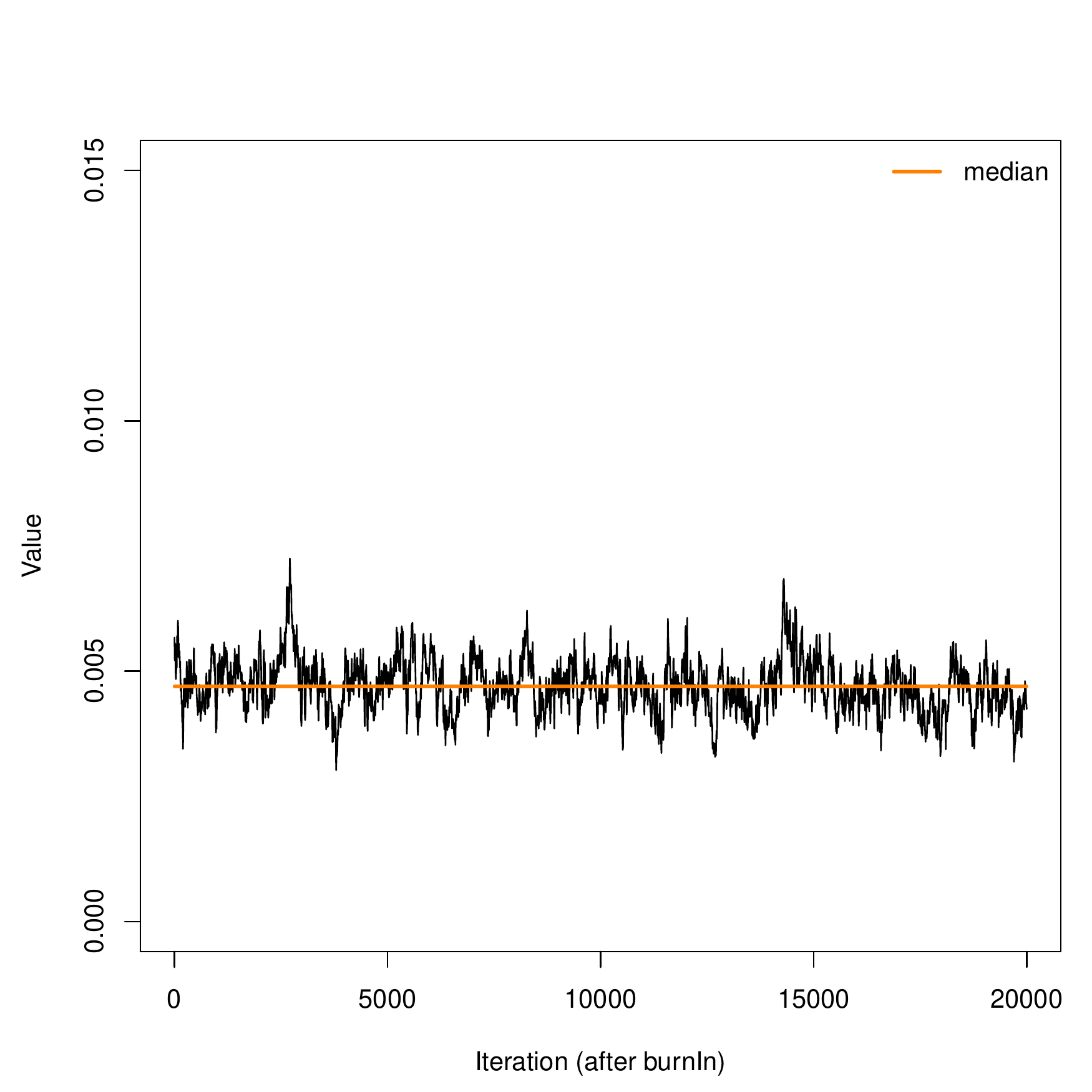}}
\quad 
\subfloat[][\emph{$(4-3-[2-(2,4-dihydroxyphenyl)-2-oxoethyl]-DHMPMB-SA)$}.]
{\includegraphics[width=7cm,height=7cm,]{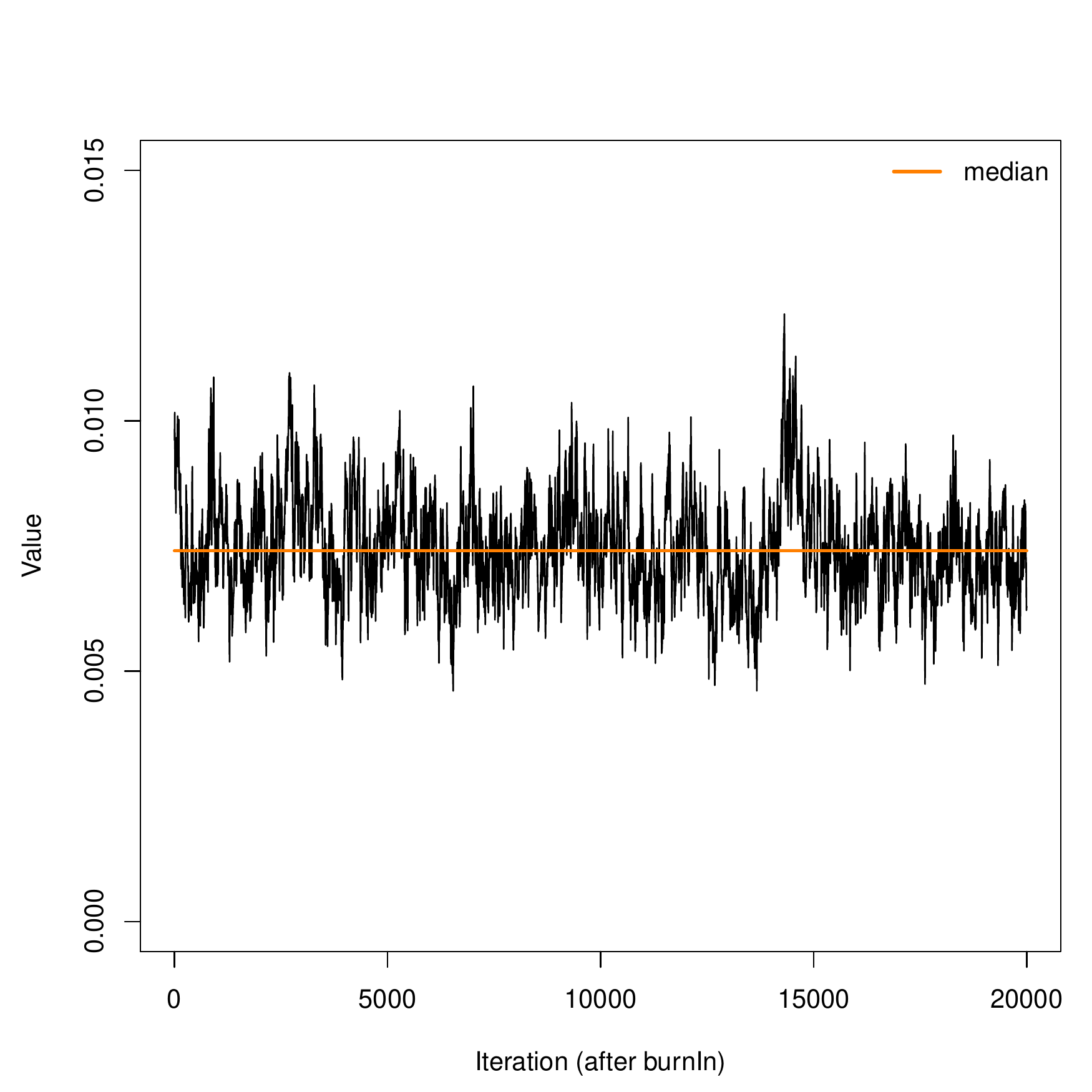}}\quad
 \subfloat[][\emph{Glucodistylin}.]
 {\includegraphics[width=7cm,height=7cm,]{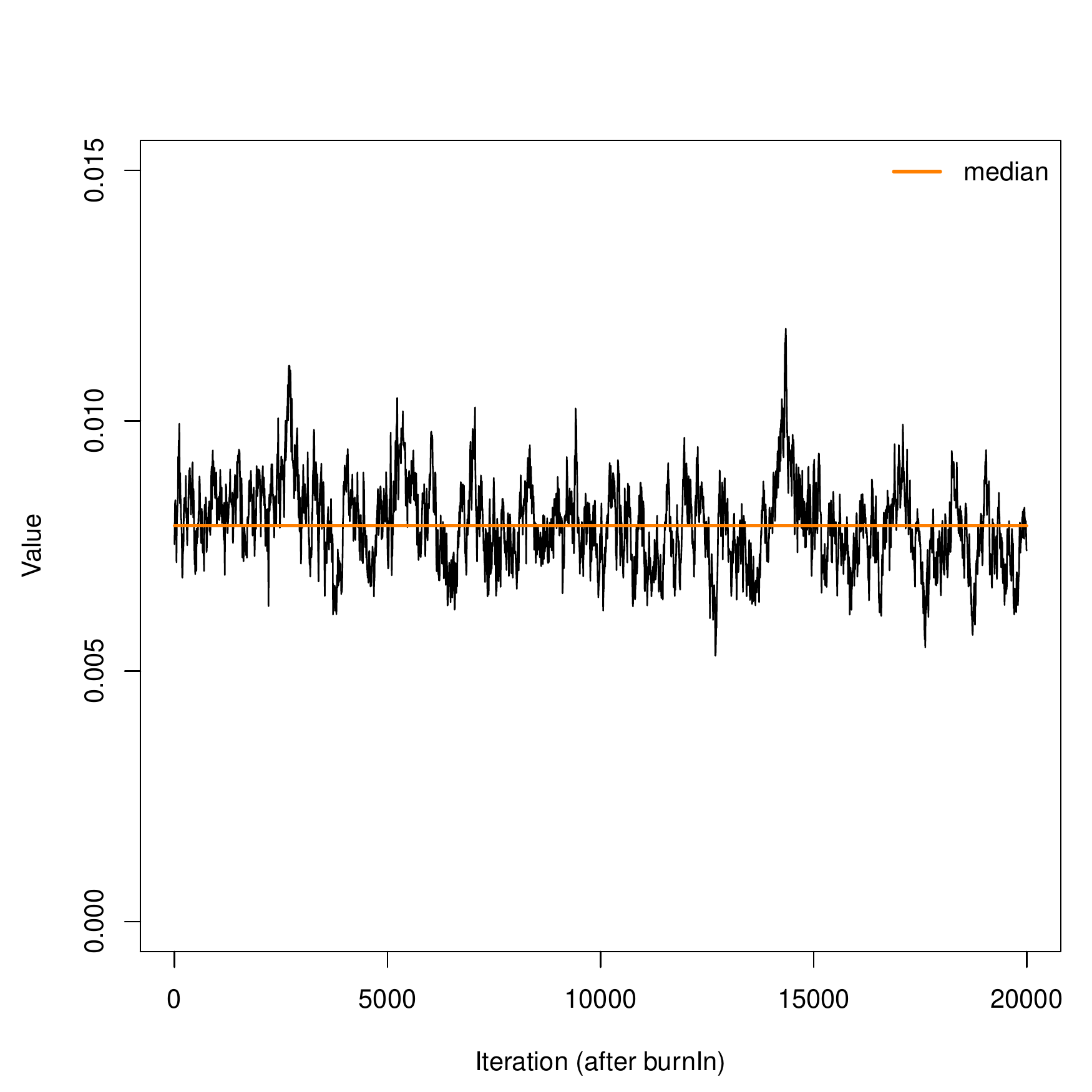}}
 \caption{Apple data. Trace plots for the estimated $\beta_p$ parameters, $p=1,\dots,P$.}
  \label{fig:applesdata1-extra4}
\end{figure}

Table \ref{tab:ESSApp} reports a summary of the ESS (Expected Sample Size) values for the model parameters (30000 MCMC iterations considered). Figures \ref{fig:applesdata1-extra1} and \ref{fig:applesdata1-extra2} report, respectively, the estimated posterior distributions for the $\alpha_p$ and $\beta_p$ parameters, $p=1,\dots,P$, in the apple intake data. In addition, Figures \ref{fig:applesdata1-extra3} and \ref{fig:applesdata1-extra4} report, respectively, the trace plots for the estimated $\alpha_p$ and $\beta_p$ parameters, $p=1,\dots,P$, in the apple intake data. 

\begin{table}[!h]
\begin{center}
   \caption{Apple data. Summary of the ESS (Expected Sample Size) values for the model parameters ($30000$ MCMC iterations considered).}
   \label{tab:ESSApp}
 \begin{tabular}{|ccccccccccccc|}
  \toprule
   & &  \multicolumn{10}{c}{Parameter} & \\
 & & $\mu_{\alpha}$ & $\mu_{\beta}$ & $\sigma_{\beta}^2$ &$\alpha$ & $\beta$ & $\sigma_p^2$ & $\theta_d^2$ &  $z$ & $\gamma$ & $\eta$ &\\
\midrule
min    & & 21673 & 17584 & 13399  & 1162  & 830  & 724 & 906 & 450 & 220&  2014 & \\
median && 24162 & 22262 & 21464 & 3308  & 1192 & 7213 & 2170 & 1159 & 870 &  2219 & \\
Max    && 27129 & 25890 & 24935 & 6607 & 2941 & 22843 & 15717 & 29174 & 1119 &  2312& \\
\bottomrule
  \end{tabular}
  \end{center}
\end{table}

\end{document}